\documentclass[fleqn,usenatbib]{mnras}

\usepackage{newtxtext,newtxmath}
\usepackage{subcaption}

\usepackage[T1]{fontenc}
\usepackage{pdflscape}
\DeclareRobustCommand{\VAN}[3]{#2}
\let\VANthebibliography\thebibliography
\def\thebibliography{\DeclareRobustCommand{\VAN}[3]{##3}\VANthebibliography}

\usepackage{graphicx}	
\usepackage{amsmath}	
\usepackage{orcidlink}

\usepackage[T1]{fontenc}

\title[UV Study of Star Forming Dwarf Galaxies]{A Comparative Study of Star Forming Dwarf Galaxies using the UVIT}

\author[S Amrutha]{
S Amrutha\orcidlink{0009-0005-6072-9252},$^{1,2}$\thanks{E-mail:amrutha.s@iiap.res.in, amrutharao99@gmail.com}
Mousumi Das\orcidlink{0000-0001-8996-6474},$^{1}$
Jyoti Yadav\orcidlink{0000-0002-5641-8102},$^{1,2}$
\\
$^{1}$Indian Institute of Astrophysics, Koramangala II Block, Bangalore 560034, India\\
$^{2}$Pondicherry University, R.V. Nagar, Kalapet, 605014, Puducherry, India\\
}

\date{Accepted XXX. Received YYY; in original form ZZZ}

\pubyear{2023}

\begin{document}
\label{firstpage}
\pagerange{\pageref{firstpage}--\pageref{lastpage}}
\maketitle

\begin{abstract}
We present a Far-Ultraviolet (FUV) study of sixteen star-forming dwarf galaxies (SFDGs) using the Ultraviolet Imaging Telescope (UVIT). Morphologically, SFDGs are classified as dwarf spirals, dwarf irregulars, and blue compact dwarfs (BCDs). We extracted the star-forming complexes (SFCs) from the sample galaxies, derived their sizes, and estimated the FUV+24$\mu m$ star formation rates (SFRs). We also determined the approximate stellar disk mass associated with the SFCs using IRAC 3.6-micron images. We derived the specific SFRs (sSFRs), as well as the SFR densities ($\Sigma(SFR)$) for the SFCs. We find that the lower $\Sigma(SFR)$ for each type is different, with the dwarf irregulars having the lowest $\Sigma(SFR)$ compared to others. However, the median size of the SFCs in the dwarf irregulars is the largest compared to the other two types when compared at roughly the same distance. We have derived the star-forming main sequence (SFMS) on the scale of SFCs for all three classes of SFDGs. We find that although all SFDGs approximately follow the global SFMS relation, i.e. $SFR\propto {M_{*}}^{\alpha}$ (where globally $\alpha\approx1$ for low surface brightness galaxies and $0.9$ for SFDGs), on the scale of SFCs the $\alpha$ value for each type is different. The $\alpha$ values for dwarf spirals, dwarf irregulars, and BCDs are found to be 0.74\(\pm\)0.13, 0.87\(\pm\)0.16, and 0.80\(\pm\)0.19, respectively. However, the age of all SFCs approximately corresponds to 1 Gyr. Finally, we find that the outer SFCs in most galaxies except BCDs have a high sSFR, supporting the inside-out model of galaxy growth.

\end{abstract}

\begin{keywords}
galaxies: dwarf  -- galaxies: star formation  -- ultraviolet: galaxies 
\end{keywords}



\section{Introduction}

Dwarf galaxies are low mass (\(M_{*}<10^{10}\ M_\odot\)) galaxies that have smaller sizes and lower dynamical masses compared to normal galaxies on the Hubble sequence \citep{lelli.2022}. Due to their low stellar masses, they usually have shallower disk potentials. Hence, dwarfs often lose a significant amount of metal content via ejection by stellar winds and supernova bursts \citep{Bergvall_2011}, making them often metal-poor and gas-poor \citep{mac-low_ferrara.1999}. Most dwarfs do not appear to have molecular hydrogen ($H_2$) gas as they are not detected in CO emission \citep{Leroy_2005,das.etal.2006,schruba.etal.2012}. They also have thick/flared disks, resulting in a low mid-plane pressure in the disk gas. Hence, they are fairly extreme environments for star formation as it is difficult for gas to cool and even be retained within the disk \citep{elmegreen.hunter.2015}. 

It is well known that dwarf galaxies dominate the galaxy population in the universe \citep{loveday.1997}. The $\Lambda$CDM cold dark matter models and cosmological simulations of the large-scale structure show that they are important for the hierarchical growth of galaxies through star formation and mergers \citep{1978MNRAS.183..341W, deason.etal.2014,wheeler.etal.2019}. Due to their abundance, most mergers at all redshifts are between dwarf galaxies \citep{2006MNRAS.366..499D}. The nearest example of such a merger is the infall of the Large Magellanic Cloud (LMC) and Small Magellanic Cloud (SMC) into our Galaxy \citep{Nidever.etal.2017}. The two galaxies are connected by a gaseous bridge composed of HI complexes, and the entire structure is called the Magellanic Stream \citep{donghia.fox.2016}. Such studies of nearby dwarfs in our Local Group, combined with simulations (as in the TiNy Titans (TNT) program), help to understand the redistribution of stellar and gaseous material in interacting/merging dwarf galaxies \citep{Besla_2010,2015ApJ...805....2S}. 

\begin{figure}
\includegraphics[width=\columnwidth]{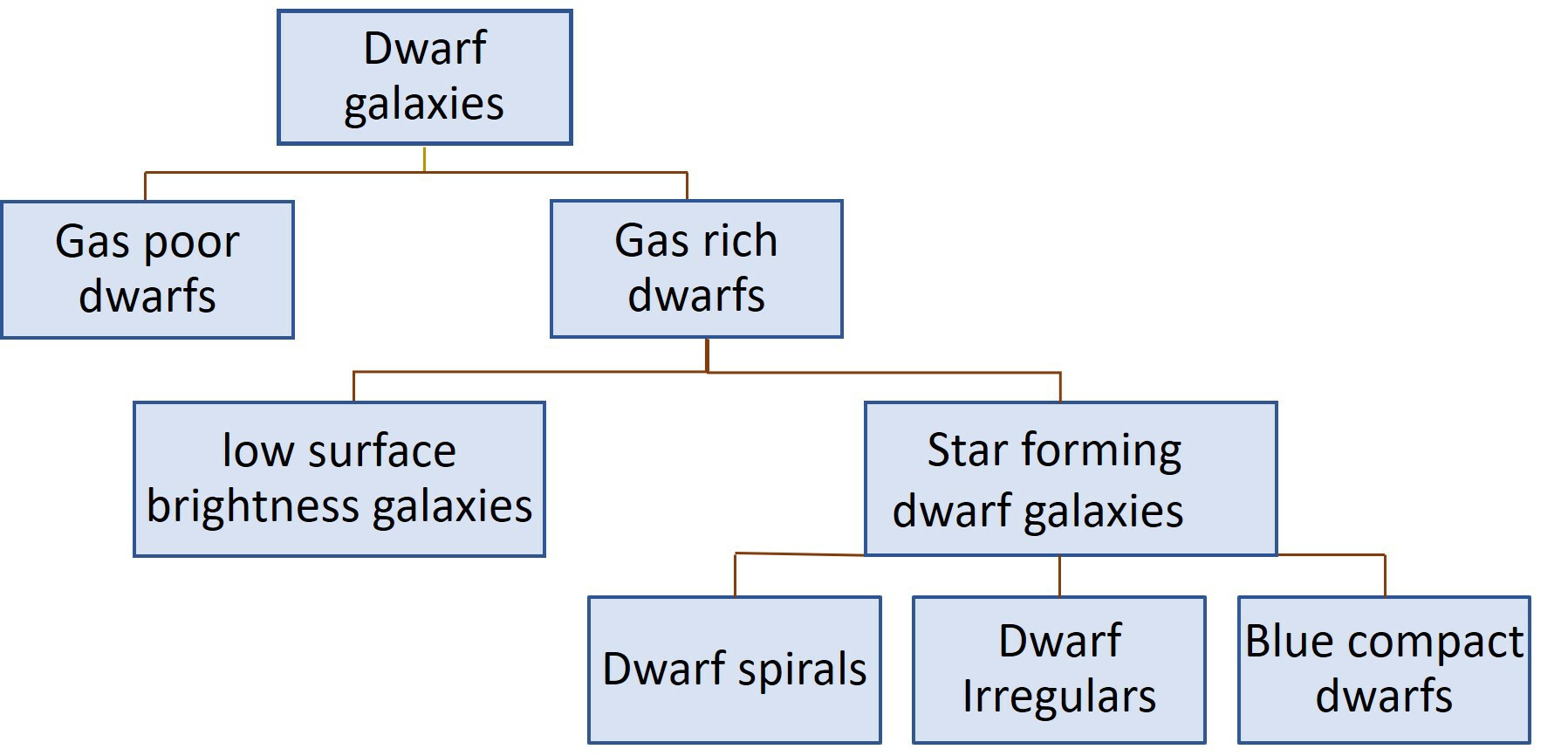}
\caption{Classification of dwarf galaxies based on the presence of gas and star formation.
\label{fig1}}
\end{figure}

Surveys over the past two decades show that dwarf galaxies are not only abundant, but they also span a wide variety of galaxy types. Broadly speaking, dwarfs can be classified into gas-poor and gas-rich based on the presence of gas, as shown in Figure \ref{fig1}. The gas-poor dwarf galaxies consist of early-type dwarf spheroidals, dwarf ellipticals, and gas-free ultra-diffuse or ultra-faint galaxies (UDGs). Most of these dwarf galaxies are satellites of giant host galaxies and are very faint. Their low luminosity makes them hard to detect, and it is only recently that they have been detected in large numbers \citep{koda.etal.2015}. There are many theories explaining why satellite galaxies are gas-poor, but the most probable explanation is gas stripping due to interaction with massive galaxies \citep{putman.etal.2021}. 

A large fraction of the gas-rich dwarfs are low surface brightness (LSB) galaxies. Such galaxies have low luminosities, diffuse stellar disks, and high dark matter content \citep{honey.etal.2018}. LSB dwarfs include both disky and irregular dwarfs. Their HI masses are very high, and some of them have HI gas mass fractions that are close to 100\% of their dynamical mass \citep{10.1111/j.1365-2966.2008.13150.x,Schombert_2001}. They have very low star formation rates (SFRs) and consume gas at a very slow pace \citep{2009ApJ...692.1305L}. However, studies show that a significant fraction of LSB dwarfs have some star formation and follow a star-forming main sequence \citep{2007ApJS..173..538T,McGaugh_2017}.  

Although dwarfs present extreme environments for star formation, about 70\% of them are star-forming \citep{2004AJ....127.2031K} and are called star-forming dwarf galaxies (SFDGs). Their small sizes, slow rotational velocities, and hence low shear, as well as overall low metallicities, result in star formation being very distinct from that observed in regular disk galaxies \citep{Bergvall_2011}.  Feedback and the environment of SFDGs also contribute to the differences \citep{1995AJ....110.2665M}. 

SFDGs can be classified into dwarf spirals, dwarf irregulars, and blue compact dwarfs. The dwarf spirals often have spiral arms and small bars or oval bulges (e.g. NGC4136), which are common in normal disk galaxies and are important for triggering star formation \citep{das.etal.2019}. However, the spiral arms are not as well-defined as in regular spirals, probably because the stellar disk is not dense enough. Or, as in some cases, the spiral arms can be due to tidal interactions or dwarf-dwarf mergers \citep{stierwalt.etal.2015}. On the other hand, dwarf irregulars have a sub-threshold environment for star formation, as they generally do not have bars or spiral arms nor the critical gas density to form stars \citep{Melena_2009}. Star formation is distributed over their disks and is due to local disk instabilities \citep{2010AJ....139..447H}.
The third class of SFDGs are the blue compact dwarfs (BCDs), which have small, compact star-forming inner disks and diffuse LSB outer disks. The BCDs tend to form superclusters \citep{8224322,1995AJ....110.2665M}. Although the origin of the starbursts in BCDs is unknown, one of the main possibilities is that they are merger-driven \citep{2015ApJ...805....2S}. These galaxies are very metal-poor, which gives them a blue appearance. 

The main tracers of star formation in galaxies are H\(\alpha\) and Ultraviolet (UV) emission. The H\(\alpha\) is emitted by hot O-type stars, and the emission lasts for 1 to 10 Myr. Far UV (FUV) and near UV (NUV) emission arise from O and B type stars, and the emission lasts for a longer timescale (\(\leq 100-200 Myr\)) compared to H\(\alpha\) \citep{Boissier_2007,koda.etal.2012}. In many galaxies, the UV emission extends far beyond the H\(\alpha\) disk emission \citep{das.etal.2019}, the extreme examples being the extended UV (XUV galaxies) \citep{2007ApJS..173..538T}. Sensitive observations can trace the  H\(\alpha\) and UV emission from the star-forming XUV disks of massive glaxies \citep{das.etal.2021,Yadav_2021}, as well as from nearby dwarf galaxies \citep{lee.etal.2009}. 

In this study, we use UV to trace recent massive star formation in nearby SFDGs. We use the Ultra Violet Imaging Telescope (UVIT) observations of 16 nearby SFDGs to extract and characterize their star-forming regions. The UVIT has a spatial resolution advantage over the previous space UV telescope GALEX. It has a spatial resolution of \(\approx 1.2"\) with a plate scale of 0.41 arcsec per pixel. Previously, \cite{Melena_2009} have used GALEX images to characterize the star-forming regions of nearby dwarf irregulars. But as shown by \cite{2021JApA...42...50M}, the UVIT can extract many more regions than GALEX, giving a better and more accurate idea of the size and distribution of star forming complexes (SFCs). Our study uses FUV filters and NUV filters for only a few galaxies, as NUV has not been in operation since 2018.

In the following sections, we present our sample of SFDGs and compare their star forming properties by extracting the SFCs. In section \ref{sec:sample selection}, we explain the criteria for choosing our sample, and section \ref{observations} provides information about the data. In section \ref{sec:Data Analysis}, we discuss the data analysis, including image analysis, source extraction, UV photometry, finding the total star formation rate using UV, and 24 \(\mu m\) MIPS images, and modelling the mass of the SFCs using 3.6 \(\mu m\) MIPS images. Section \ref{sec:Results} describes our results, including the observed radial trends in \ref{subsec: Observed Radial trends of SFCs} and the star-forming main sequence for SFCs in \ref{subsec:Star-forming main sequence for SFCs }. In the Discussion section \ref{sec:Discussion}, we discuss our results, including how interactions affect SFDGs and how their star formation main sequence compares with massive galaxies.  

\begin{figure*}
\includegraphics[scale=0.8]{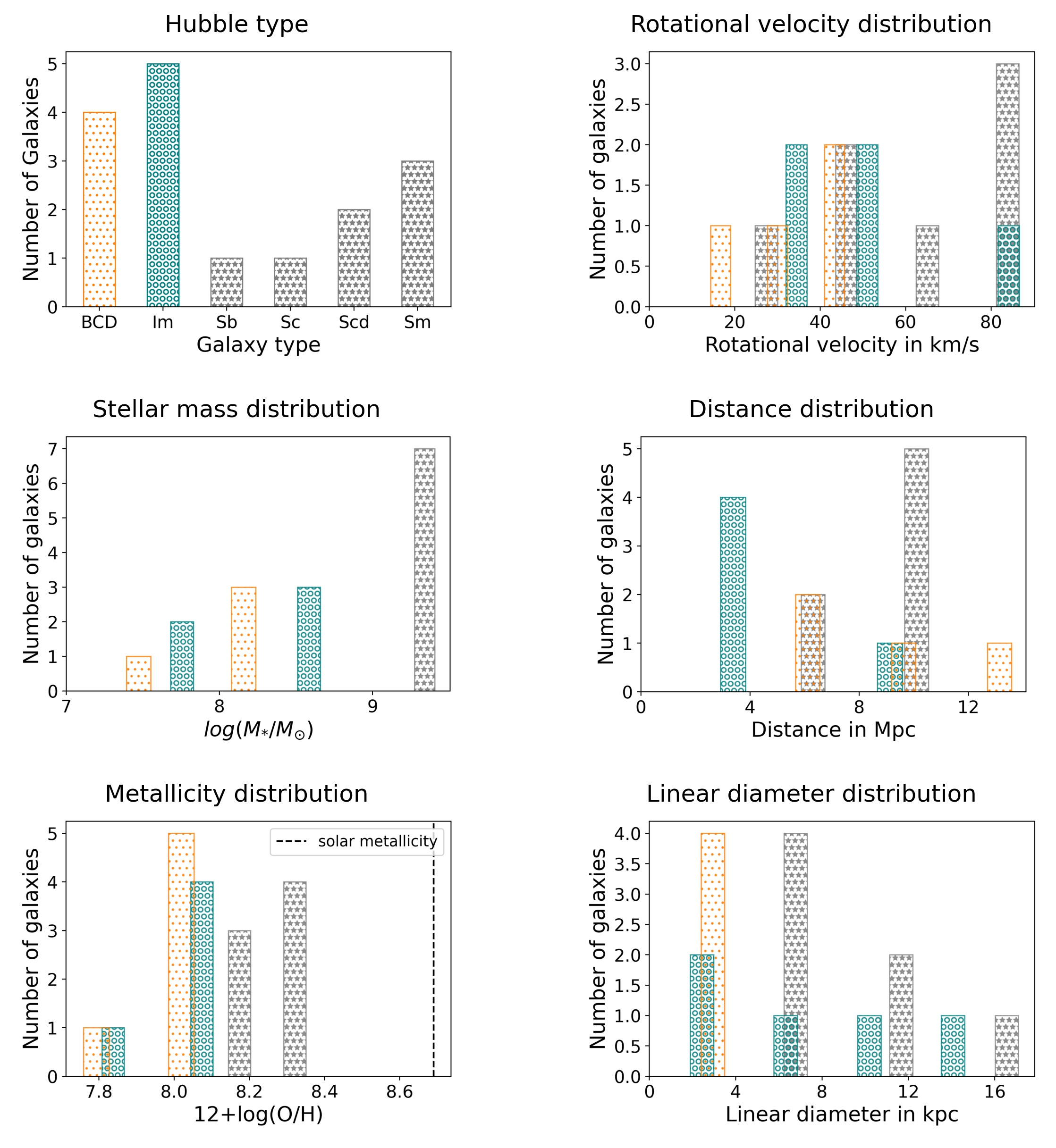}
\caption{Properties of the sample galaxies. The grey colour with '*' as a pattern represents dwarf spirals, the teal with the pattern 'o' represents dwarf irregulars, and the orange with pattern dots represents BCDs. In these histograms, we see that the dwarf spirals have relatively higher mass and metallicity compared to other types, But they all have similar rotational velocities. All galaxies have subsolar metallicities.}
\label{fig2}
\end{figure*}

\newcounter{mpFootnoteValueSaver}
    \setcounter{mpFootnoteValueSaver}{\value{footnote}} 
\begin{table*}





    \centering
    \caption{Properties of the sample galaxies.
    \label{table1}}
        \begin{tabular}{ccccccccccc}
    \hline
    Galaxy   & Type & D & Method& Vrot & i& P.A. & \(r_{25}\) & E(B-V)& \(log(SFR_{FUV})\) & \(log(M_{HI})\) \\ 
     & & (Mpc)&  & (\(kms^{-1}\)) & (\textdegree) & (\textdegree) & (')& (mag) &\(M_{\odot}yr^{-1}\) & \(M_{\odot}\)\\
    \hline
    NGC 4136 & SAB(r)c & 10.9 & virgo infall & 93.3 & 0 & 179 & 1.17 & 0.016&-0.80&8.98\\
    NGC 4618 & SB(rs)m & 9.43 & TRGB & 67.6 & 39 & 25 & 1.74 & 0.018&-0.51&9.11\\
    NGC 4625 & SAB(rs)m pec & 9.43 & TRGB & 47.4 & 21 & 90 & 0.79 & 0.016&-1.19&8.68 \\
    NGC 5474 & SA(s)cd pec & 6.8 & TGRB & 22.1 & 35 & 90 & 1.2 & 0.01&-0.67&8.96 \\
    NGC 5832 & SB(rs)b & 11 & Virgo infall & 74.5 & 47 & 45 & 0.93 & 0.022&-0.96&8.88 \\
    NGC 4395 & SA(s)m & 4.41 & TGRB & 57.4 & 35.6 & 147 & 6.7 & 0.015&-0.49&- \\
    NGC 2541 & SA(s)cd & 12 & Cepheids & 93.1 & 64.4 & 169 & 1.51 & 0.043&0.09&9.39 \\
    \hline
    UGC 7608 & Im & 10.64 & virgo infall & 46.6 & 28.9 & 90 & 1.73 & 0.015&-1.22&-\\
    WLM & IB(s)m & 0.98 & TGRB & 26.2 & 70 & 1 & 5.75 & 0.031&-2.24&7.85 \\
    UGC 4305 & Im & 3.37 & TGRB & 35.1 & 52.6 & 15 & 3.95 & 0.027&-1.06&8.85 \\
    NGC 6822 & IB(s)m & 0.5 & TGRB & 92.4 & 68 & 10 & 7.74 & 0.203&-1.82&8.11 \\
    IC 2574 & Irregular & 3.9 & TRGB  & 46.9 & 53.1 & 50 & 6.44 & 0.031&-0.78&9.12 \\
    \hline
    NGC 3738 & BCD & 5.1 & TGRB & 52.3 & 31 & 155 & 1.3 & 0.009&-1.33&8.06\\
    VIIZw403 & BCD & 4.35 & TGRB  & 14.8 & 55.8 & 14 & 0.75 & 0.032&-1.94&7.69 \\
    Haro 36 & BCD & 10.53 & virgo infall & 51.7 & 43.7 & 105 & 0.65 & 0.013&-1.41&8.16 \\
    UGCA 130 & BCD & 14.9 & Virgo infall & 35.2 & 47 & 8 & 0.3 & 0.074&-1.65&7.86 \\
    \hline
    
	\end{tabular}
       \begin{minipage}{160mm}
        \textit{Note:} Distance (D), Method and $r_{25}$ (radius of galaxy at 25 B-mag $arcsec^{-2}$) are from Hyperleda \footnotemark[1] and NED \footnotemark[2].\\
        Rotational velocities (Vrot) are from \cite{Leroy_2005} and Hyperleda.\\
        Inclination (i) and Position angle (P.A.) are from NED and Hyperleda, respectively.\\
        Galactic reddening E(B-V) is calibrated using \cite{2011ApJ...737..103S} and taken from IRSA \footnotemark[3].\\
        $\log(SFR_{FUV})$ is FUV star formation rate as listed in \cite{2013AJ....146...46K} for all galaxies, except NGC 5832 and UGCA130, for which it is taken from \cite{2009ApJ...706..599L}, and \cite{L_pez_S_nchez_2010} respectively.\\ $M_{HI}$ is neutral hydrogen mass taken from \cite{2007ApJS..173..538T} for spirals and \cite{Hunter_2012} for the irregulars and BCDs except NGC 6822 and UGCA 130, for which it is taken from \cite{de_Blok_2000} and \cite{L_pez_S_nchez_2010} respectively.
        \end{minipage}%
      
\end{table*}

\section{Sample Selection} \label{sec:sample selection}
Our study aims to understand star formation in dwarf galaxies by resolving their SFCs. Our definition of dwarf galaxies is derived from the study of \cite{Leroy_2005}, where dwarfs have a stellar mass of \(M_{*}<10^{10} M_\odot \), a typical physical diameter of approximately \( 10 kpc\),  and flat rotation velocities of \(V_{rot} \leq 100 kms^{-1}\). To resolve the SFCs, we need high-resolution UV observations of nearby galaxies. For example, galaxies at distances $<20$Mpc are suitable because SFCs of sizes $\sim 100$pc in their disks can be resolved. All our sample dwarf galaxies are star-forming and have UVIT data. The sample includes both isolated and interacting galaxies. We proposed some of the UVIT observations, and some were obtained from archival data. In the following paragraphs, we describe our sample, which was mainly derived from the surveys of \cite{Leroy_2005} and \cite{Hunter_2012}. All of them follow the chosen criteria, and the distribution of their properties is shown in Figure \ref{fig2}.

\subsection{Dwarf Spirals}\label{subsec:Dwarf Spirals}
We have seven star-forming dwarf spirals in our sample, and they are all selected from \cite{Leroy_2005}. These are the most massive dwarfs with a mass range of $10^{9}\ M{_\odot}<M_{*}<10^{10} M_{\odot}$, the most massive being NGC 4618 with a mass of \( M_{*}=10^{9.56}\ M_\odot \) (see Table \ref{table5}). Most of these dwarfs are late-type spiral galaxies, and their metallicity is higher than the other two classes. Some are fairly extended. NGC 4395 has the largest diameter (\(D_{25}=16.9 kpc\)) in our sample. Also, most of our dwarf spirals are classified as Extended Ultraviolet galaxies \citep{2007ApJS..173..538T}. Some of them are also interacting, such as NGC 4625 and NGC 4618, which are interacting with each other, and with NGC 4625A that lies between them \citep{2007ApJS..173..185G}. The galaxy NGC 5474 is probably interacting with the giant spiral M101 \citep{2013ApJ...762...82M, 2020A&A...634A.124B}. For the other 4 spirals, we do not find any sign of interaction.

\subsection{Dwarf Irregulars} \label{subsec:Dwarf Irregulars }
We have 5-star-forming dwarf irregulars in our sample. Most of these are from the Local Irregulars That Trace Luminosity Extremes, The HI Nearby Galaxy Survey (LITTLE THINGS \citep{Hunter_2012}). Besides this, we also included UGC 7608  from \cite{2021MNRAS.505.3998S}. Dwarf irregulars have stellar masses  \(\leq 10^{9}\ M_\odot\) and are generally smaller than dwarf spirals. The largest in this group is IC 2574, which has a diameter of \(D_{25}=14.61 kpc\); this is high compared to the typical sizes of dwarf galaxies. The irregulars are at closer distances compared to the other two types of dwarfs. For example, NGC 6822 and WLM are at distances of 0.5 Mpc and 0.98 Mpc, respectively. NGC 6822 is one of the nearest dwarf galaxies around the Milky Way, and it may have interacted with the Milky Way as well as with other dwarf galaxies \citep{Zhang_2021}. The irregular galaxy UGC 4305 is part of the M81 group and is infalling towards M81 \citep{2002AJ....123.1316B}.  

\subsection{Blue Compact Dwarfs} \label{subsec:Blue Compact Dwarfs}
We have 4 Blue Compact Dwarfs (BCDs). NGC 3738, VIIZw403, and Haro 36 are from LITTLE THINGS. We included UGCA 130 from \cite{Leroy_2005}. BCDs have smaller diameters compared to dwarf spirals and irregulars. In our sample, their diameters are less than \(1.5'\). Hence, they appear to be very compact and very bright in UV. The BCD UGCA 130 is the farthest galaxy in our sample. We do not see any sign of interactions in the BCD sample.

\section{Observations} \label{observations}

\subsection{UV Data} \label{subsec:UV Data}

We performed deep UV imaging observations of the galaxies NGC 7608, NGC 4136, NGC 4625, NGC 4618, NGC 2541, and NGC 5832 using the UVIT on board the AstroSat telescope \citep{2012SPIE.8443E..1NK}. For all other galaxies, we used archival UVIT data. The details of the UVIT filter and exposure time of galaxies are given in Table \ref{table2}. The Ultraviolet imaging telescope (UVIT) is an instrument with twin telescopes and coaligned Ritchey-Chrétien (RC) optics. One of the telescopes is in the FUV (1300–1800Å) waveband, and the other is in the near-UV (NUV; 2000–3000Å) and visible (VIS) bands. The instrument can simultaneously observe in all three bands with a field of view of 0.5\textdegree. The drift correction is done using the VIS channel or else by using the NUV imager. The FUV and NUV telescopes work in photon counting mode, whereas the VIS channel works in the integrated mode. Hence, the FUV and NUV telescopes have sub-pixel resolutions of \(\approx 1.2"\), which is three times better than the Galaxy Evolution Explorer (GALEX) \citep{2005ApJ...619L...1M}. The FUV and NUV telescopes also have multiple narrow-band photometric filters \citep{rahna.etal.2018}. Since the year 2018, the UVIT NUV filter stopped working due to payload-related issues. Hence, only a few galaxies have NUV images.

\subsection{Infra-red Data}
The UV radiation emitted by the young stars is absorbed by interstellar dust and re-emitted in the infrared (IR) band, mainly around 24 \(\mu\)m.  Hence,  we used  24\(\mu\)m images from the Multiband Imaging Photometer for Spitzer (MIPS) instrument in our study \citep{2004ApJS..154...25R}. The resolution of this 24\(\mu\)m image is \( 6"\). The data reduction procedure of the MIPS image is mentioned in \cite{2005PASP..117..503G}. All the images we used here are reduced archival images. We used background-subtracted images for our analysis.

Emission from older stellar photospheres is studied using 3.6\(\mu\)m wavelength images. As the stellar mass of a galaxy is mainly due to an older population, we used the 3.6\(\mu\)m luminosity to calculate the stellar mass density of the galaxies. The 3.6-micron images were from one of the four cameras, the Infrared Array Camera (IRAC) cameras on Spitzer \citep{2004ApJS..154...10F}. It has an angular resolution of \( 2"\) and has a field of view of \( 5'\). It has a good sensitivity compared to the other three cameras in the instrument.

    \stepcounter{mpFootnoteValueSaver}%
    \footnotetext[\value{mpFootnoteValueSaver}]{%
     https://leda.univ-lyon1.fr/}%
  \stepcounter{mpFootnoteValueSaver}%
    \footnotetext[\value{mpFootnoteValueSaver}]{%
      https://ned.ipac.caltech.edu/}
        \stepcounter{mpFootnoteValueSaver}%
    \footnotetext[\value{mpFootnoteValueSaver}]{%
      https://irsa.ipac.caltech.edu/applications/DUST/}

\begin{table*}
    \centering
        \caption{UVIT Filter and Image Information.}
            \label{table2}
            \begin{tabular}{ccccccccc}
        \hline
        No & Galaxy & Other names & FUV filter & Exptime & PSF & NUV filter & Exptime & PSF \\ 
         &  &  &  & (s) &  &  & (s)& \\
        
        \hline
        
        1 & NGC 4136 & UGC7134 & CaF2 & 3640.447 & 1 & - & - & -\\
        2 & NGC 4618 & IC3667; UGC7853 & CaF2 & 5893.002 & 1 & Silica15 & 1576.123 & 1.2 \\
         &  & ARP23; VV73 &  &  &  &  &  &  \\
        3 & NGC 4625 & IC3675; UGC7861 & CaF2 & 5893.002 & 1 & Silica15 & 1576.123 & 1.2 \\
        4 & NGC 5474 & UGC 09013; VV 344b & BaF2 & 1500.408 & 1 & NUVB4 & 763.004 & 1.1 \\
        5 & NGC 5832 & UGC 09649 & CaF2 & 1976.753 & 1.1 & NUVB13 & 1691.195 & 1.3 \\
        6 & NGC 4395 & UGC 07524 & CaF2 & 709.967 & 1.1 &NUVB15&6386.25 &1.2 \\
        7 & NGC 2541 & UGC 04284 & BaF2 & 2660.16 & 1.1 &NUVB15&1759.857& 1.2\\
        \hline
        8 & UGC 7608 & DDO 129 & CaF2 & 7543.797 & 1 & - & - & - \\
        9 & WLM & UGCA 444; DDO 221 & CaF2 & 5078.986 & 1.2 & Silica15 & 2706.837 & 1.1 \\
        10 & UGC 4305 & Holmberg II & CaF2 & 18229.966 & 1.1 & NUVB13 &  9607.65  & 1.1\\
         &  & ARP 268; DDO 050 &  &  &  &  &  &  \\
        11 & NGC 6822 & IC 4895; DDO 209 & CaF2 & 4085.606 & 1.1 & Silica15 & 315 & 0.9 \\
        12 & IC 2574 & UGC 05666 & CaF2 & 9871.635 & 1.1 &            -                            &           -                  &       - \\
         &  & DDO 081; VII Zw 330 &  &  &  &  &  &  \\
         \hline
        13 & NGC 3738 & UGC 06565; ARP 234 & BaF2 & 2659.343 & 1 & - & - & - \\
        14 & VIIZw403 & UGC 06456; VV 574 & BaF2 & 2649.006 & 1 & Silica15 & 2792.765 & 1 \\
        15 & haro 36 & UGC 07950 & BaF2 & 1584.729 & 1 & - & - & - \\
        16 & UGCA130 & MRK 0005 & CaF2 & 4617.873 & 1 & - & - & - \\
        \hline
            \end{tabular}
            \begin{minipage}{130mm}
            \textit{Note}: Other names are as listed in NED. Only few galaxies have UVIT NUV filter information as it has stopped working since 2018.
            \end{minipage}
\end{table*}

\section{Data Analysis} \label{sec:Data Analysis}

\subsection{UV data reduction} \label{subsec:UV data Reduction}

The UVIT level 1 data is raw data and was downloaded from the Indian Space Science Data Centre (ISSDC) website for all the sample galaxies. We used CCDLAB \citep{2017PASP..129k5002P}, a graphical user interface, to reduce the data and produce science-ready images. It corrects for field distortions, flat-fielding, and drift. CCD lab also combines and aligns the orbit-wise UVIT images to create a final deep image. We did the astrometry on these images using a tool in CCDLAB that can match GAIA DR3 sources with UVIT sources and do the astrometry. Then, these images were background subtracted using IRAF \citep{1993ASPC...52..173T} by finding the counts of a few regions around the galaxy and then subtracting the mean of these counts from the scientific image.

\subsection{Source Extraction} \label{subsec:Source Extraction}

Using Python, we cut out the image of the galaxies from the entire field of view of UVIT. Then we used the SExtractor \citep{1996A&AS..117..393B, 2016JOSS....1...58B} to extract the SFCs from the cutout image. SExtractor (Source Extractor) is one of the most commonly used command-line programs to perform multiple tasks, like background estimation, source detection, deblending, and even aperture photometry. Since we already did the background subtraction, we estimated the global background noise (\(\sigma\)) from the image. We set the detection threshold to a few times \(\sigma\)  such that the pixels with values higher than the given threshold are detected as objects. Then we gave a deblending threshold and deblending ratio so that the detected objects are deblended and seen as separate sources. We had set the detection threshold to 8\(\sigma\) and minimum area to 10 \(pixels\) as recommended by \cite{1996A&AS..117..393B}  for clean and confident detection. SExtractor fits an ellipse around the source, giving us the position, position angle, and size parameters of the SFCs. 

While detecting sources, we may even detect some foreground sources. To eliminate such sources, we found bright sources around a galaxy centre in Gaia, within a radius comparable to the field of view of the cutout image. These foreground sources have parallax and high proper motion. We matched the RA and Dec of the SFCs detected with SExtractor with the RA and Dec of the sources obtained from the Gaia catalogue using TOPCAT \citep{2005ASPC..347...29T}. If the matched sources had high parallax and high proper motion, we masked those sources using Python. 

We found the deprojected area of the remaining SFCs, which is given by
\begin{equation} \label{eq1}
area=\frac{\pi \times a \times b}{ cos(i)}
\end{equation}
with a and b as the semi-major and semi-minor axes of an ellipse fitted to the SFCs, respectively, and 'i' is the angle of inclination of the galaxy. We found the approximate size or radius of the SFCs by equating this area to the area of a circle.

\begin{figure}
\includegraphics[width=0.9\columnwidth]{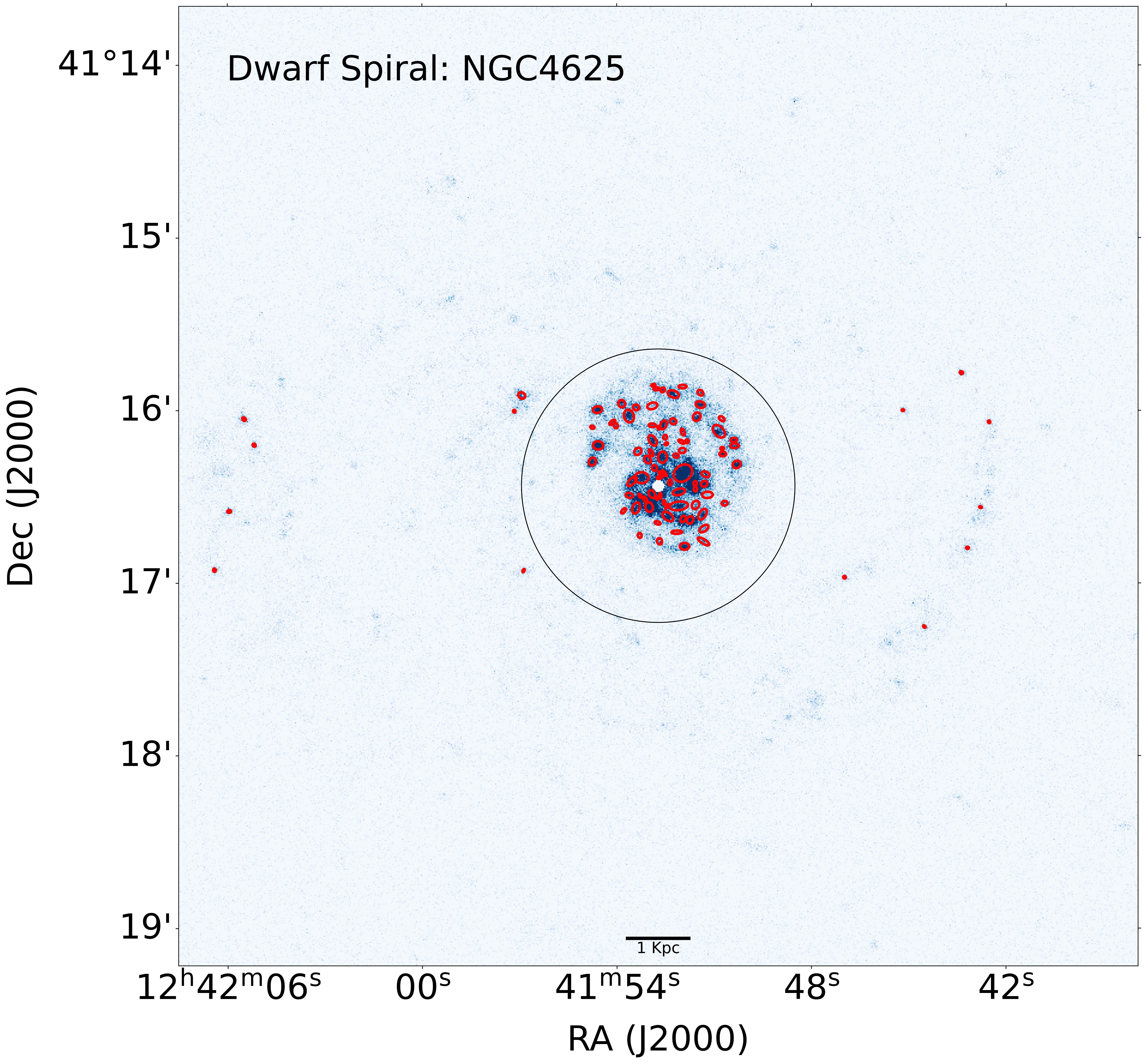}
\caption{UVIT image of NGC 4625. The red ellipses are fitted to sources using SExtractor \citep{1996A&AS..117..393B}. The detection threshold is 8\(\sigma\). Black circles have a radius of \(r_{25}\), which is mentioned for this galaxy in Table \ref{table1}.}
\label{fig3}
\end{figure}

\subsection{UV Photometry} \label{subsec:UV Photometry}

We performed elliptical aperture photometry on the SFCs using the Python astropy package Photutils \citep{2013A&A...558A..33A, 2018AJ....156..123A}  and obtained the counts per second CPS from the SFCs. We found the Galactic extinction corrected apparent AB magnitude \(m_{AB}\) for each SFC using the formula,
\begin{equation} \label{eq2}
 m_{AB}=-2.5 \times log10(CPS) + ZP-A_{FUV}
 \end{equation}
Where ZP is zero point magnitude, and \(A_{FUV}\) is FUV Extinction, which is calculated using Galactic reddening E(B-V) \citep{2011ApJ...737..103S} from IRSA and using extinction laws \citep{2012ivoa.rept.1015R}. We found the corrected CPS with this corrected magnitude. We calculated Flux (\(ergs^{-1}cm^{-2}Hz^{-1}\)) from the corrected CPS using the formula,
\begin{equation} \label{eq3}
 F_\nu=\frac{\lambda^{2}}{c} \times UC \times CPS_{corr}
 \end{equation}
 where \(\lambda\) is the mean wavelength of the FUV filter, c is the speed of light, UC is the unit conversion factor in \(ergs^{-1}cm^{-2}A^{-1}\) for the filter used as given in \cite{2017JApA...38...28T}, and \(CPS_{corr}\) is corrected CPS for the source.

We calculated the extinction-corrected star formation rate (SFR) using the FUV luminosities in \(ergs^{-1}Hz^{-1}\) and the ratio of 24\(\mu\)m Flux to FUV flux, using the formula from \cite{2008AJ....136.2782L}. This is explained in detail in the next section. For most galaxies, we obtained the FUV fluxes using the CaF2 filter, which is similar to GALEX. The SFR calculations apply to the CaF2 filter. However, for five galaxies, the filter used is BaF2. We mentioned filter information in Table \ref{table2}. Hence, we calibrated the Flux in the BaF2 filter with the GALEX observations using five stars in the field of the galaxy.

\subsection{The Host galaxy extinction and Total Star Formation Rate }\label{subsec: The Host galaxy extinction and Total Star Formation Rate }

\begin{figure}
\includegraphics[width=0.9 \columnwidth]{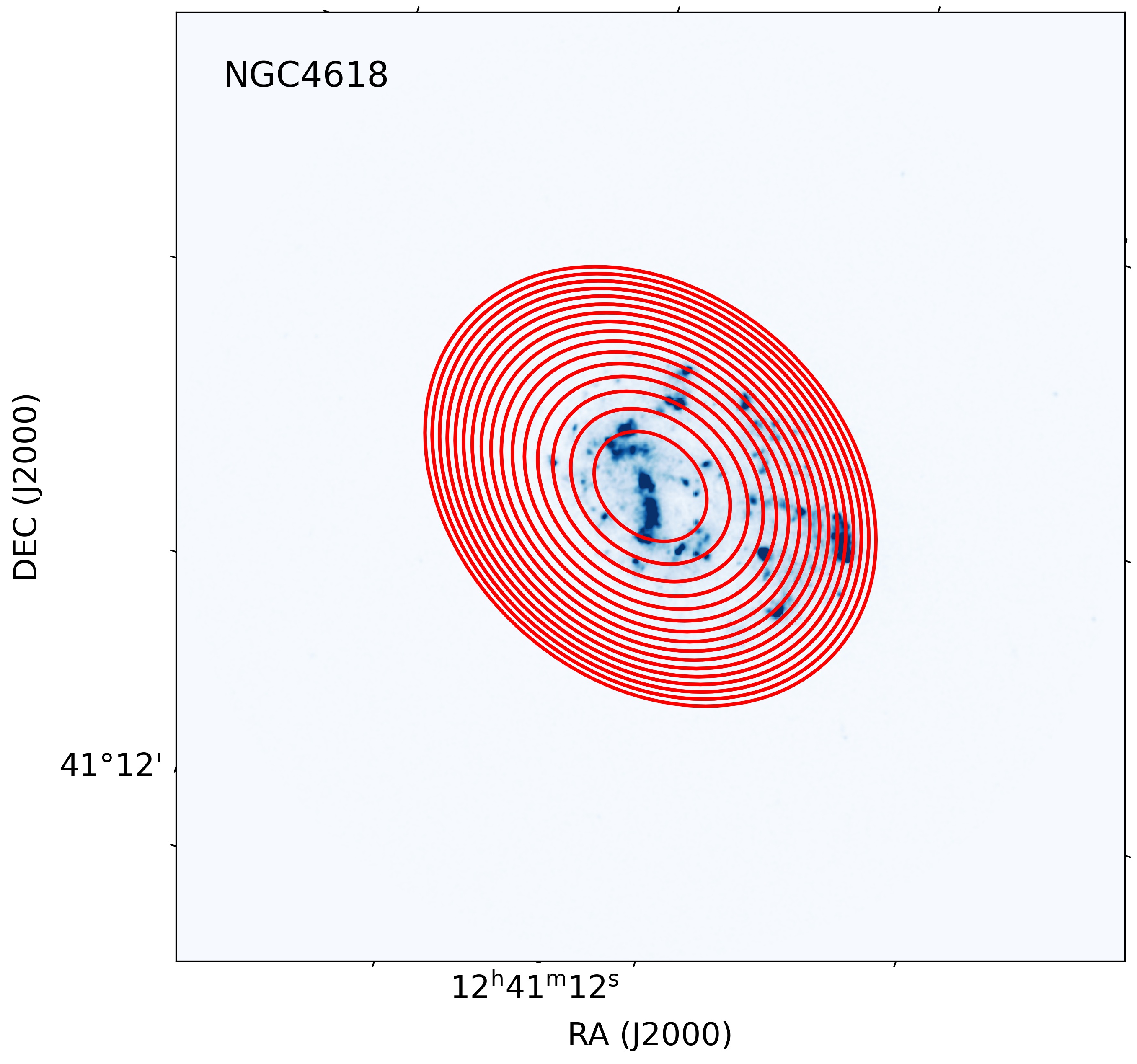}
\caption{Elliptical annuli of equal area are fitted to the convolved UVIT image of NGC 4618, with the position angle as listed in Table \ref{table1}.
\label{fig5}}
\end{figure}

We calculated the extinction corrected total SFR using 24\(\mu\)m MIPS observations. As the  UVIT resolution is almost five times greater than 24\(\mu\)m MIPS, we convolved the UVIT image to 24\(\mu\)m MIPS image resolution. Then we fitted elliptical annuli to the galaxy's 24\(\mu\)m MIPS image and convolved UVIT image, such that the annular regions' area should be the same as the inner ellipse, as shown in Figure \ref{fig5}. The outermost ellipse has a semimajor axis value equal to the distance of the farthest SFC, as seen in the UVIT image. Then, we performed aperture photometry on the 24\(\mu\)m images using Photutils and obtained the flux in \(MJysr^{-1}\) and converted it to Jy. Similarly, We performed aperture photometry on the convolved UVIT images and followed the same procedure as we followed on 24\(\mu\)m images to obtain the CPS. As mentioned in the previous section, we calculated the UV Flux in Jy from the CPS.

We calculated the IR flux corresponding to one Jy of UV flux for each annulus by dividing the integrated 24\(\mu\)m flux by the integrated UV flux obtained from that annulus. We denote the ratio as \(\alpha\), and we plotted the variation of \(\alpha(r)\) for each galaxy  (see Figure \ref{alpha}). Since we know the deprojected galactocentric distance of the SFCs from the galaxy centre in the UVIT image, we assumed the corresponding radial distance from the plot to determine \(\alpha(r)\). In this way, we associated  \(\alpha\) with all the SFCs in the galaxy.

We found the total SFR for each SFC using the formula mentioned in \cite{2008AJ....136.2782L}, where the FUV SFR is taken from \cite{Salim_2007} for sub-solar metallicities (0.8 $Z_{0}$) and considered the Chabrier Initial Mass Function (IMF) \citep{2003PASP..115..763C}.
\begin{equation} \label{eq4}
SFR_{Total}=(0.68 \times 10^{-28}+\alpha \nu_{24}  2.14 \times 10^{-42})L_{FUV} 
\end{equation}
We thus obtain the SFR in \(M_\odot yr^{-1}\), where \(L_{FUV}\) is the Galactic extinction corrected FUV luminosity in \(ergs^{-1}Hz^{-1}\), and \(\nu_{24}= 1.25 \times 10^{13}\)Hz. We calculated the SFR density \(\Sigma(SFR)\) by dividing \(SFR_{Total}\) by the deprojected area.

In previous studies like \cite{Belfiore_2023, Leroy_2023}, it is mentioned that the mid-IR flux also comes from the local conditions in a galaxy disk, like reflection heating due to sources other than star formation. However, \cite{Leroy_2023} showed that 22 \(\mu\)m or 24 \(\mu\)m is more associated with star-forming regions than any other MIR bands. But the conversion factor ($C_{24}$), which converts UV or H$\alpha$ flux to 24 \(\mu\)m flux, is higher for the whole galaxy when compared to intense star-forming regions. Hence, the equation \ref{eq4} slightly overestimates the total SFR for the SFCs as it averages the dust from molecular gas, diffuse emissions and emission from star-forming regions.

Figure \ref{alpha} shows that the radial variation of $\alpha(r)$ is significant (Red dashed line). For most of the dwarf spirals, $\alpha(r)$ is radially decreasing, with some showing higher values at the end. For irregulars, there is no particular trend. Surprisingly, BCDs show a radially increasing trend. The $\alpha$ considered for the whole galaxy (black dashed line) is in the range of varying $\alpha(r)$. For most of the galaxies, it falls almost at the mean value. But for some, it is either at the high or low end of varying $\alpha(r)$ (See NGC 6822). We used convolved UV and IR images to plot the variation of $\alpha(r)$, which has a resolution of 6". Hence, there is a resolution difference as SFCs have UVIT resolution ($\approx$ 1.2"), which is another factor that adds to the uncertainty. Although considering $\alpha(r)$ for SFCs by this method may not be accurate, it shows how $\alpha(r)$ varies for each type of SFDG and gives approximate $\alpha$ values. 

To find the global extinction-free SFR of the galaxies, we took integrated FUV flux and integrated the 24\(\mu\)m flux from the convolved UVIT images and 24\(\mu\)m MIPS images, respectively, within the outermost ellipse around the galaxies. Then, we obtained the $\alpha$ value from those fluxes. We have tabulated the fluxes along with $\alpha$ for each galaxy in Table \ref{table A2}.

We compared our global SFRs from Table \ref{table5} with the tabulated global SFRs and stellar masses of \cite{Leroy_2019}. Although our methods overlap, there is a significant difference in total SFR values. We see that our values are much higher than \cite{Leroy_2019}. The main reason is the differences in the aperture sizes to obtain the integrated luminosities. \cite{Leroy_2019} have considered the optical radius $r_{25}$ as the semimajor axis for the elliptical aperture around galaxies, which is much less than ours. Since there are galaxies with extended disks (and XUV disks), considering $r_{25}$ as the semimajor axis will significantly decrease the total SFR of the galaxies. Also, it should be noted that for some galaxies, the distances considered are slightly different.  

\subsection{Mean Disk Mass of the SFCs} \label{subsec: Mass of SFCs }
We found the underlying disk masses in the areas covered by the SFCs using 3.6 IRAC images. We masked all foreground sources using GAIA data. We used the same method of fitting elliptical annuli to the galaxy as in the subsection \ref{subsec: The Host galaxy extinction and Total Star Formation Rate }.

We determined the 3.6\(\mu\)m flux from each annulus. We converted the flux to AB absolute magnitude using distance and then to luminosity. We calculated disk mass \( M_{*}\) using 3.6 luminosity \(L_{*}\) in \(L_{\odot}\), with mass to light ratio \(\Upsilon_{3.6}\) in \(M_{\odot}L_{\odot}^{-1}\) using relation,
\begin{equation} \label{eq5}
 M_{*}=\Upsilon_{3.6}L_{*}
\end{equation}

Mass to light ratio \(\Upsilon_{3.6}\) is the relationship between the mass and luminosity of a disk galaxy. It depends on the FUV-NUV colour for dwarf galaxies. We took approximate \(\Upsilon_{3.6}\) from \cite{Schombert_2022} based on the colour of the galaxy. We took \(\Upsilon_{3.6}=0.4 M_{\odot}L_{\odot}^{-1}\) for galaxies with \(0.15<FUV-NUV <0.4\) and \(\Upsilon_{3.6}=0.35 M_{\odot}L_{\odot}^{-1}\) for galaxies with \(FUV-NUV <0.15\).

\begin{figure}
\includegraphics[width=0.9 \columnwidth]{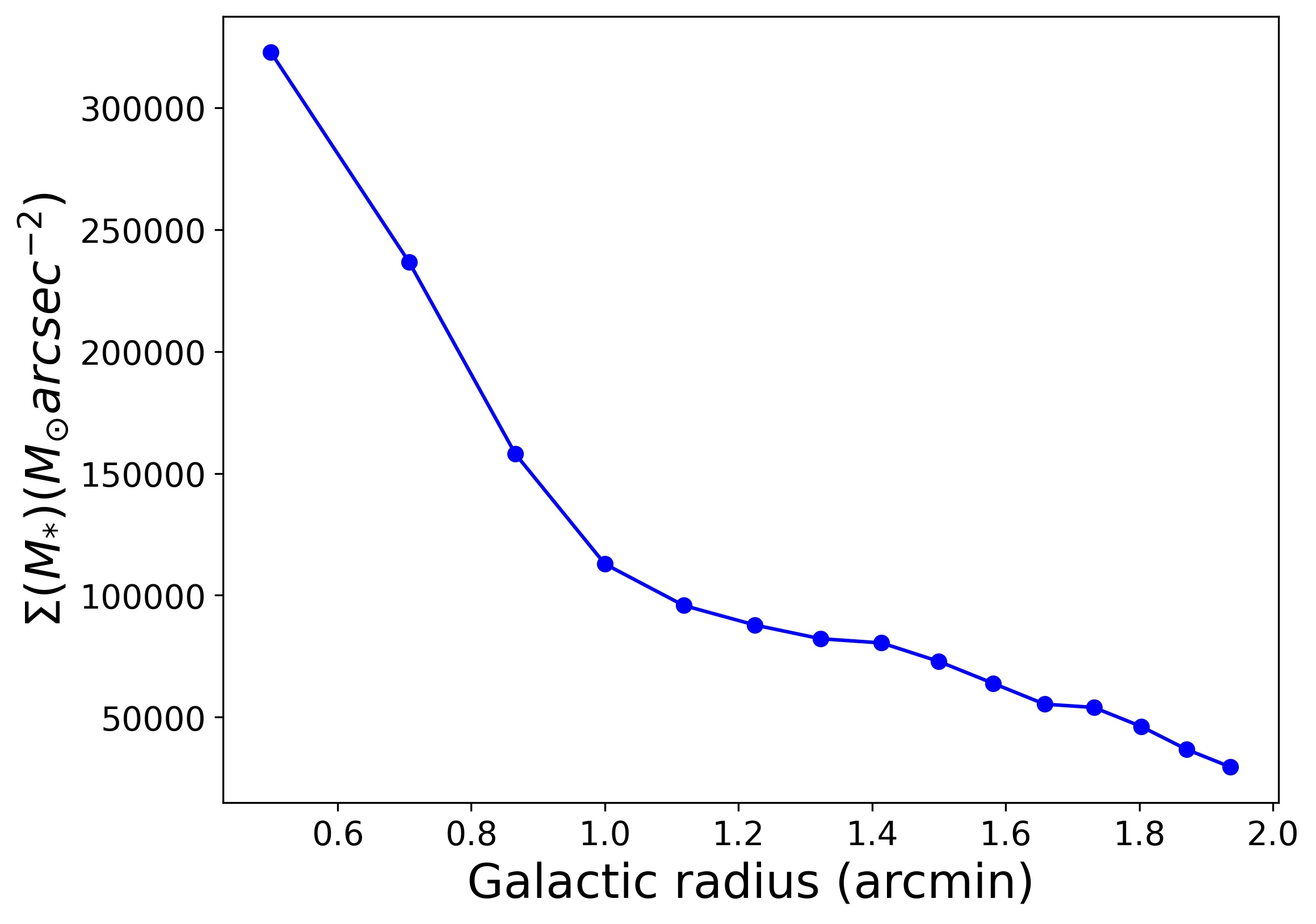}
\caption{Radial variation of Stellar mass density for galaxy NGC 4618. From this, we associated the stellar mass density with the SFCs by looking at their galactocentric distance. 
\label{fig6}}
\end{figure}

We derived the stellar mass density \(\Sigma(M_{*})\) by dividing the stellar disk mass by the area of the bounding ellipse. Then, we plotted the radial variation of \(\Sigma(M_{*})\) for each galaxy as shown in Figure \ref{fig6}. Then we associated  \(\Sigma(M_{*})\) with all the SFCs in the galaxy, as explained in the previous section. To find the mean disk mass, we multiplied \(\Sigma(M_{*})\) with its area. We used the total 3.6-micron luminosity within the outermost ellipse to calculate the total stellar mass of the galaxies.

We compared the total stellar mass of the galaxies from Table \ref{table5} with the corresponding values in \cite{Leroy_2019}. Our estimated mass is higher than theirs. Some of the reasons are mentioned in subsection \ref{subsec: The Host galaxy extinction and Total Star Formation Rate } where we compare the global SFRs. Another important reason is \cite{Leroy_2019} have considered the value of \(\Upsilon_{3.6}\) from 0.2 to 0.5 \(M_{\odot}L_{\odot}^{-1}\), and some of our sample galaxies have \(\Upsilon_{3.6}\) value lesser than 0.35 \(M_{\odot}L_{\odot}^{-1}\) and the least value we have taken is 0.35 \(M_{\odot}L_{\odot}^{-1}\). Hence, this also affects the estimated mass value.

\section{Results} \label{sec:Results} 

\subsection{ Morphology of Dwarfs in UV emissions } \label{subsec: Morphology of Dwarfs in UV emissions }
To understand star formation in dwarfs, it is important to compare their UV morphology with the stellar disk at other wavelengths, such as near-infrared (NIR), which traces the old stellar population and hence the main stellar mass distribution, and the optical B band which traces the overall stellar light distribution  from star formation. The FUV, NUV, NIR, and Optical B or g band images of our sample galaxies are shown in Appendix \ref{sec:appendix}. As mentioned earlier, young massive stars emit in the UV band, which can be traced even in distant galaxies \citep{2011Ap&SS.335...51B}. So, by comparing the FUV and NUV emission with the older stellar population traced in NIR and the intermediate stellar population traced in the optical band, we can understand how star formation has propagated in these galaxies. 

\textit{Dwarf Spirals:} We see strong UV emission from the spiral arms of the dwarf spirals, which appear to be very flocculent in nature. However, disk emission in the 3.6 \(\mu\)m NIR band images is poor for most of them, and no well-defined spiral arms are detected. We can see some emission in the optical B or g band images, but it is generally not as extended or bright as the FUV and NUV emission. Hence, these galaxies are forming stars in regions where the stellar density is low \citep{2007ApJS..173..538T}. Overall, the NIR emission is strong only in the central disk, bar, and bulge regions for some dwarf spirals.

\textit{NGC 4136:} This galaxy is nearly face-on and has multiple flocculent arms seen in the UV and B band images. The arms are more extended and clear in the UV images, especially the southern arm. The galaxy also has a compact oval-shaped bulge \citep{2002ApJS..143...73E}. There is no UV emission associated with the short, strong bar, which can be seen in the NIR image.  But there is UV emission at the bar ends, and it is loosely associated with the spiral arms since not all the arms are exactly coming from the ends of the bar. This galaxy has a strong inner ring visible in UV emission and a co-rotation ring around a weak bar that is visible in both the B and UV band images. The spiral arms appear to be arising from this ring. There is a bright star-forming nucleus at the centre.

\textit{NGC 4618:} This galaxy appears very compact in UV. It shows strong UV emission from the bar and has a single spiral arm extending towards the south. The bar is brighter in the NIR and B band images compared to UV, but the spiral arm appears surprisingly faint. The optical centre is slightly offset from the kinematical centre due to the presence of the strong southern spiral arm \citep{2008MNRAS.388..500E}. Although NGC 4618 is classified as a single-armed galaxy, there is a small arm-like structure that can be seen as faint FUV emission extended towards the north and perhaps associated with the bar end. This arm also appears as a faint emission in NUV. The interaction with NGC 4625 could be the reason for the high star formation rate in the bar and the southern arm, and the arm probably formed due to the tidal interaction.

\textit{NGC 4625:} NGC 4625 shows very bright FUV and NUV emission concentrated towards its centre. This region is also brighter in the NIR and B bands, indicating that the stellar mass is also concentrated in the centre. We see a ring structure in the centre due to the tightly wound spiral arms and also a small bar. The most striking feature in this galaxy is the very extended spiral arms that are bright in FUV and NUV emission and can be traced well beyond the central bright disk. The multiple faint arms extend out to several kiloparsecs around the galaxy but are surprisingly not visible in the NIR image and barely detected in the g band image. The \(r_{25}\) of this galaxy is 0.79' as per literature. But the UV extends out to 3.2$^{\prime}$, which is 4 times more than the \(r_{25}\) radius of the galaxy. This was first noted in the GALEX survey by \cite{2007ApJS..173..185G}, which classified it as a Type 1 XUV galaxy. We see some bright regions in the faint arm of this galaxy, and the extended arms seen in UV could be due to the interaction with NGC 4618 and NGC 4625A.

\textit{NGC 5474:} This galaxy has a peculiar shape due to its highly asymmetric structure. It has a bright central region, which can be seen in UV and the other bands. There are multiple spiral arms arising from the centre, and they all appear distorted in UV emission. To the west and southeast of the galaxy centre, some arms in UV are not connected to the centre and appear to be arising from the other arms. Such bifurcation in spiral arms has been observed in grand design spiral structures \citep{rahna.etal.2018}. In the northern region, we see an arc-like structure. All these peculiar structures are due to interaction with NGC 5457, which may have caused the formation of flocculent spiral arms. We also observe bright star-forming regions or SFCs along these arms. However, in the NIR image, which traces the old stellar population, we can see only one southern arm, and in the B band, only two arms can be detected. This galaxy is classified as a mixed type XUV galaxy \citep{2007ApJS..173..538T}. 

\textit{NGC 5832:} In this galaxy, there is significant bright UV emission along the bar, but the spiral arms appear flocculent and poorly defined. The multiple arms show diffuse FUV, NUV emission, and a few bright regions  with compact UV emission. Some of the spiral arms are connected to the ends of the bar, and some are clearly not. In this galaxy, there are more compact, bright star forming regions in FUV than in NUV. However, the NUV is more extended and smoother. In NIR, we can detect a weak bar with a bright bulge in the centre. The outer disk is barely visible. Hence, this galaxy is classified as a mixed XUV galaxy. In the B band, the galaxy has an S shape with a bright bar and bulge. There are two prominent arms arising from the ends of the bar, but they become fainter at larger radii. There are also more fainter arms in the southwestern part of the galaxy. 

\textit{NGC 4395:} This is the largest galaxy in our sample, with several bright regions in UV emission. The UV-bright spiral arms are extended but very distorted in shape and do not follow the traditional spiral arm structure. There are multiple arms, at least 3 major ones, all supporting UV bright SFCs. The arm extending southeast of the centre is the brightest in UV emission. We can detect its counterpart in both the NIR and B-band images. There is an oval bulge that is prominent in NIR and hosts a Seyfert 1 type AGN \citep{2008MNRAS.388..500E}. There is an association of bright UV regions around it, and it also appears bright in the B band. This galaxy is brighter in the B band than the other galaxies, but some arms appear fainter than what we observe in UV.

\textit{NGC 2541:} This galaxy has a flocculent multi-armed structure that is bright in UV emission but does not appear to be connected to the inner disk. At least 3 arms are extended into the outer LSB disk and support many SFCs. The galaxy has a prominent bulge that is bright in the NIR image, but the outer disk is faint and similar to an LSB disk. This galaxy has been classified as a Type 2 XUV galaxy.

\textit{Dwarf irregulars:} They have very faint and diffuse NIR disks, but show very extended UV emission. We find that some irregulars have spiral arm-like features that are relatively bright in FUV and NUV emission but are not detected or are faint in NIR and B or g band images. Hence, star formation is very recent in almost all of these irregular galaxies, as their old stellar population is so faint.

\textit{UGC 7608:} A bar-like elongated structure is visible in the FUV, NUV, and B bands, but there is no sign of a bar in the NIR image. There are some distinct spiral arm-like structures in the disk that have FUV, NUV bright SFCs distributed along them. The NUV emission appears smoother and has comparatively fewer bright star forming regions compared to FUV.

\textit{UGC 4305:} This galaxy is bright in FUV and NUV emission, but the star forming regions are irregularly spread over the galaxy disk. There are four ill-defined arm-like structures in the north and west regions that could be due to clumpy local star formation resulting from the effect of some interaction in the M81 group. There is an extended distribution of faint FUV and NUV emission over the galaxy in addition to bright, compact UV arising from the SFCs. Surprisingly, there is hardly any emission in the B and NIR images, suggesting that the old stellar population is very low.

\textit{WLM:} This is the most inclined galaxy in the sample with an inclination of 70 \textdegree. As it is nearby, we can detect many small UV-bright SFCs in this galaxy. The FUV and NUV emission is spread over the disk. The galaxy is classified as barred, but we do not detect a bar in any band (FUV, NUV, B, NIR). However, we detected some disk emission in the NIR, which is also bright in the B band.

\textit{NGC 6822:} This is the nearest galaxy in the sample; hence, we can detect a larger number of small, UV-bright regions. The tidal features in the southeast region, as well as both northeast and northwest regions, are very prominent in UV, which clearly shows that there have been episodes of interactions associated with this galaxy. This galaxy also shows strong IR and B-band emissions from the central disk.

\textit{IC 2574:} The galaxy has a very irregular structure. We detect more emission in the UV and B bands. The bright bar in the northeastern region is not visible in the NIR image. A tail-like structure in the southeastern region suggests that the galaxy is tidally interacting with another galaxy. However, we do not detect any close companions.

\textit{BCDs:} All the BCDs are compact and show strong central emission in all four bands (FUV, NUV, NIR, B). We find a clear UV emission gradient from the centre to the outside BCD, as they are UV bright in the centre but not in the outer LSB disks. The BCDs do not have many SFCs, and we detect no complexes in their outer disks.

\subsection{FUV and NUV SFCs} \label{subsec: FUV and NUV complexes }

The number of FUV and NUV SFCs detected depends on the parameters given in the source extractor, such as exposure time, distance of the galaxy, and the size of the galaxy. We have taken a 8\(\sigma\) threshold for all our galaxies. The other parameters in the source extractor were chosen to deblend the sources. The galaxies with higher exposure time are brighter, and we are likely to detect more SFCs from them. Also, due to the blending effect, we detect more complexes in nearby galaxies than farther ones.
We have summarized the number and parameters of FUV and NUV sources that are detected in Table \ref{table4}. The high resolution of UVIT enables us to detect relatively small SFCs present in the low gas threshold outer regions. For example, NGC 6822, WLM, and UGC 4305 were studied by \cite{Melena_2009} using GALEX data, and they detected only 713, 165, and 139 complexes, respectively, which is approximately 78\%, 30\%, and 7.5\% of regions detected using the UVIT. 

The optical radius (\(r_{25}\)) differentiates most of the stellar disk from the outer disk. After this radius, the stellar disk surface density, and usually the gas surface density both fall rapidly in both massive galaxies and dwarfs \citep{2020ApJ...889...10D}. Hence, we expect low star formation rates (SFRs) in these regions and the properties of SFCs in the outer and inner regions to be different \citep{Yadav_2021}. All the sample galaxies except some BCDs have complexes outside the optical radius. The outer complexes are more than the inner ones in the XUV dwarf spiral NGC 5474 and in NGC 2541. Other than these dwarfs, NGC 6822 has almost the same number of complexes in both the inner and outer regions. These galaxies must have interacted recently or undergone rapid gas accretion to support such high star formation outside the optical radius. The number of outer complexes are much less than the inner ones for all other galaxies. We even find that for most of the galaxies with UVIT NUV data, the number of FUV complexes detected in these galaxies is more than the NUV complexes in both inner and outer regions, while we expect the reverse as NUV can detect complexes till 200 Myrs, whereas FUV only till 100 Myrs. This could be due to the narrower UVIT NUV bands in some cases, or because the SFCs contain stars that are very young and emit more FUV flux than NUV flux in these galaxies. It could also be a combination of both reasons. We have detected more NUV complexes than FUV ones in only the nearby galaxies WLM and UGC 4305. We have detected similar numbers of FUV and NUV complexes in the dwarf irregular UGC 4305. It has 1842 FUV complexes and 1860 NUV complexes. We should note that this galaxy has the highest exposure time (18.23 ks) in our sample.

\begin{table}
     \centering
                \caption{Number of FUV and NUV bright SFCs inside and outside optical radius \(r_{25}\).}
                    \label{table4}
                    \begin{tabular}{cccccccccc}
                \hline
                Galaxy & \(FUV_{in}\) & \(FUV_{out}\)&\(NUV_{in}\)  & \(NUV_{out}\)&  \\ 
                 &  &  & & \\
                \hline
                NGC 4136 &122& 24&-&-\\
                NGC 4618 &500&12&227&7\\
                NGC 4625 &81&11&52&0\\
                NGC 5474 &40&58&39&46 \\
                NGC 5832 &24&8&23&9 \\
                NGC 4395 &473&3&369&4 \\
                NGC 2541 &43&49&14&17\\
                UGC 7608 &56&2&-&-\\
                WLM &533&10&541&14\\
                UGC 4305&1536&306&1613&247\\
                NGC 6822 &507&411&311&348 \\
                IC 2574 &780&56&-&-\\
                NGC 3738 &28&1&-&-\\
                VIIZw403 &38&0&29&0\\
                Haro 36 &39&0&-&- \\
                UGCA 130&1&1&-&- \\
                \hline

                    \end{tabular}
                    \begin{minipage}{80mm}
                   \textit{Note}:  \(FUV_{in}\) and \(FUV_{out}\) are inner and outer FUV bright SFCs respectively. \(NUV_{in}\) and \(NUV_{out}\) are inner and outer NUV bright SFCs respectively.  
                    \end{minipage}
\end{table}

When we compared the number of complexes in NGC 5474 with the number of complexes in massive galaxies such as NGC 5457 and NGC 0628 from \citep{Yadav_2021}, which are at almost similar distances, both NGC 5457 and NGC 0628 have more complexes( 1425 and 706 complexes respectively) than NGC 5474 (98 complexes). But NGC 5474 has more outer complexes than inner complexes, whereas the other two have more inner complexes than outer. 

In contrast to dwarf spirals and irregulars, BCDs have only one or no complexes outside their optical radius. Even the number of inner complexes is much lower than the other two types. UGCA 130 is the farthest BCD in our sample, which has only two complexes.

\begin{figure*}
 \includegraphics[width=1\linewidth]{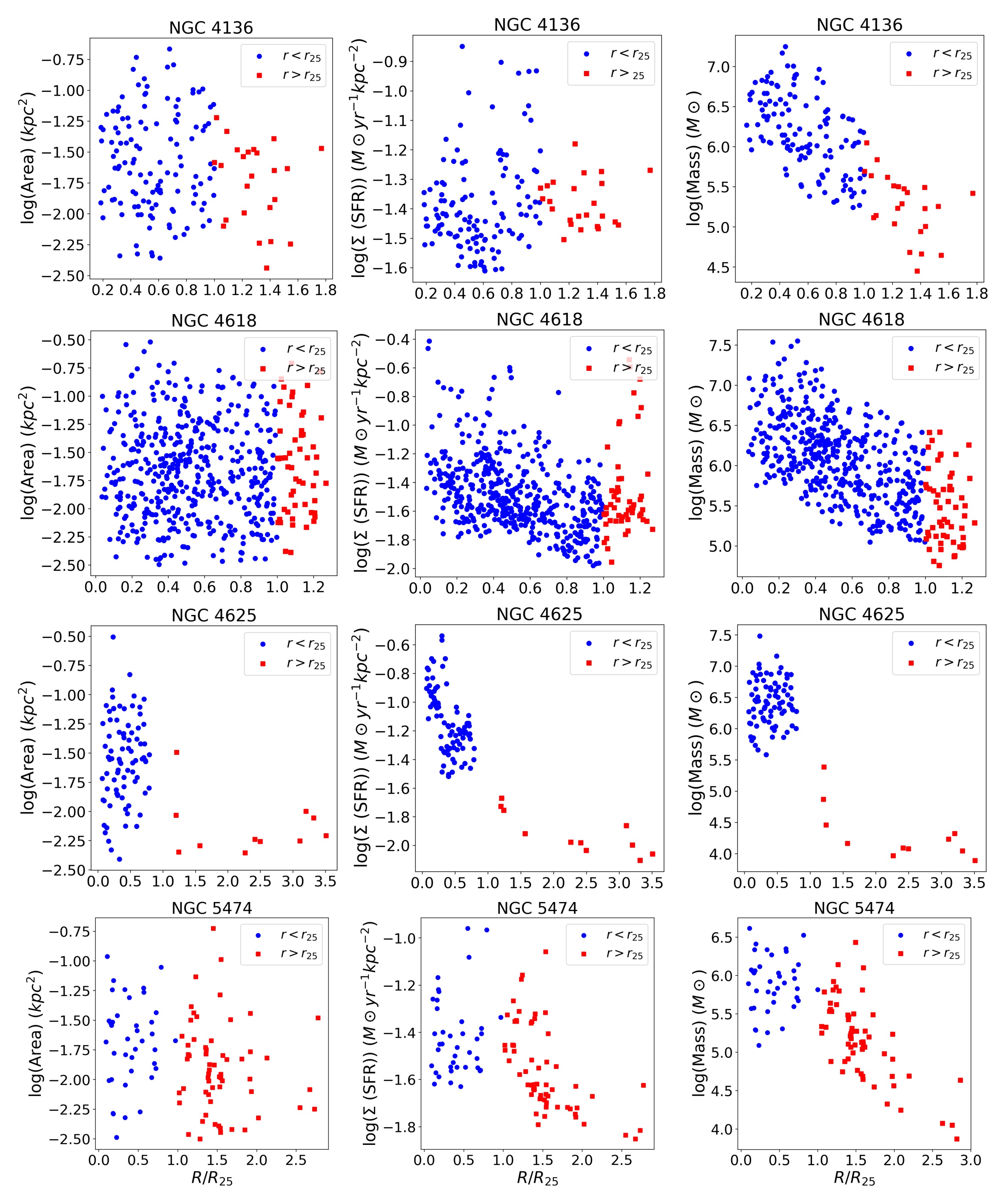} 
 \caption{First, second, and third column in Figure \ref{fig7} is a variation of area, star formation rate density, and the mass of the SFCs, respectively, with respect to the deprojected galactocentric radius scaled with the optical size of the galaxy.}
 \label{fig7}
\end{figure*}

\begin{figure*}
 \includegraphics[width=1\linewidth]{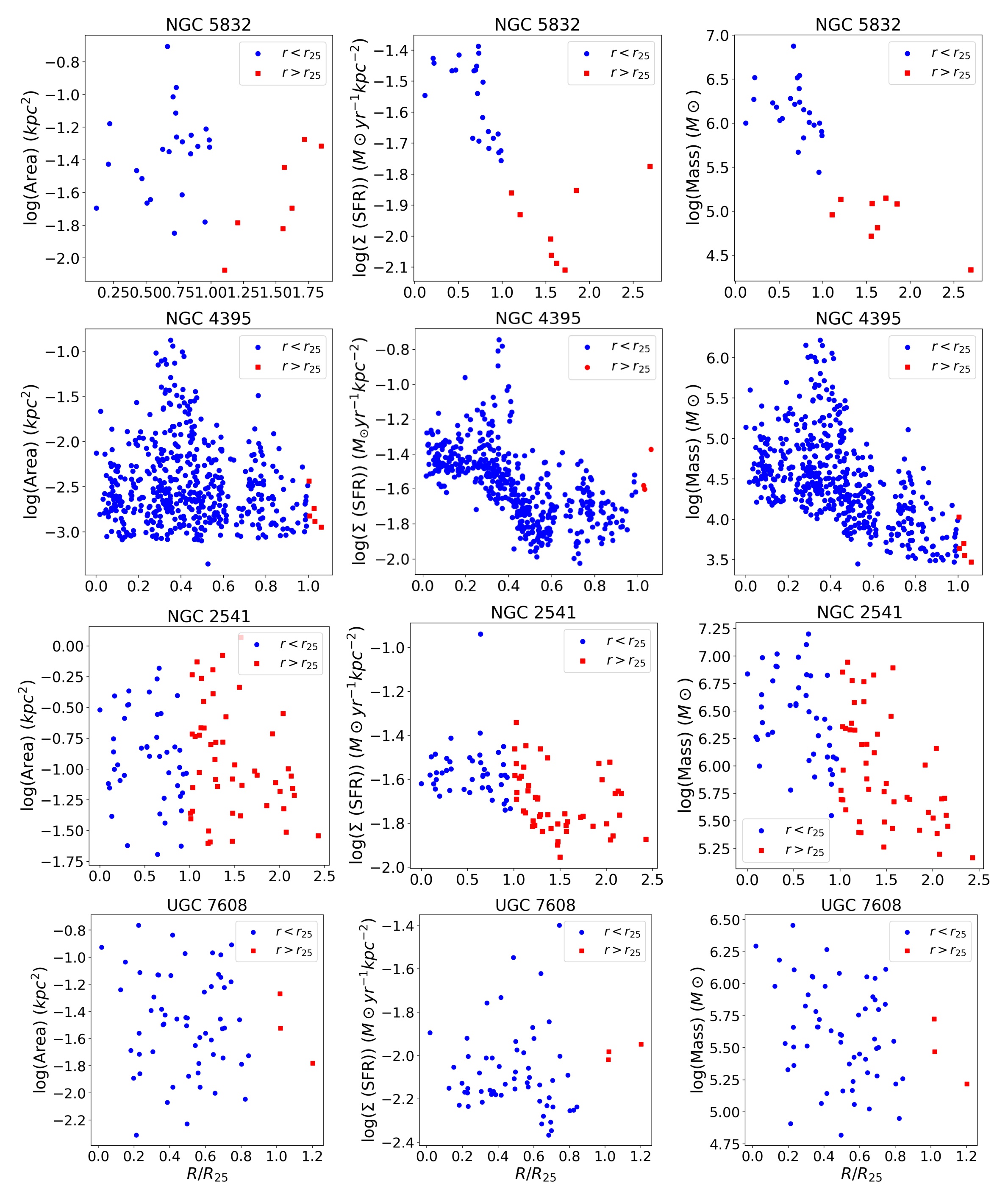} 
 \contcaption{} 
\end{figure*}
\begin{figure*}
 \includegraphics[width=1\linewidth]{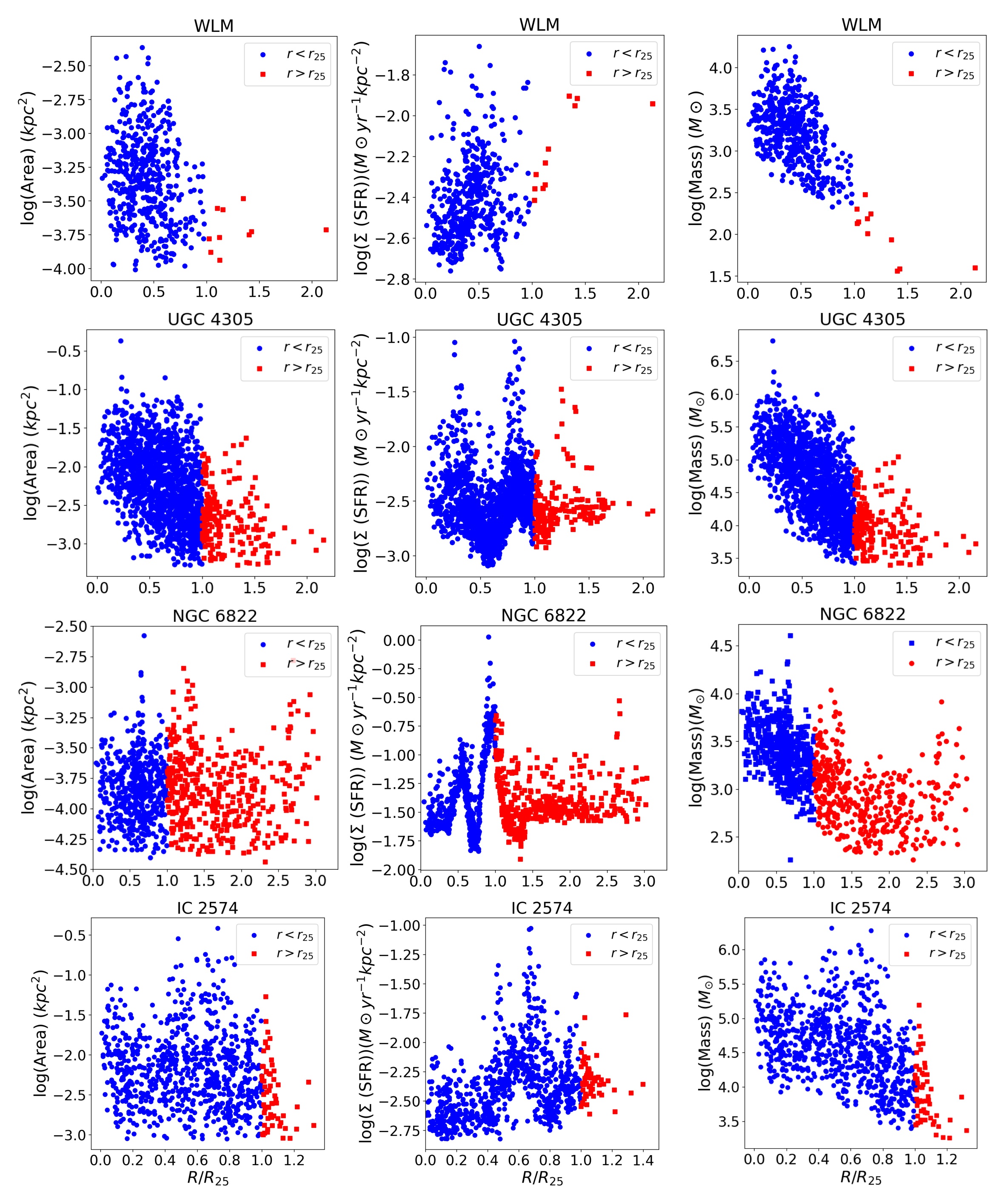} 
 \contcaption{} 
\end{figure*}
\begin{figure*}
 \includegraphics[width=1\linewidth]{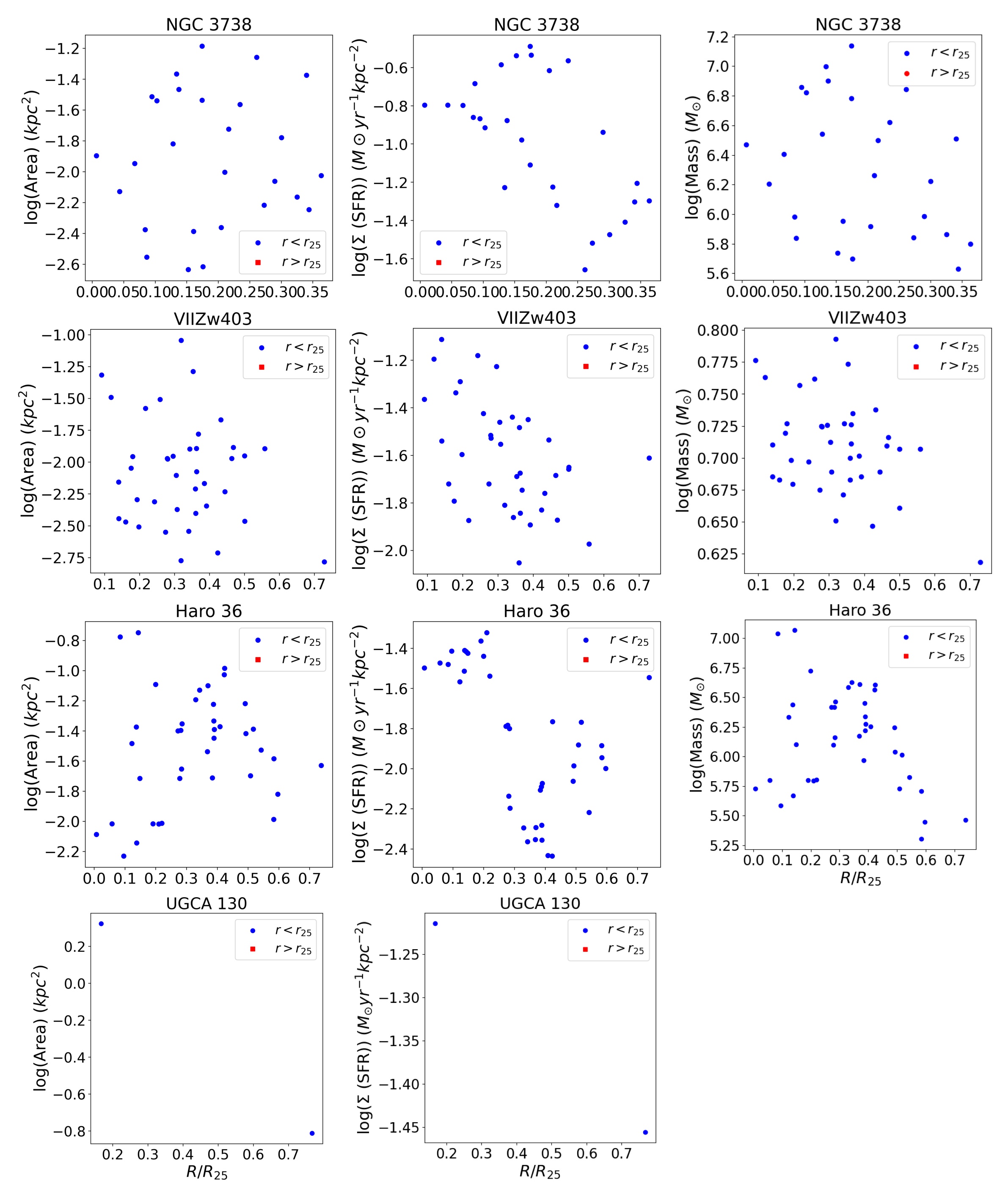} 
\contcaption{}
\end{figure*}

\begin{figure*}
     \centering
     \begin{subfigure}[b]{0.3\textwidth}
         \centering
         \includegraphics[scale=0.3]{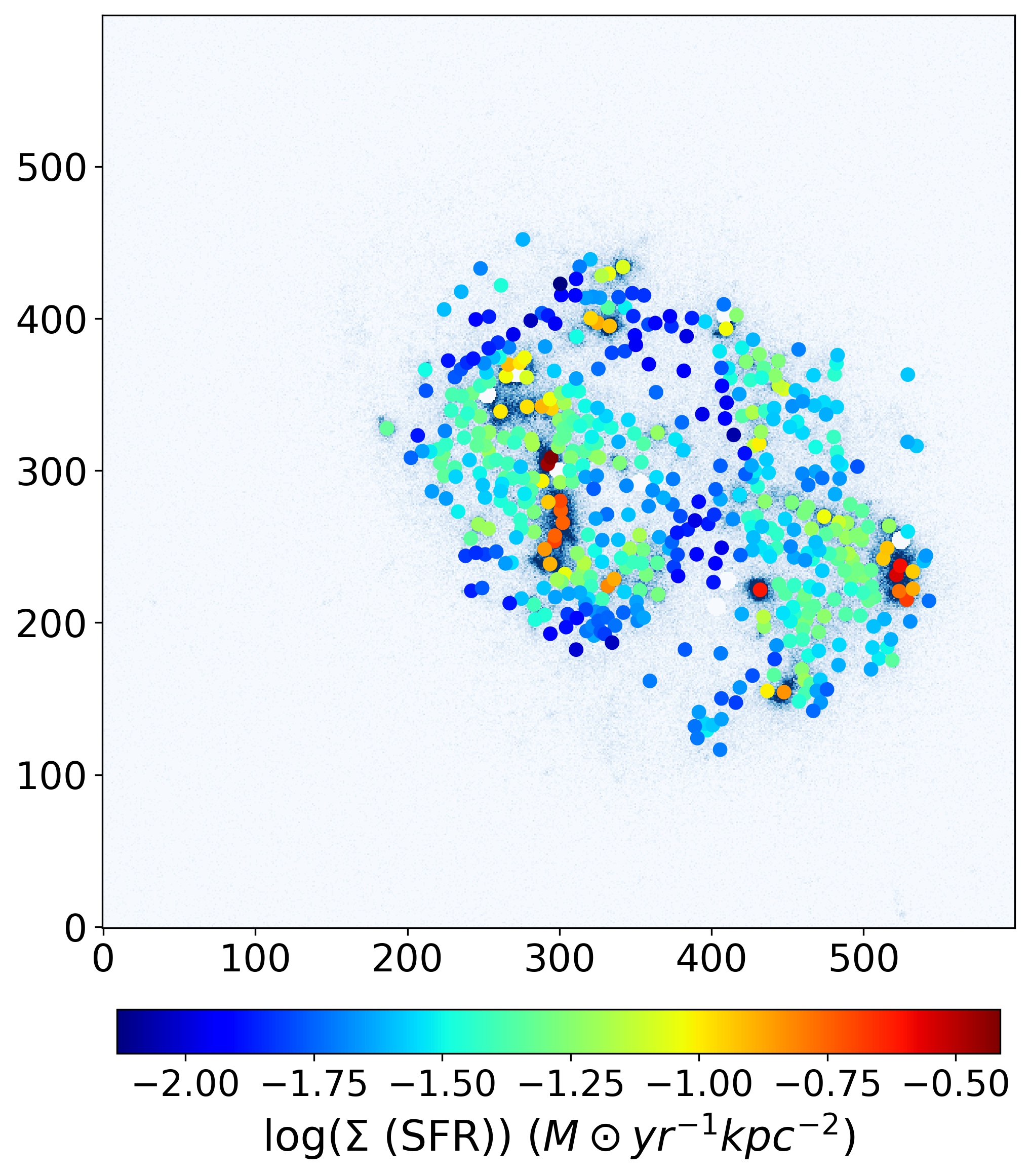}
         \caption{NGC 4618}
         \label{SFR4618}
     \end{subfigure}
     \hfill
     \begin{subfigure}[b]{0.3\textwidth}
         \centering
         \includegraphics[scale=0.3]{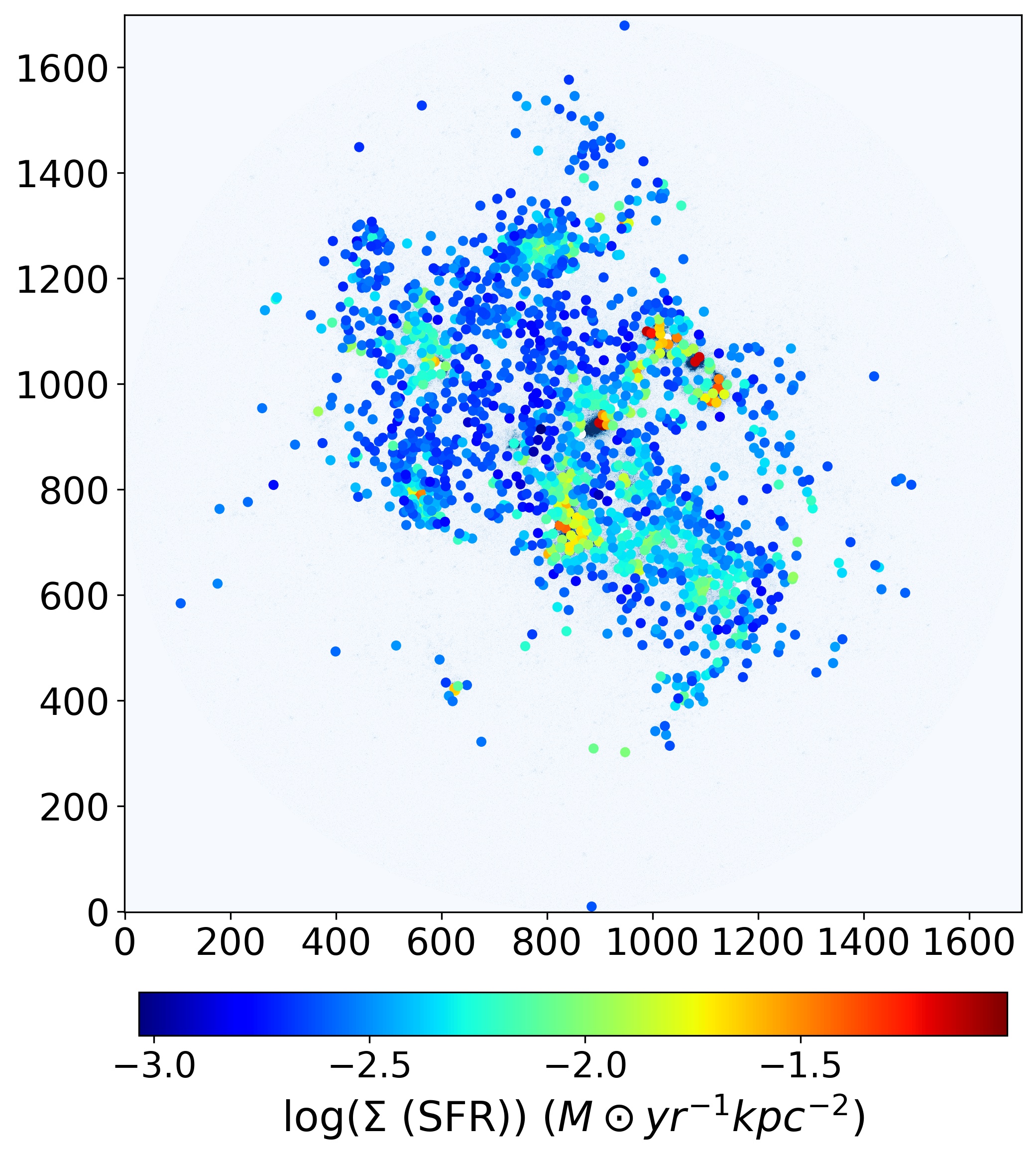}
         \caption{UGC 4305}
         \label{SFR4305}
     \end{subfigure}
     \hfill
     \begin{subfigure}[b]{0.3\textwidth}
         \centering
         \includegraphics[scale=0.3]{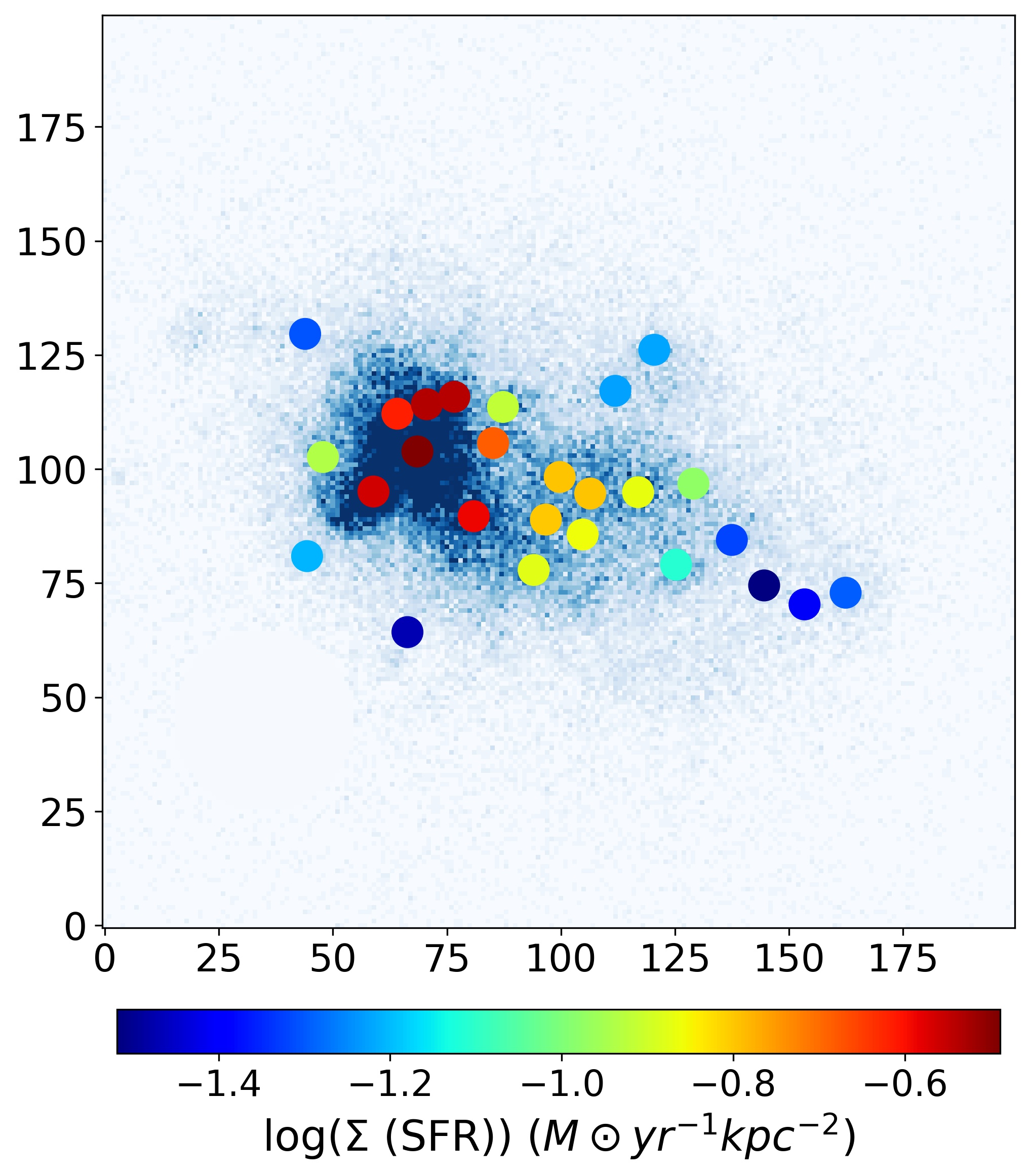}
         \caption{NGC 3738}
         \label{SFR3738}
     \end{subfigure}
        \caption{The above figures show the scatter plot of \(log(\Sigma(SFR))\) for dwarf spiral (NGC 4618), dwarf irregular (UGC 4305), and a BCD (NGC 3738). }
        \label{fig:three figure}
\end{figure*}

\subsection{Observed Radial trends of SFCs} \label{subsec: Observed Radial trends of SFCs}

The radial trends of SFC properties are very important for understanding how a galaxy grows via star formation. Hence, we plotted the quantities like SFC area, star formation rate density ($\Sigma(SFR)$), and corresponding disk stellar disk masses of SFCs with respect to the deprojected galactocentric radius scaled with the optical size of the galaxy. The deprojected galactocentric distance R is  
\begin{equation} \label{eq6}
 R=\frac{R_{obs}}{\sqrt{cos^{2}(\phi)+sin^{2}(\phi)cos^{2}(i)}}
\end{equation}
Where \(R_{obs}\) is the observed galactocentric distance, and i is the inclination of the galaxy.\(\phi\) is the angle between the radial distance to the SFC and the semi-major axis in the galaxy's plane.
\begin{equation} \label{eq7}
 tan(\phi')=tan(\phi)cos(i)
\end{equation}
where \(\phi'\) is the angle between the radius vector to the SFC and the semi-major axis projected on the sky \citep{1995AJ....110..199A}.

The area or the size of the complexes detected depends on the distance of a galaxy. The minimum resolvable area will be smaller for the nearby galaxies and increases with the galaxy distance due to the blending effect. The minimum resolvable radius of SFCs in our sample is for the galaxy NGC 6822, and the value is 2pc. Whereas for UGCA 130, which is much farther away, it is 87pc. Hence, the number of complexes detected will decrease with increasing galaxy distance. And we cannot compare the area of the complexes for the galaxies at different distances. This effect is tested in \cite{Melena_2009}, and it is observed that the general trend remains.

\begin{figure*}
 \includegraphics[scale=0.6]{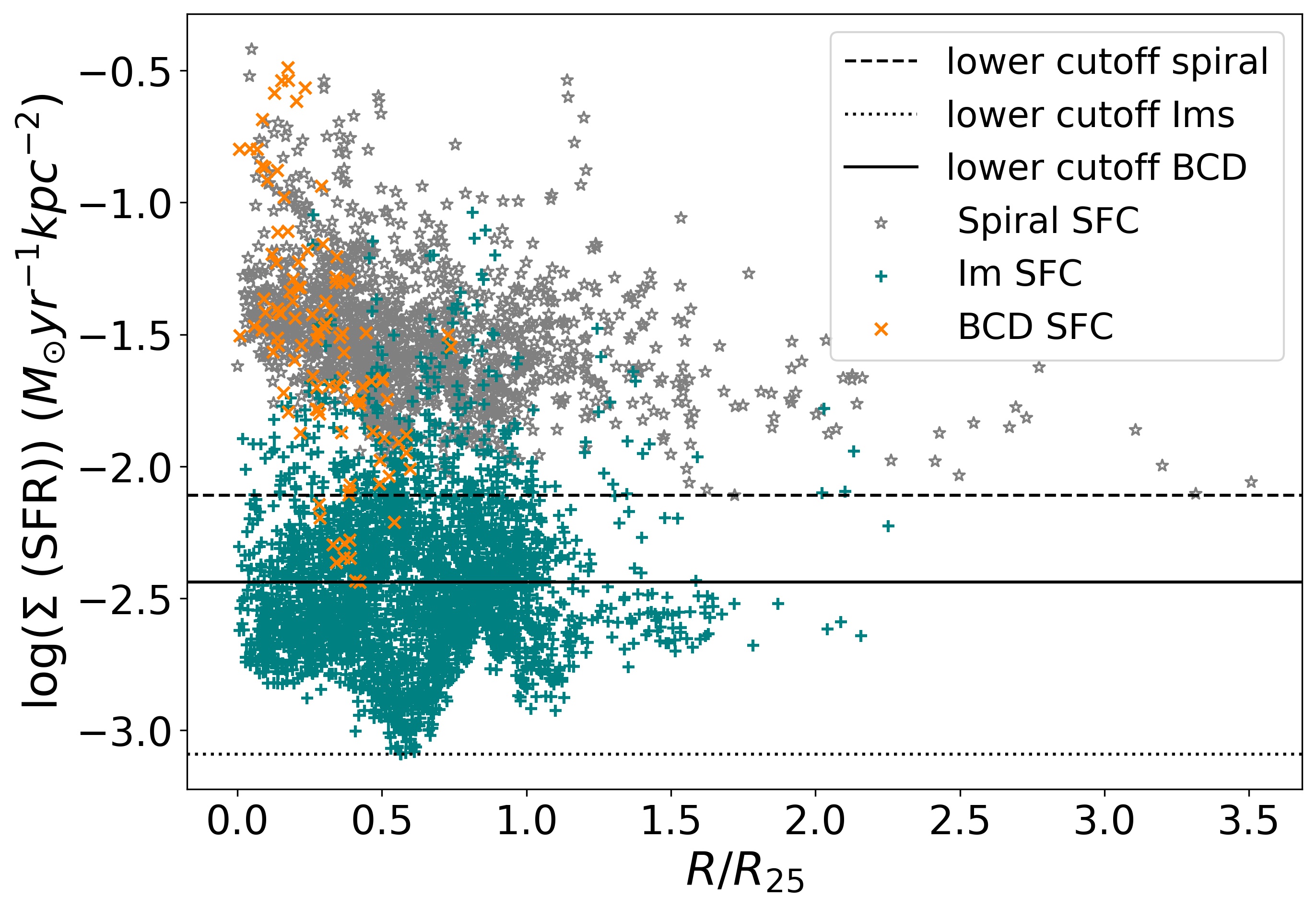}
\caption{Radial variation of the star formation rate density of SFCs for all the sample galaxies, except NGC 6822.
\label{fig8}}
\end{figure*}

\begin{figure*} 
 \includegraphics[scale=0.8]{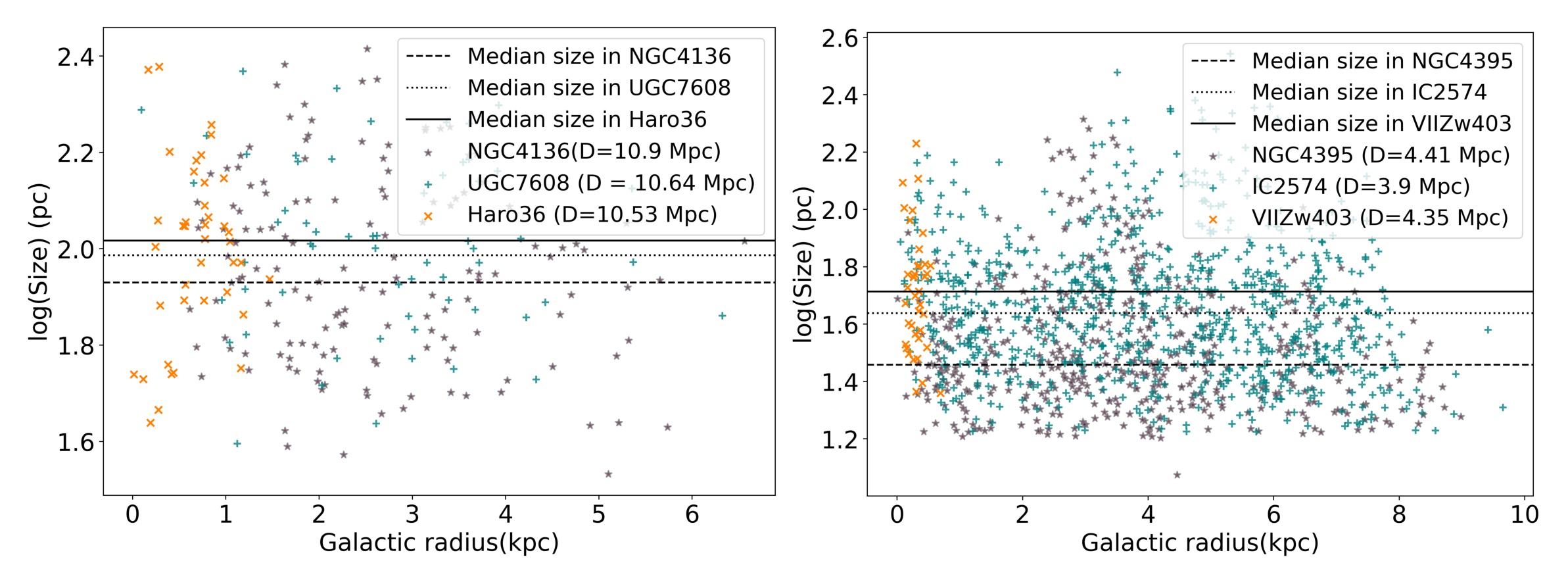}
\caption{Radial variation of the log(size) of SFCs for some galaxies with approximately equal distances. The median size value for dwarf spiral (NGC 4136, NGC 4395), irregular (UGC 7608, IC 2574), and BCDs (Haro 36, VIIZw403) at a distance D \(\approx\) 10.5 Mpc is 85.11 pc, 96.96 pc, and 104 pc, respectively. And at distance D \(\approx\) 4.1 Mpc, it is 27.9 pc, 43.54 pc, and 51.29 pc, respectively.
\label{fig16}}
\end{figure*}

\begin{figure}
\includegraphics[width=1 \columnwidth]{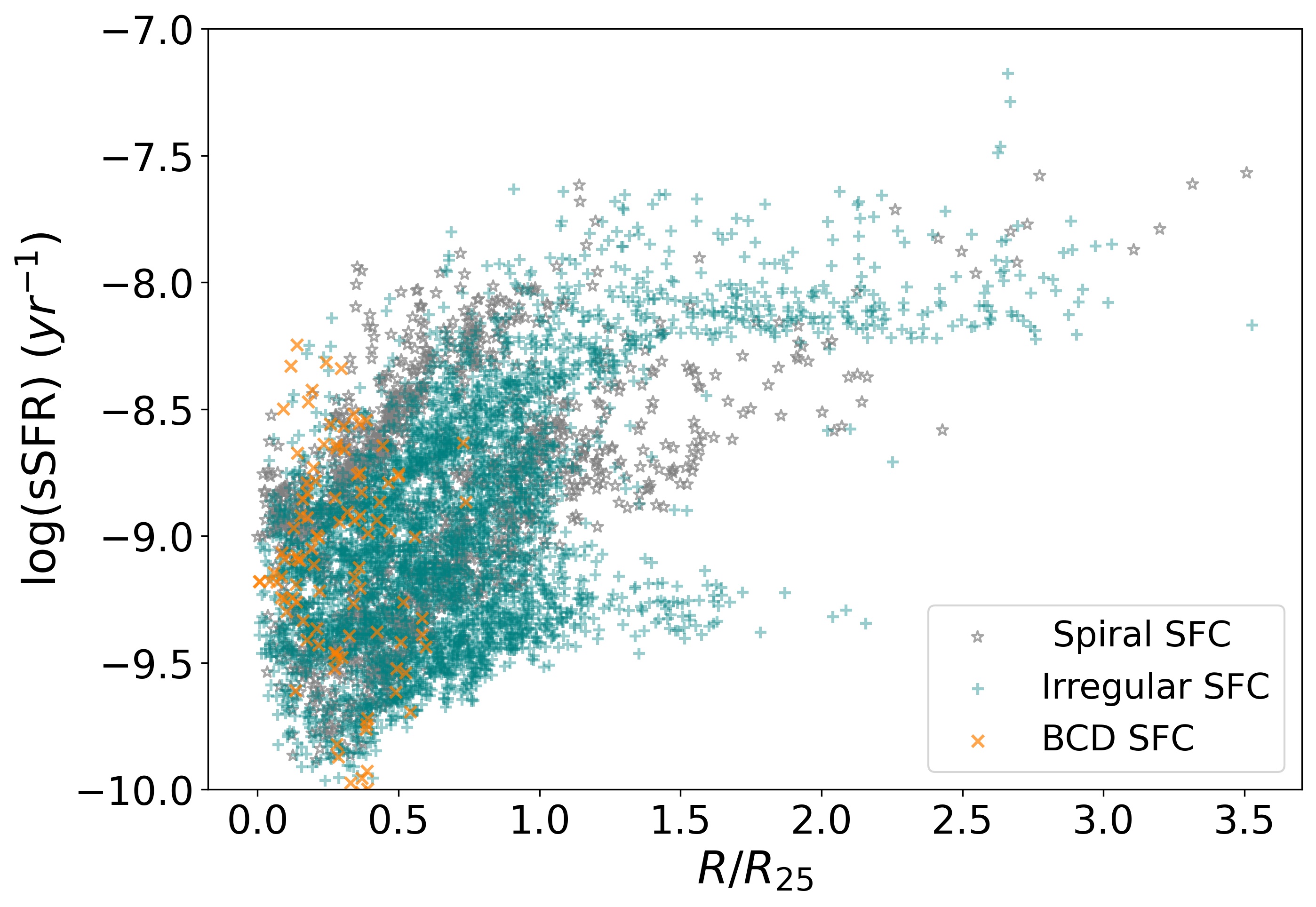}
\caption{Radial variation of the specific star formation rate of SFCs for all the sample galaxies.}
\label{ssms_radial_sfcs}
\end{figure}

Figure \ref{fig7} shows the radial variation of the area of the SFCs for the sample. In general, the SFCs have a smaller area beyond the optical radius for most galaxies, except for some galaxies that are either interacting (for example, NGC 4618 and NGC 6822) or are highly extended (NGC 2541), whereas, within the optical radius, both small and relatively larger complexes can be seen. The large SFCs in dwarf spirals, away from the centre, must be associated with structures due to global instabilities such as spiral arms, bars, and bar ends for spiral galaxies. The larger complexes are irregularly spread within the optical radius or associated with ill-defined arms in dwarf irregulars. The distribution of BCDs is entirely irregular.

Unlike the area of the complexes, extinction-corrected \(\Sigma(SFR)\) can be compared between galaxies at different distances. The radial trend here is either oscillatory or radially decreasing. Dwarf spirals show a prominent radially decreasing trend along with some SFCs with high \(\Sigma(SFR)\) values at some radius, primarily associated with spiral arms ends of the bars. In some galaxies, SFCs associated with bar have higher \(\Sigma(SFR)\) (See NGC 4618 in Figure \ref{fig:three figure}). Even for irregulars, the clumpiness is more only at a certain radii. Hence, they also show the oscillatory trend. Most SFCs with higher \(\Sigma(SFR)\) are associated with larger complexes. The radial trend of \(\Sigma(SFR)\) for SFCs in BCDs mainly drops with increasing galaxy radius. We have shown for each SFDG type an example log(\(\Sigma(SFR)\)) scatter plot over the galaxy disk in Figure \ref{fig:three figure}. This plot shows how the \(\Sigma(SFR)\) distribution differs for the three types of SFDGs.

The stellar disk mass of the SFCs is distance-dependent, and the blending effect is prominent because of it. The larger the complex, the more disk stellar mass it will correspond to. Even the mass radial trend is mainly similar to log(\(\Sigma(SFR)\)) for some galaxies, but the radially decreasing trend is clearly evident for all the galaxies. However, this could also be due to the radial profile of the stellar mass density, which falls off with radius. Hence, we do not find any massive complex in the outer region.

We also plotted the \(\Sigma(SFR)\) for SFCs of all types in a single plot, as shown in Figure \ref{fig8}, and found that the \(\Sigma(SFR)\) of the dwarf spirals is much higher than the other two types. We have not included data from NGC 6822 as it shows unusually high \(\Sigma(SFR)\) values compared to other dwarf irregulars, and the reason for it is discussed in section \ref{Difference in Star formation rate density for different Dwarf galaxy types}. The lowest \(\Sigma(SFR)\) measured for SFC in dwarf spirals is 0.0078 \(M_{\odot}yr^{-1} kpc^{-2}\). 0.0008 \(M_{\odot}yr^{-1} kpc^{-2}\), and 0.0036 \(M_{\odot}yr^{-1} kpc^{-2}\) for dwarf irregulars and BCDs respectively. There is minimal overlap exists between the dwarf spirals and the dwarf irregulars \(\Sigma(SFR)\)s.

Figure \ref{fig16} shows the comparison of the sizes of the SFCs in three different types of SFDGs, which are at roughly the same distance. The left and right panels show the comparison at 10 Mpc and 4 Mpc distances, respectively. The solid, dotted, and dashed lines show the median size of the SFCs in BCDs, irregulars, and spirals. The median sizes of the SFCs are the largest in BCDS. The irregulars host larger SFCs than spirals. One reason for this could be that irregular galaxies lack spiral density waves, and their kinematics are dominated by random gas motions. The discussion on the difference in size and \(\Sigma(SFR)\) of SFDGs is discussed in section \ref{Difference in Star formation rate density for different Dwarf galaxy types}. 

We even plotted variations of specific star formation rates (sSFRs) of SFCs along the galaxy radius. We plotted the SFCs of all the sample galaxies in the same plot as shown in Figure \ref{ssms_radial_sfcs}. The trend is radially increasing till a little beyond $r_{25}$, and later it becomes nearly constant. 

\subsection{Star forming main sequence for SFCs} \label{subsec:Star-forming main sequence for SFCs }

The star-forming main sequence (SFMS) is a relation between the SRF and mass of the galaxy, and has important implications for understanding the age and evolutionary states of star-forming galaxies \citep{McGaugh_2017}. The equation of the SFMS has a form 
\begin{equation} \label{eq8}
log(SFR)=\alpha log(M_{*})+ \beta
\end{equation}
with \(\alpha\) and \(\beta\) likely to be functions of the time \citep{2014ApJS..214...15S}. Slope \(\alpha\) is measured between 0 to 1 \citep{2009MNRAS.393..406C} .

\begin{figure}    
\includegraphics[width=\columnwidth]{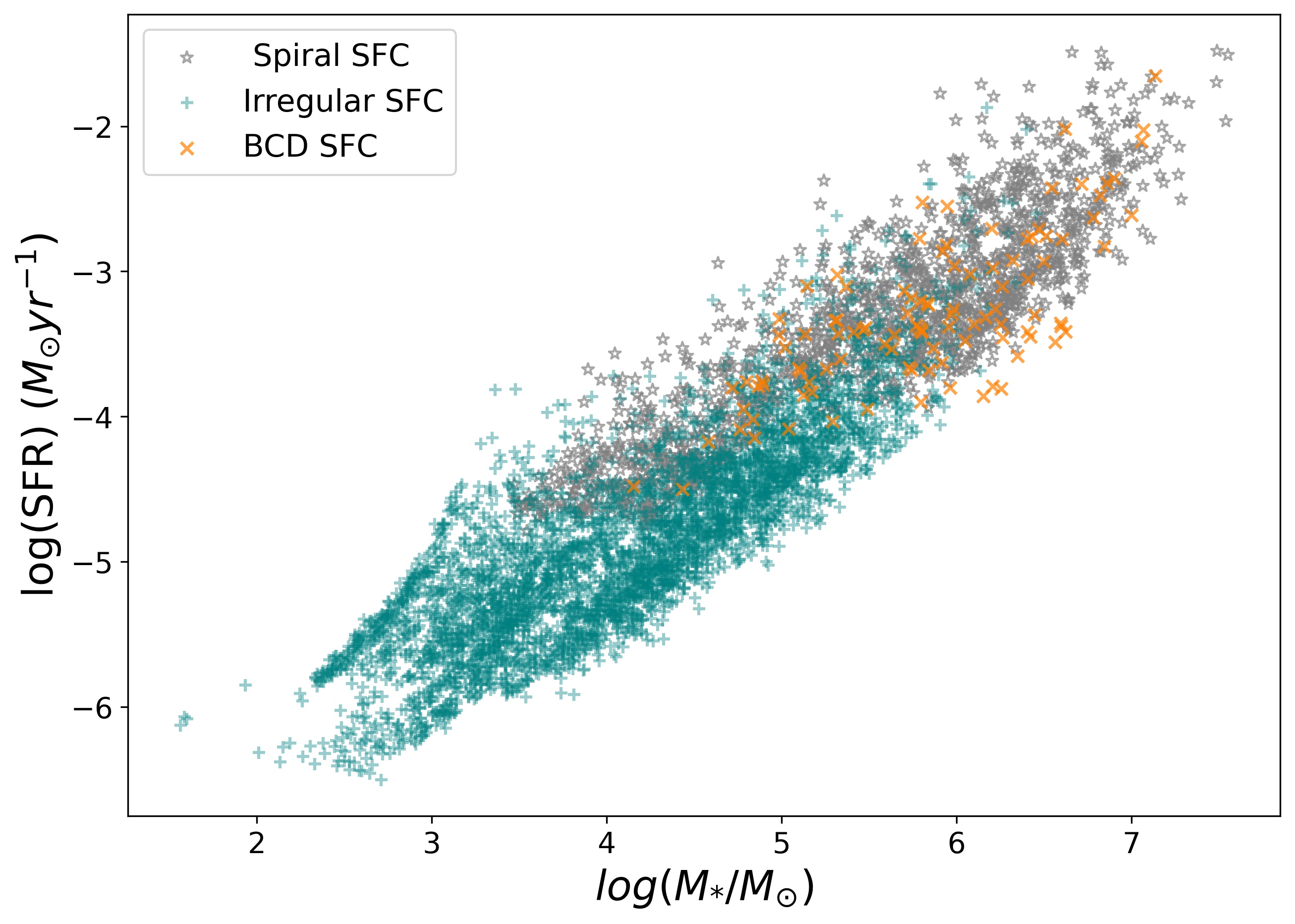}
\caption{ SFMS of SFCs for all the sample galaxies.
\label{fig10}}
\end{figure}

 In section \ref{sec:Data Analysis}, we have extracted SFCs and derived their properties, including the corresponding stellar disk mass for each SFC. We used them to construct the SFMS but on the scale of SFCs for SFDGs. So, we plotted the log of SFR versus the log of corresponding stellar disk mass for the SFCs. We find that they are very well co-related. This tells us that galaxies follow the SFMS globally and even at the level of SFCs, which is on scales well below a kpc. The parameters of the SFMS for the SFCs are given in Table \ref{table6}. However, the slopes of the three types of dwarf galaxies are slightly different. It is clear that the dwarf spirals have a slightly shallower slope than the other two types. The slope of SFMS for dwarf spiral SFCs has a wide range from 0.93 to 0.59, and the mean of the slope is around 0.74. NGC 4136 has the lowest slope of 0.59. The mean slope of the dwarf irregulars is 0.87, and for BCDs, it is 0.8.

\begin{figure*}
\includegraphics[width=1\linewidth]{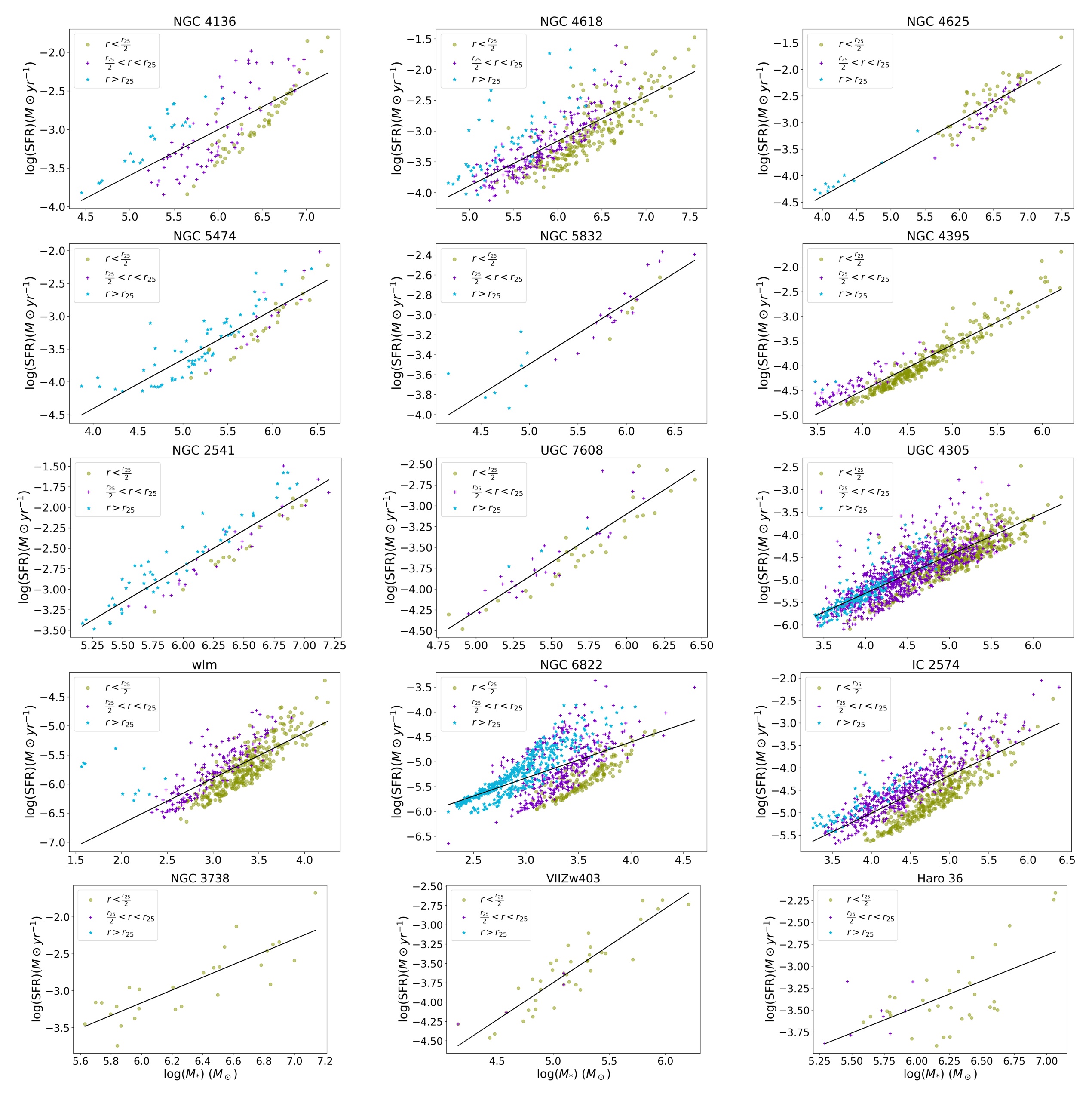} 
\caption{Star forming the main sequence of SFCs for all of our sample galaxies except UGCA 130, as it has only 2 complexes. We see that the inner SFCs with yellow circles ( r\(< r_{25}/2\)), middle SFCs with purple plus ( \(r_{25}/2\)\(<\)r \(<\) \(r_{25}\)), and outer SFCs with blue stars (r  \(> r_{25}\)) have different SFMS for many galaxies. Outer SFCs have higher specific star formation rates, indicating the inside-out evolution of the galaxies}.
\label{fig11}
\end{figure*}

From Figure \ref{ssms_radial_sfcs}, the sSFR (i.e. SFR/mass) of the SFCs is increasing with galactic radius for some galaxies and becomes nearly constant at the outer radius. To understand how it varies for each galaxy, we divided the SFCs for all the sample galaxies into inner SFCs ( r\(< r_{25}/2\)), middle SFCs ( \(r_{25}/2\)\(<\)r \(<\) \(r_{25}\)), and outer SFCs (r  \(> r_{25}\)), where \(r_{25}\) is the optical radius. For a given galaxy at any fixed $log(M_{*})$, the log(SFR) increases from the inner to outer SFCs for some galaxies. Hence, the sSFR increases as well. This is very clear in Figure \ref{fig11} and is more prominent in dwarf spirals, although some irregulars also show this trend. The BCDs do not follow any trend. 
It is important to note that most of the outer complexes correspond to low stellar mass, and the inner complexes generally correspond to higher stellar mass. But the outer disks have lower stellar surface densities, so for SFCs at outer radii, the sSFR will naturally appear higher. The possible reasons for this gradient is discussed further in subsection \ref{Difference in Star-forming main sequence in Dwarf types}.
\begin{table}
    \centering
                    \caption{Star forming main sequence parameters for sample galaxies.}
                        \label{table6}
                        \begin{tabular}{cccccccccc}
                    \hline

                    Galaxy & \(\alpha\) & \(\alpha_{mean}\)  & \(\beta\) & \(\sigma\) \\ 
                    \hline
                    NGC 4136 &0.59\(\pm\) 0.05&&-6.55\(\pm\) 0.29&0.32&\\
                    NGC 4618 &0.73\(\pm\) 0.03&&-7.53\(\pm\) 0.16&0.32&\\
                    NGC 4625 &0.71\(\pm\) 0.03&&-7.24\(\pm\) 0.19&0.23&\\
                    NGC 5474 &0.75\(\pm\) 0.04&0.74\(\pm\) 0.13&-7.4\(\pm\) 0.24&0.26& \\
                    NGC 5832 &0.61\(\pm\) 0.05&&-6.54\(\pm\) 0.29&0.17 & \\
                    NGC 4395 &0.93\(\pm\) 0.02& &-8.22\(\pm\) 0.08&0.2 &\\
                    NGC 2541 &0.88\(\pm\) 0.04&&-7.98\(\pm\) 0.25&0.25&\\
                    \hline
                    UGC 7608 &1.16\(\pm\) 0.07&&-10.07\(\pm\) 0.39&0.2&\\
                    WLM &0.78\(\pm\) 0.03&&-8.25\(\pm\) 0.1&0.29&\\
                    UGC 4305 &0.84\(\pm\) 0.01&0.87\(\pm\) 0.16&-8.66\(\pm\) 0.06&0.32&\\
                    NGC 6822 &0.73\(\pm\) 0.03&&-7.5\(\pm\) 0.1&0.38 \\
                    IC 2574 &0.84\(\pm\) 0.02&&-8.35\(\pm\) 0.1&0.37&\\
                    \hline
                    NGC 3738 &0.86 \(\pm\)0.1 &&-8.34\(\pm\) 0.71&0.26&\\
                    VIIZw403 &0.96\(\pm\) 0.07&0.80\(\pm\) 0.19&-8.5\(\pm\) 0.38&0.2&\\
                    Haro 36 &0.59\(\pm\) 0.12&&-7.01\(\pm\) 0.74&0.31& \\
                    \hline
                        \end{tabular}
                        \begin{minipage}{80mm}
                   \textit{Note}: \(\alpha\) and \(\beta\) represent as mentioned in SFMS equation \ref{eq8}. \(\alpha_{mean}\) is the mean of slope for each type of SFDGs, and \(\sigma\) is the Standard deviation of the data points.
                    \end{minipage}

\end{table}

\subsection{ Other results of SFDGs} \label{subsec: Macroscopic results for SFDGs}

\begin{table*}
     \centering
                    \caption{Calculated parameters of Sample galaxies.}
                        \label{table5}
                        
                        \begin{tabular}{cccccccccc}
                    \hline
    
                    Galaxy & \(m_{B}\) & 12 +log (O/H) & FUV-NUV & \(log(M_{*})\)&  \(log(SFR)\)  & \\ 
                     & (mag) &  &(mag) & \(M_{\odot}\) & \(M_{\odot}yr^{-1}\) &\\
                    \hline
                    NGC 4136 &11.69\(\pm\)0.17 &8.38\(\pm\)0.02&0.26\(\pm\)0.01  &9.38& -0.04 & \\
                    NGC 4618 &11.31\(\pm\)0.16 &8.39\(\pm\)0.02&0.25\(\pm\)0.01 &9.56 &0.1 &\\
                    NGC 4625 &12.85\(\pm\)0.18 &8.17\(\pm\)0.03&0.3\(\pm\)0.01  & 9.15&0.01&\\
                    NGC 5474 &11.49\(\pm\)0.17 &8.27\(\pm\)0.02&0.12\(\pm\)0.01 &9.12& -0.46& \\
                    NGC 5832 &12.9\(\pm\)0.3 &8.21\(\pm\)0.04&0.39\(\pm\)0.01 &9.2&-0.6 & \\
                    NGC 4395 &10.3\(\pm\)0.54 &8.21\(\pm\)0.08&0.12\(\pm\)0.01  &9.14&-0.38&\\
                    NGC 2541 &12.02\(\pm\)0.22 &8.37\(\pm\)0.03&0.09\(\pm\)0.01 &9.49 &0.2&\\
                    \hline
                    UGC 7608 &13.6\(\pm\)0.19 &8.1\(\pm\)0.03&0.11\(\pm\)0.01 &9 &-0.84&\\
                    WLM &11\(\pm\)0.08 &7.72\(\pm\)0.01&0.12\(\pm\)0.01  &7.34& -2.2&\\
                    UGC 4305 &10.9\(\pm\)0.2 &8.13\(\pm\)0.03&0\(\pm\)0.01 &8.64&-0.96 &\\
                    NGC 6822 &7\(\pm\)0.3 &8.09\(\pm\)0.04&0.4\(\pm\)0.08 &8.03 &-1.66  &\\
                    IC 2574 &10.8\(\pm\)0.3 &8.19\(\pm\)0.04&0.03\(\pm\)0.01 & 9 &-0.63 &\\
                    \hline
                    NGC 3738 &11.79\(\pm\)0.19&8.13\(\pm\)0.03&0.37\(\pm\)0.01 &8.31& -0.98 &\\
                    VIIZw403 &14.5\(\pm\)0.19 &7.68\(\pm\)0.03&-0.05\(\pm\)0.01& 7.13&-1.71&\\
                    Haro 36 &14\(\pm\)0.18 &8\(\pm\)0.03&0.27\(\pm\)0.01 &8.26&-1.26 & \\
                    UGCA 130 &15.1\(\pm\)0.04 &7.99\(\pm\)0.01 &0.21\(\pm\)0.06 &8.1&-1.49& \\
                    \hline
                    
                        \end{tabular}
                        \begin{minipage}{110mm}
                   \textit{Note}: Apparent B magnitude \(m_{B}\) is taken from NED.  
                    \end{minipage}

\end{table*}

\subsubsection{ The Global Star forming main sequence for SFDGs}\label{Star forming main sequence for SFDGs}

\begin{figure}
\includegraphics[width=\columnwidth]{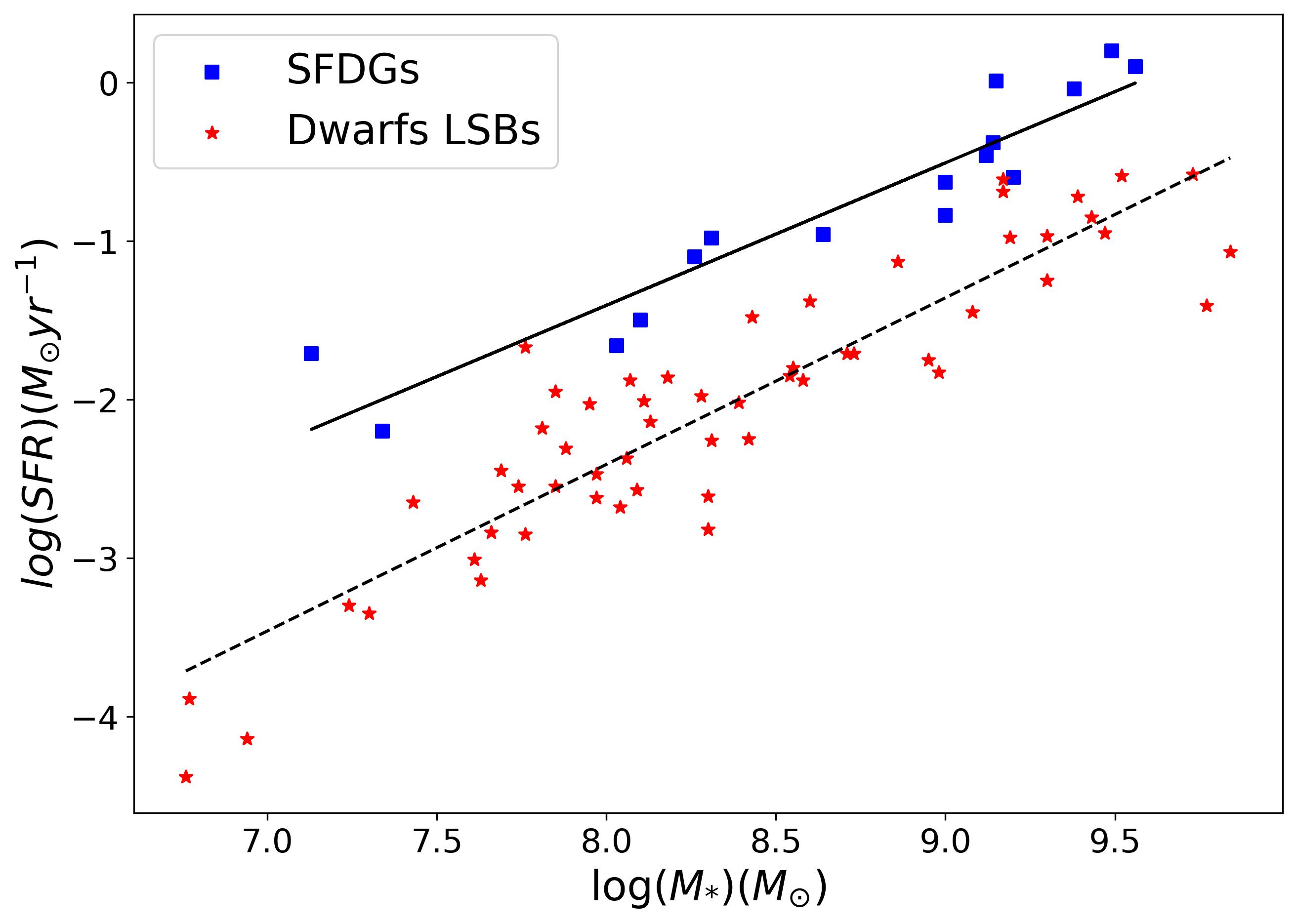}
\caption{Blue squares are our sample SFDGs, and red stars are dwarf LSBs from \citet{McGaugh_2017}. The black solid line is a linear fit for SFMS of SFDGs with the slope of \(\alpha =0.8989\). The dashed line is the linear fit for dwarf LSBs with the slope of \(\alpha =1.05\).
\label{fig9}}
\end{figure}

We took the global SFR and total mass of the SFDGs from Table \ref{table5}. We plotted the SFMS for our sample galaxies and determined the slope of the linear regression fit to our sample. We found that it is shallower than the dwarf LSBs and steeper than that for the massive galaxies, which will be discussed in section \ref{Difference in Star-forming main sequence}. The fit for SFDGs,
\begin{equation} \label{eq9}
log(SFR)=(0.899\pm 0.087) log(M_{*})+(-8.59 \pm 0.75)
\end{equation}
   
   with slope \(\alpha =0.8989 \). We see very little deviation of our sample from the fit. The internal scatter of the relation is \(\sigma=0.25\). As the sample has only 16 galaxies, we do not see much deviation between the dwarf types in Figure \ref{fig9}. But most of the irregulars lie below the fit as they have lower SFRs. Some high-mass dwarf spirals and BCDs are above the fit. 

\subsubsection{Mass Metallicity relation for SFDGs}

\begin{figure}
\includegraphics[width=0.9 \columnwidth]{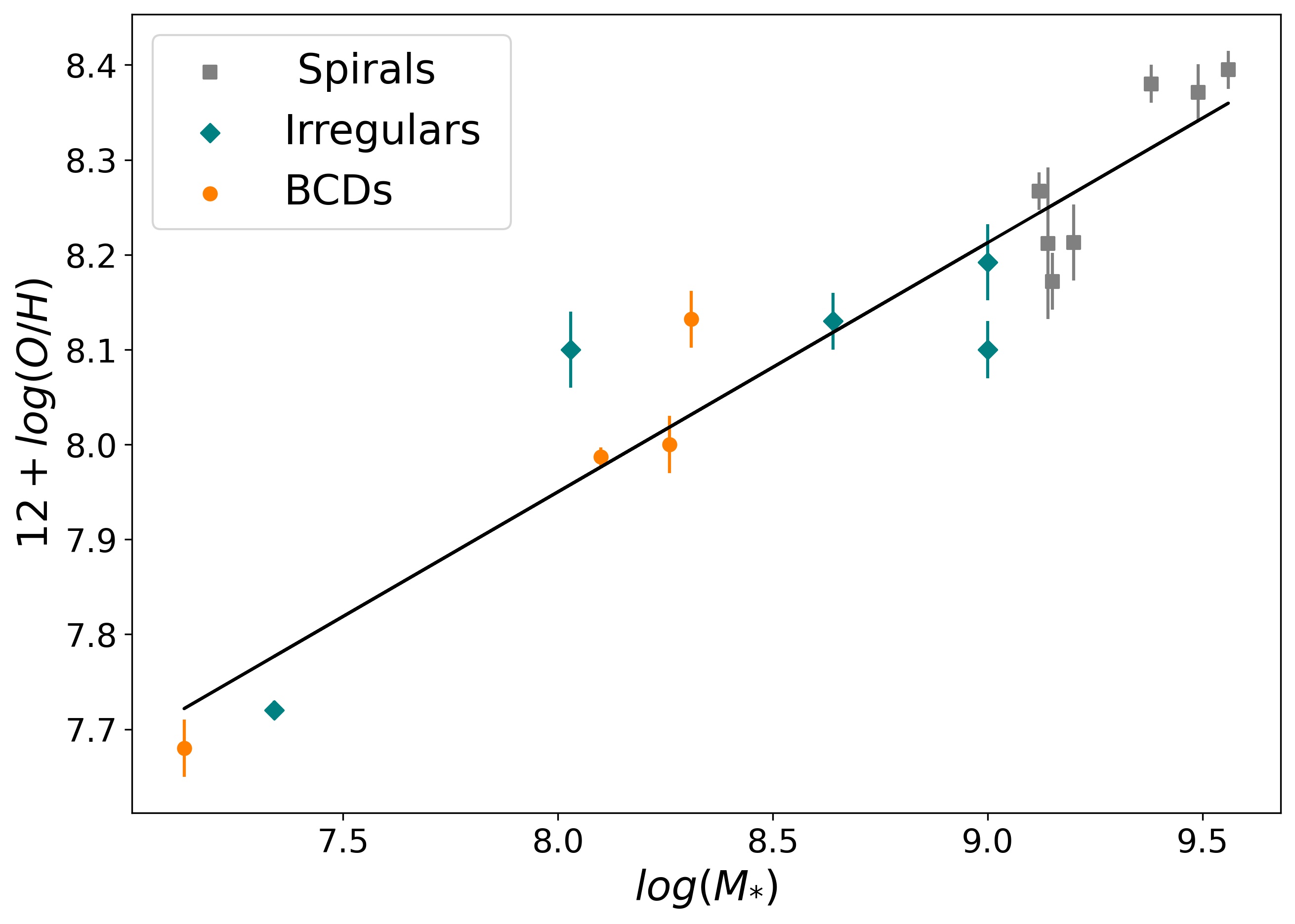}
\caption{ Metallicity-mass plot for SFDGs. Values of errors associated with metallicity can be found in Table \ref{table5}.
\label{fig12}}
\end{figure}

It is found that for dwarf galaxies with known metallicity, the metallicity strongly correlates with the B-band luminosity \citep{1995ApJ...445..642R}. Hence, B-band luminosity is considered a better predictor of metallicity for dwarf galaxies \citep{2003AJ....125..610S}. We found the metallicity for our sample galaxies using a fit derived by \cite{1995ApJ...445..642R} from local group dwarf irregulars.
\begin{equation} \label{eq10}
12 +log (O/H) = (5.67\pm 0.48) +(-0.147\pm 0.029)M_{B}
\end{equation}

 We took apparent B magnitudes from NED, and using distances as listed in Table \ref{table1}, we calculated the absolute B magnitude \(M_{B}\). The \(M_{B}\) and metallicity of our sample galaxies are listed in Table \ref{table5}. 

 We plotted the independently calculated metallicity and mass for SFDGs in Figure \ref{fig12}. We find that the linear fit for metallicity and mass is given by,
 \begin{equation}\label{eq11}
   12 +log (O/H) =(5.85\pm 0.208)  +(0.26\pm 0.024) log(M_{*})    
 \end{equation}
This mass-metallicity relation matches very well with the previous study by \cite{Berg_2012}, where they found a similar relation for low luminosity galaxies. Our relation lies within its error bars. The B magnitude of his sample has a range (\( -10.8 \geq M_{B} \geq -18.8 \)). We find that our galaxies also have a B- magnitude in that range. This indicates that our galaxies have low luminosity in the B band. The intrinsic scatter of the data points from our fit is \(\sigma=0.07\), which is very small. We see that all our sample SFDGs follow this relation irrespective of their type. This supports our mass and metallicity calculations.

 \subsubsection{(FUV-NUV) Colour - Metallicity plot for SFDGs}
 \begin{figure}
\includegraphics[width=0.9 \columnwidth]{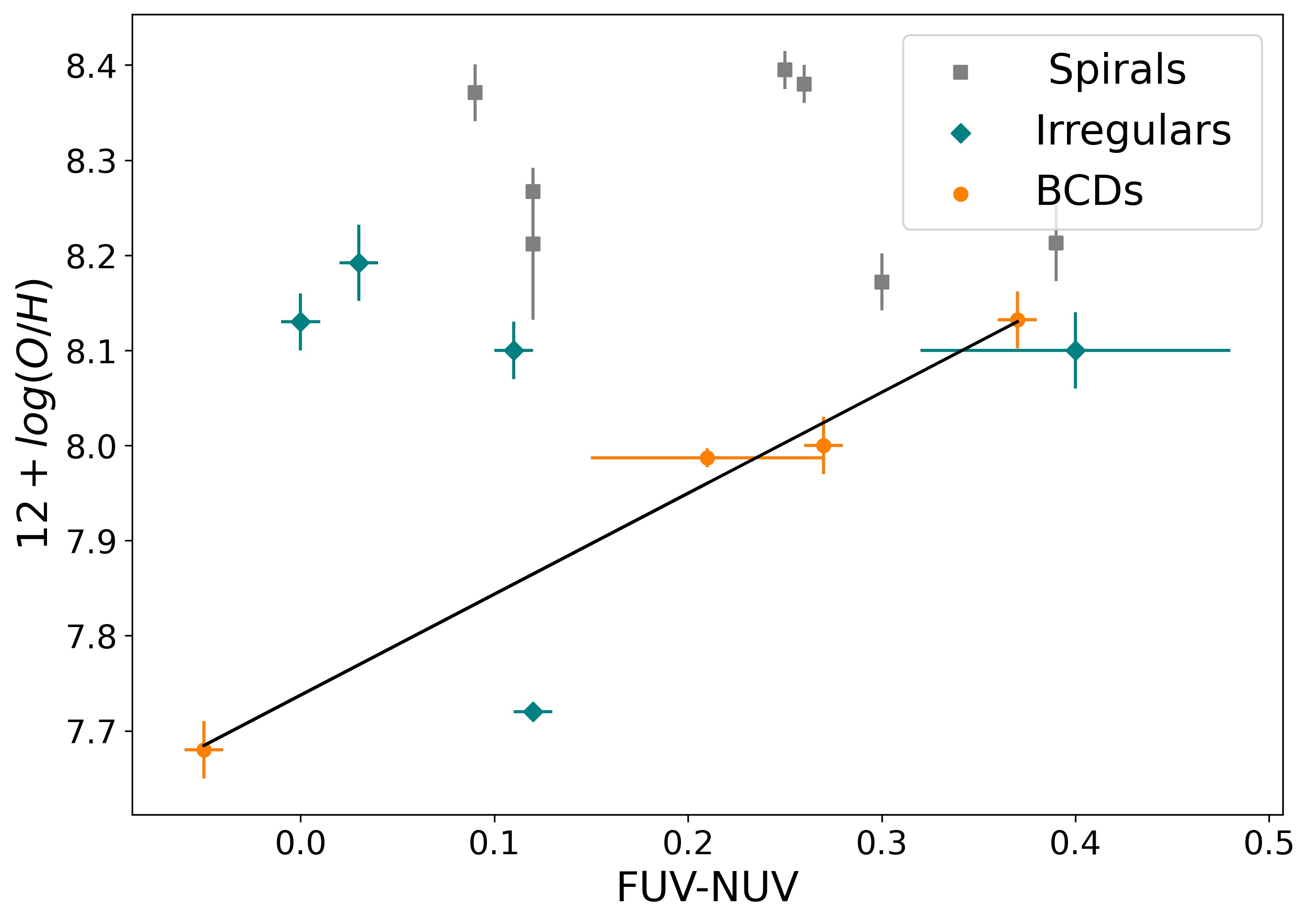}
\caption{ Metallicity-colour plot for SFDGs. We find a positive correlation only for BCDs.}
\label{fig13}
\end{figure}
The (FUV-NUV) colour traces the age of the star-forming galaxies up to a few hundred Myrs. As they get older, the colour reddens \citep{2021JApA...42...50M}. The redder galaxies are expected to be comparatively metal-rich as they are more evolved than the bluer ones \citep{Hogg_2004, 2010MNRAS.407..937P}.\\
We plotted the metallicity against (FUV-NUV) colour, as shown in Figure \ref{fig13}. We calculated the (FUV-NUV) colour for GALEX data from NED. We neglected the host galaxy extinction as we are calculating the colour, and the difference in the extinction for FUV and NUV bands is negligible. We see a proper co-relation only for BCDs, where we find bluer galaxies are more metal-poor, and as they get redder, they turn metal-rich. The other two types, i.e. dwarf spirals and irregulars, do not show any kind of co-relation. 

In Figure \ref{fig13}, the dwarf spirals have [12 +log (O/H)] \(>\)8.1, and their colour is spread between 0.1 to 0.4. For the galaxies with the same colour, galaxies with higher stellar mass have higher metallicity than those with lower stellar mass. The irregulars and BCDs have 12 +log (O/H) \(<8.2\), where all dwarf irregulars except NGC 6822 have a colour lesser than 0.2, and BCDS except  VIIZw403 have a colour greater than 0.2. This indicates that although they are redder than dwarf irregulars, the BCDs are metal-poor compared to dwarf spirals and dwarf irregulars \citep{Kunth_2000}. The BCD VIIZw403 has a colour lesser than 0. This indicates that the galaxy is very young and could be a starburst galaxy. 

 \subsubsection{The radial variation of sSFR measured over annuli}

\begin{figure*}
\includegraphics[width=1\linewidth]{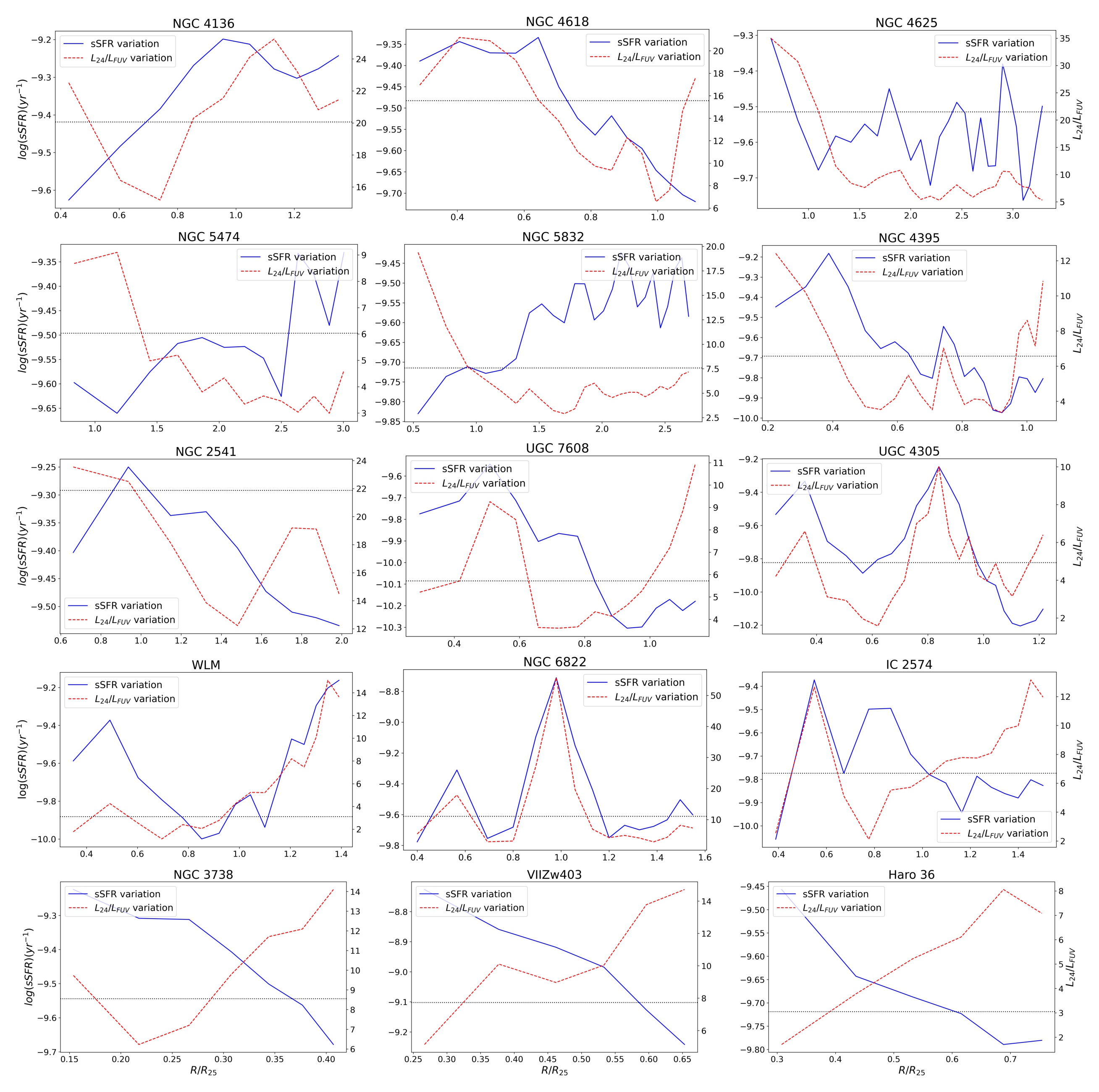} 
\caption{The radial variation of the sSFR is represented as a blue solid line. Whereas 24\(\mu\)m and convolved UV flux ratio ($\alpha$ as in subsection \ref{subsec: The Host galaxy extinction and Total Star Formation Rate }) is represented as a red dashed line. The black dotted line represents the average $\alpha$ calculated for each galaxy, taking integrated 24\(\mu\)m and convolved UV flux within the outer ellipse. The galaxy UGCA 130 is not included.}.
\label{alpha}
\end{figure*}

  We calculated the \(\Sigma(SFR)\) from each fitted elliptical annulus of the UVIT image as explained in subsection \ref{subsec: The Host galaxy extinction and Total Star Formation Rate }. Similarly, we estimated \(\Sigma(M_{*})\) from each fitted elliptical annulus of the 3.6 $\mu m$ image. We plotted the radial variation of sSFR for each galaxy by dividing the estimated \(\Sigma(SFR)\) by \(\Sigma(M_{*})\). We have shown this variation in Figure \ref{alpha}.

 We find that the variation is different for each galaxy, and except for BCDs, we don't see any particular trend for types of SFDG. For BCDs, it clearly decreases radially. However, for dwarf spirals and irregulars, we see a lot of variation in the plots. Galaxies like NGC 4618, NGC 2541, and NGC 6822 show a kind of opposite trend to what we see in their SFCs, which will be briefly discussed in the subsection \ref{Difference in Star-forming main sequence in Dwarf types}.

\section{Discussion} \label{sec:Discussion}
\subsection{Difference in Sizes and Star formation rate densities of SFCs in the different SFDG types} \label{Difference in Star formation rate density for different Dwarf galaxy types} 

In previous studies, it has been shown that galaxies show hierarchical star formation, and in general, the SFCs correspond to the largest stellar groups with sizes on the scales of 100s of parsec \citep{1995AJ....110.2757E, 1999AJ....117..764E, 2014ApJ...787L..15E}. But with UVIT resolution, our study resolves smaller star forming units like SFCs and OB associations in nearby galaxies on the scales of a few tens of parsec \citep{rahna.etal.2018}. In section \ref{subsec: Observed Radial trends of SFCs}, we find that although the sizes of SFCs are fairly similar over the SFDG types, the SFCs in dwarf irregulars are slightly larger than the SFCs in dwarf spirals. The BCDs have the largest SFCs compared to the dwarf spirals and irregulars. As previously mentioned, random gas motions can significantly contribute to the pressure in irregulars, and in some cases, rotation can be insignificant compared to the gas velocity dispersion. As a result, for effective star formation to occur within irregular galaxies, a higher gas surface density is required to overcome the pressure. This leads to the formation of larger SFCs within irregular galaxies compared to spirals. 

 However, in general, many factors play a role in the variation of SFC size across different morphological galaxy types, such as the interplay between their inherent kinematic characteristics, feedback from massive stars, the presence of spiral arms and the gas pressure in the disk. A previous study by \cite{1996ApJ...467..579E} mentions that the late-type galaxies (mostly dwarf spirals) and dwarf irregulars have almost similar sizes. And since we have compared galaxies at only two distances, the result is statistically insignificant.

Figure \ref{fig7} shows the distribution of the properties of the SFCs, including the radial variation of the \(\Sigma(SFR)\) for the SFDG sample. Although we do not see much difference between the radial trends in dwarf spirals and dwarf irregulars, there is a marked difference in their \(\Sigma(SFR)\) lower cutoff as shown in Figure \ref{fig8}. The SFCs of dwarf spirals show a higher \(\Sigma(SFR)\) than dwarf irregulars. For BCDs, it varies over a wide range. 

\begin{figure}
\includegraphics[width=\columnwidth]{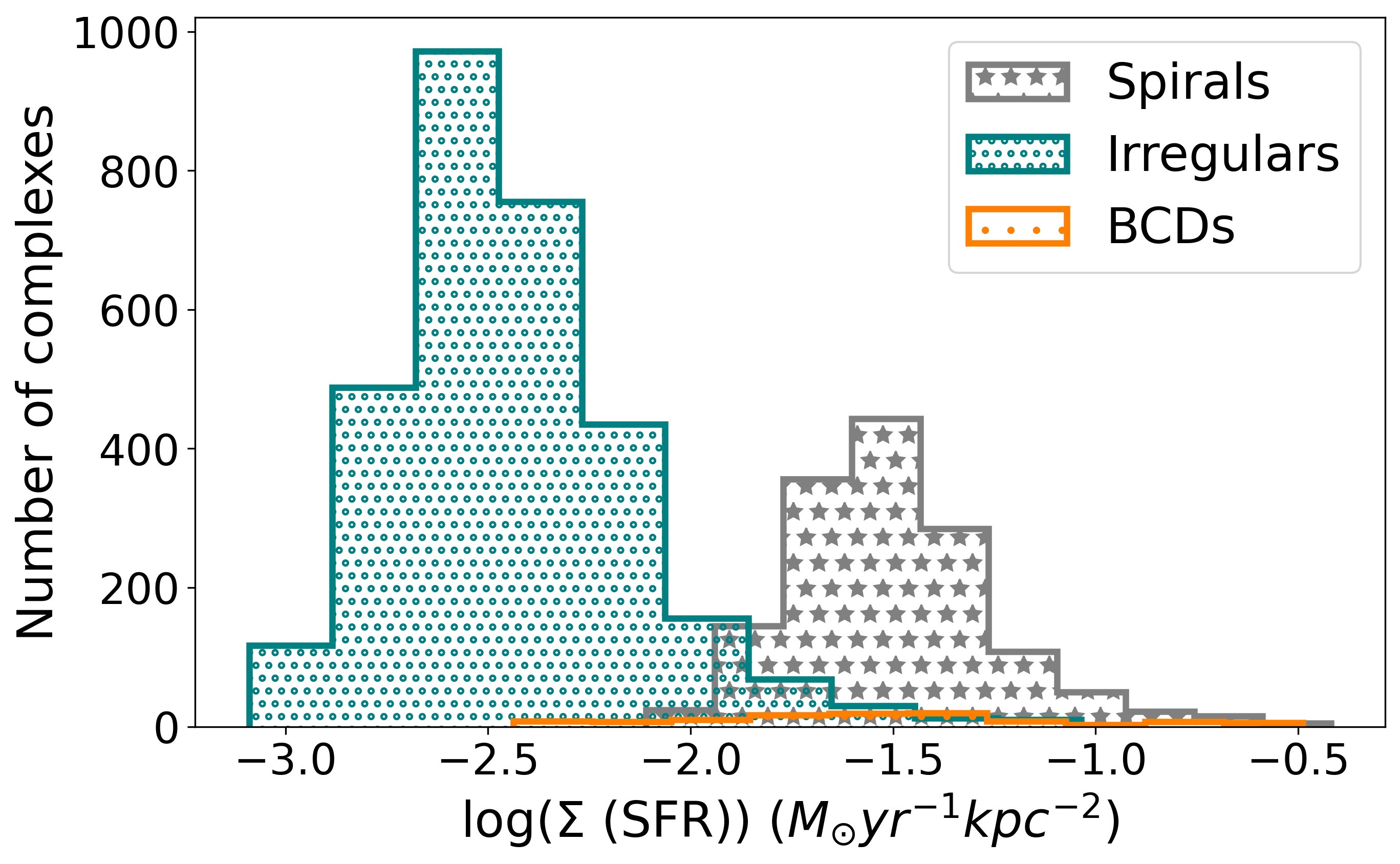}
\caption{Distribution of log(\(\Sigma(SFR)\)) for SFCs of all sample galaxies (except NGC 6822). The mode of log(\(\Sigma(SFR)\)) for dwarf spirals, dwarf irregulars, and BCDs is -1.84 \(M_{\odot}yr^{-1}kpc^{-2}\), -2.09 \(M_{\odot}yr^{-1}kpc^{-2}\), and -1.5 \(M_{\odot}yr^{-1}kpc^{-2}\) respectively}. 
\label{fig14}
\end{figure}

To understand this further, we plotted the histogram of \(\Sigma(SFR)\) for the SFCs of the three classes of SFDGs separately, as shown in Figure \ref{fig14}. The mode of log(\(\Sigma(SFR)\)) for dwarf spirals, dwarf irregulars, and BCDs is -1.84 \(M_{\odot}yr^{-1}kpc^{-2}\), -2.09 \(M_{\odot}yr^{-1}kpc^{-2}\), and -1.50 \(M_{\odot}yr^{-1}kpc^{-2}\) respectively. Hence, most of the SFCs of dwarf irregulars have a lower \(\Sigma(SFR)\) compared to dwarf spirals and BCDs. However, the SFCs with higher \(\Sigma(SFR)\) means that they either have a higher star formation rate than other SFCs in the galaxies or have a smaller area. Hence, the high \(\Sigma(SFR)\) for SFCs in the outer disk could be due to the comparable star formation rate to the inner complexes with minimal areas. But in Figure \ref{fig7}, we can see that the SFCs in the outer disk, which have higher \(\Sigma(SFR)\), have larger areas, implying a higher star formation rate in the outer disk.

It must be noted that the \(\Sigma(SFR)\) of NGC 6822 is not included as it has an unusually high \(\Sigma(SFR)\) compared to the other dwarf irregulars (See Figure \ref{fig7}). This could be due to a recent interaction that it has undergone, which is evident from its tidal features that can be seen in FUV and NUV images in Appendix \ref{sec:appendix}. Another reason could be that the NGC 6822 is very nearby. The SFCs detected in it are very small, which could be smaller unit-like clusters, and bound clusters are densest in the star groups \citep{2014ApJ...787L..15E}.

The \(\Sigma(SFR)\) distribution and lower cutoff are clearly connected to a variation in stellar disk surface densities and mean stellar masses in the three types of SFDGs  (see Table \ref{table5}). The mean stellar disk surface density is the largest for the BCDs, followed by the dwarf spirals and the irregulars. Also, the mode of \(\Sigma(SFR)\) is the highest for the BCDs and is the least for dwarf irregulars. This confirms that the stellar mass is the main factor influencing the SFRs, as shown in earlier studies of the SFMS \citep{2012ivoa.rept.1015R, 2014ApJS..214...15S}.

We have also listed \(log(M_{HI})\) in Table \ref{table1}. From those values, we can see that the dwarf irregulars have \(log(M_{HI})\) greater than \(log(M_{*})\), whereas, for dwarf spirals and BCDs, the stellar mass is greater. Hence, dwarf spirals and BCDs must have used gas to form stars at a higher rate, whereas dwarf irregular do not have stellar disk potentials strong enough to convert \(M_{HI}\) to stellar mass at such a rate \citep{2008MNRAS.385.2181F, 2012ivoa.rept.1015R}.\\

\subsection{Effect of interaction on star formation in SFDGs } \label{Effect of interaction on interacting and non-interacting dwarfs}
As our samples have both interacting dwarf galaxies and isolated dwarf galaxies, we could examine the effect of interactions on their SFRs. However, due to their low masses, the effect of interactions on their morphology or SFRs is not very clear \citep{2015ApJ...805....2S}. 
Some of the signatures of galaxy interaction are starburst activity, gaseous bridges, Magellanic-type streams, and extended tidal arms or tails. In this section, we describe some of these features in our sample.

As discussed in section \ref{subsec: Morphology of Dwarfs in UV emissions },
in the interacting dwarf galaxies NGC 4625, NGC 4618, and NGC 4625A, the galaxy NGC 4618 is more massive compared to the others. Hence, the extended arm in NGC 4625 could be due to the mass pulled out from the lower mass galaxy NGC 4625. We have even extracted some SFCs from those regions, as shown in Figure \ref{fig3}. The central disk in  NGC 4618 is very bright. This could be due to the funnelling of gas into the central disk, causing it to have enhanced star formation. In NGC 4618, the interaction has triggered star formation all over the galaxy, including along the bar. 

In NGC 5474, the interaction is with the very massive galaxy NGC 5457, with mass \(4.1\pm1.2\times 10^{10} M_{\odot} \) \citep{refId0}. Hence, the galaxy NGC 5474 is completely distorted and extended. The galaxy UGC 4305 also has an extended distorted arm-like structure in UV. NGC 6822 is thought to have undergone interaction twice with a companion galaxy and once with the Milky Way. We can distinguish tidal tails northeast and southwest of the galaxy centre. The southwestern tail looks like a stellar stream and is entirely absent in the IR image. Hence, its UV emission is probably due to a very recent star formation triggered by a recent interaction. It is suggested by \cite{10.1046/j.1365-8711.2003.06669.x} and \cite{2003ApJ...590L..17K} that NGC 6822 could be merged system, and the large cloud of gas at northwest could be the location of the other galaxy. It has also been mentioned that the interaction with the Milky Way in the southwestern region has a kinematic age of 100 Myr, and the merger in the Northeast region is 300 Myr.

\begin{figure}
\includegraphics[scale=0.6]{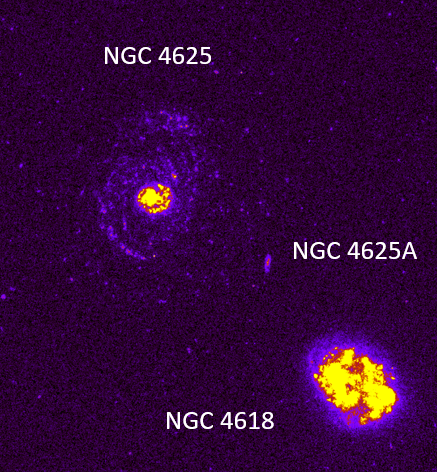}
\caption{UVIT FUV image of interacting galaxies  NGC 4625, NGC 4625A, and NGC 4618 from left .
\label{fig15}}
\end{figure}

Using New Horizon cosmological hydro-dynamical simulation, \cite{Martin_2020} has shown that interactions in dwarf galaxies enhance the overall SFR in dwarfs at low redshifts. Since we have very few galaxies that are interacting, and as we have a wide mass distribution for these galaxies, it is difficult to compare and confirm this result. But we do see clear differences in their radial trends compared to non-interacting dwarf galaxies as mentioned in \ref{subsec: Observed Radial trends of SFCs}. In NGC 4625, there are small SFCs in the tidal arms, but for NGC 4618, even outer SFCs are relatively large and have larger \(\Sigma(SFR)\). We also see this trend in NGC 5474 and NGC 6822. Hence, interaction affects the size and \(\Sigma(SFR)\) of SFCs of star-forming dwarf galaxies.

In Figure \ref{fig7} and as mentioned in sections \ref{subsec: Observed Radial trends of SFCs} and \ref{Difference in Star formation rate density for different Dwarf galaxy types}, SFCs of NGC 6822 has very high \(\Sigma(SFR)\) compared to other dwarf irregulars. The \(\Sigma(SFR)\) is comparable to dwarf spirals. It could be due to the recent interaction of the galaxy. It is believed that mergers can also drive starbursts in these galaxies \citep{Pustilnik_2001, 10.1093/mnras/stu556, 2015ApJ...805....2S}. Even in the SFMS parameters of this galaxy (Table \ref{table6}), we can see that it has a shallower SFMS slope $\alpha$ compared to other dwarf irregular galaxies. However, some outer complexes are showing a sudden increase in star formation rate after \(10^{3} M_{\odot}\). Hence, in some cases, this galaxy's SFC SFRs are not included when we have made overall conclusions about dwarf irregulars.

\subsection{The Global and local SFMS of SFDGs} \label{Difference in Star-forming main sequence}
Studies show that the SFMS slope \(\alpha\) in equation \ref{eq8} depends on time \citep{2014ApJS..214...15S}. This suggests that the galaxies where star formation appears quenched could have been actively forming stars back in time.\citep{Daddi_2007, Elbaz2007, 2014ApJS..214...15S}. Hence the Massive galaxies  (\(log(M_{*}/M_{\odot})\geq 10\)) at lower redshifts have a shallower slope of \(\alpha \approx 0.5\) as star formation is slowing in them, which explains the main sequence turnoff \citep{2014ApJS..214...15S, McGaugh_2017}. As mentioned in \cite{Leroy_2019}, the time to form the stellar disk in galaxies of mass $log(M_{*}/M_{\odot})=10$ is approximately 13-14 Gyr, which was also seen in \cite{2014ApJS..214...15S}. The galaxies with a slope of unity imply that they are formed at an early age in the universe and are forming stars at a constant rate. This has been observed in dwarf low surface brightness galaxies. They also take around 13 Gyr to build their entire stellar disk \citep{McGaugh_2017}.

The star-forming main sequence of SFDGs has a slope \(\alpha =0.899 \) as seen in equation \ref{eq9}. When we calculated the approximate age for SFDGs using global SFMS equation \ref{eq9}, as mentioned in \cite{McGaugh_2017}, we found the age to be 2-4 Gyr. This implies that the SFDGs are recently formed \citep{McGaugh_2017} or must be forming stars at a very high rate in recent times compared to dwarf LSBs.  

The higher sSFRs compared to LSB galaxies (LSBGs), as seen in Figure \ref{fig9}, could be due to a higher stellar density in SFDGs compared to LSB disks. The sudden bursts of star formation can also contribute to the stellar disk formation at a higher rate, leading to a higher gas consumption rate. Due to this, SFDGs may get quenched earlier than LSBGs but not as fast as massive galaxies. Gas can also be lost due to supernova explosions as the dwarfs do not have enough potential to retain the gas.

When we consider the local SFMS in Figure \ref{fig10}, it is remarkable that over a range of a factor of $10^{5}$ in stellar disk mass, the SFR within a complex is strongly correlated with the local stellar disk mass. We know from Toomre's theory of local disk instability that stellar mass density provides the self-gravity for cloud collapse and star formation. We see that SFCs in dwarf irregular have a mean $log(M_{*}/M_{\odot})$ around 4-4.5, whereas dwarf spirals and BCDs have around mean $log(M_{*}/M_{\odot})$ around 5.5-6. We used these masses to calculate star formation age collectively for SFCs of different SFDG types using the mean equation from Table \ref{table6}. We got an approximate age of 1Gyr for all the types, which is a little less than the global SFMS of SFDGs but is still comparable. This can also be seen in Figure \ref{fig10}, where $1/(sSFR)$ is approximately equal to 1Gyr for the entire sample.

\begin{figure}    
\includegraphics[width=\columnwidth]{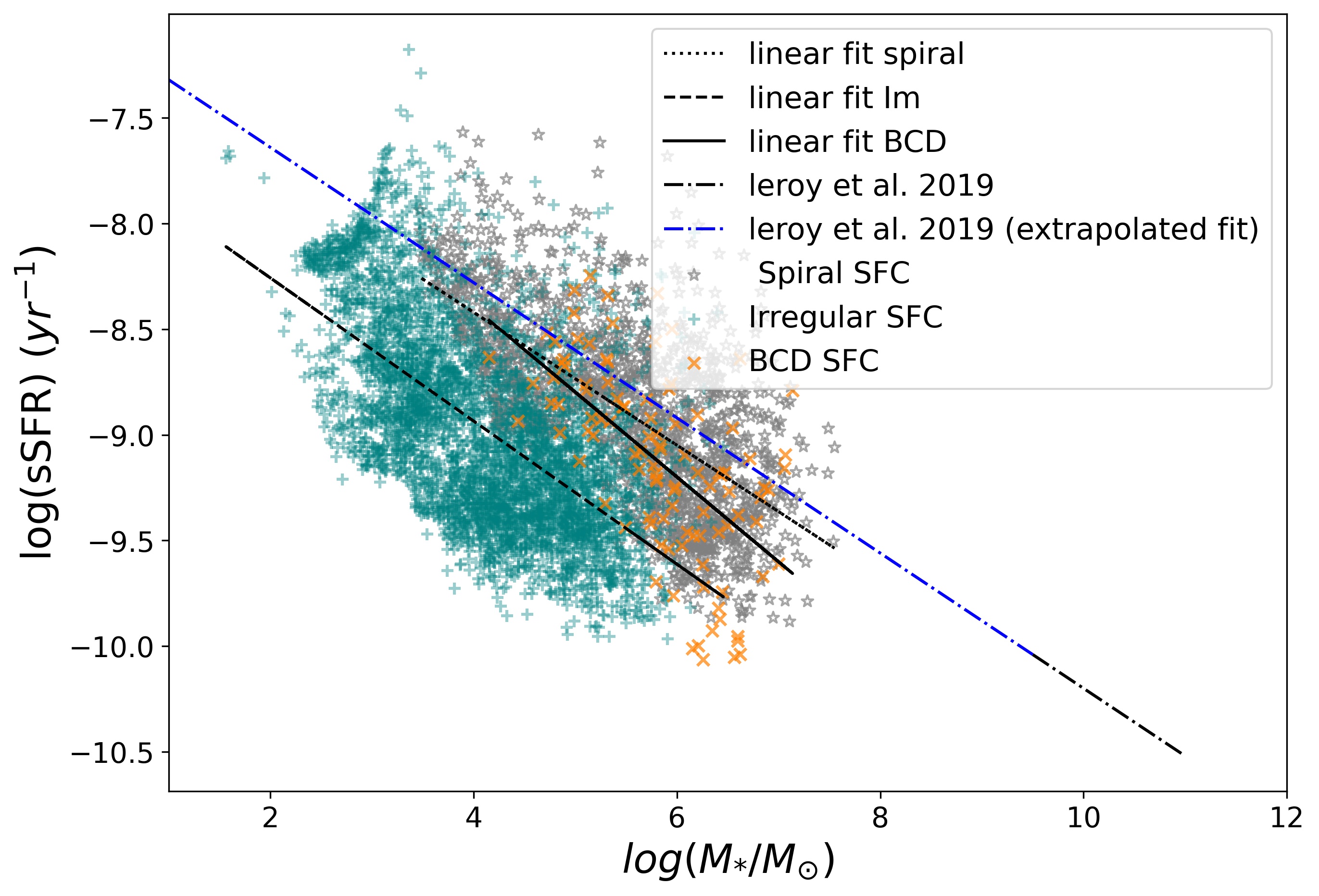}
\caption{sSFR and stellar disk mass correlation for SFCs. The black dash-dotted line is the fit from \citealt{Leroy_2019}. The blue dash-dotted line is its extrapolated fit for lower mass}.
\label{sSFRmass}
\end{figure}

In addition to Figure \ref{fig10}, we plotted the sSFR with the SFC stellar disk mass as shown in Figure \ref{sSFRmass} to further understand the importance of the local stellar disk density for star formation in SFDGs. The equation of SFCs for dwarf spirals and dwarf irregulars is very similar to the equation in \cite{Leroy_2019} where the mass range considered for the fit is $10^{9.5} M_{\odot}$ to $ 10^{11} M_{\odot}$. So, when we substitute the average mass of the SFCs, we see the difference that the age of local star formation in SFDGs is much smaller than the global stellar disk formation time for massive galaxies. It will be interesting to compare the local SFMS of massive galaxies with the global SFMS of massive galaxies.

\subsection{Difference in SFMS of the different SFDGs and inside-out Evolution of Galaxies} \label{Difference in Star-forming main sequence in Dwarf types}
The SFCs following the SFMS give us an idea of how each of our sample galaxies is evolving differently. We see in section \ref{subsec:Star-forming main sequence for SFCs }, where each SFDG type has a different average value of the SFMS slope, which indicates that the star formation process is slightly different for each type. For dwarf spirals, the slope value is closer to the slope of massive spirals, in which star formation quenches and evolves to lower SFRs as gas is depleted \citep{Oemler_2017}. But most of the irregulars have a slope closer to or a little less than unity. Galaxies with slope $\alpha=1$ are expected to evolve at a constant rate, whereas lower $\alpha$ signifies that the galaxies are forming stars at a higher rate in recent times \citep{McGaugh_2017}. The BCDs have an $\alpha$ lying somewhere between dwarf spirals and irregulars.However, the approximate time for stellar disk formation is almost the same for all three types, as seen in the previous section.

These galaxies' inner, middle, and outer SFCs, as categorized in section \ref{subsec:Star-forming main sequence for SFCs }, follow a separate trend. Outer complexes are mostly above the fit, and the inner ones are below. Hence, we find that outer complexes have higher sSFRs than the inner ones, indicating that the star-forming mechanism or evolution of the inner and outer regions of the galaxies are different and depend strongly on the stellar surface density.

Several studies have been conducted to understand the star-forming mechanisms in dwarf galaxies. Previous studies have shown that the spirals mostly have the inside-out star formation mechanism \citep{1991ApJ...379...52W, 2000MNRAS.312..497B, 2006ApJ...639..126B}. The extended UV emission discovered in outer disks by \cite{2007ApJS..173..538T, Boissier_2007} clearly indicated inside-out disk formation in those galaxies. Some studies have even shown that the late-type spirals, which are less massive, also follow this trend \citep{Dale_2020, 10.1093/mnras/stac1974}. A recent study by \cite{Mondal_2021} using UVIT on a dwarf spiral also confirmed it. But studies on dwarf irregulars, like \cite{Zhang_2012} studied 34 nearby dwarf irregulars and concluded that the radial profiles of 80$\%$ of his samples at shorter wavelengths have shorter disk scale lengths than those at longer wavelengths, indicating the outside-in disk evolution by disk shrinking. Even \cite{Dale_2020} showed that irregulars show the opposite evolution trend to the spirals.

In our sample, we have galaxies evolving via ex-situ star formation (via gas accretion or interactions)\citep{Chamba_2022}. These are expected to grow inside-out, as indicated by previous studies, which is also seen in our results. The galaxies (NGC 5474, NGC 2541, NGC 6822, etc.) have higher sSFR in the outer disk than the inner, and in Figure \ref{fig11}, we can see a clear distinction between the three regions. Some SFDGs classified as isolated, like NGC 4136 and WLM, have outer complexes with relatively higher sSFR. But in NGC 4625, NGC 4395 and in most dwarf irregulars like UGC 7608, IC 2574, and even in the galaxy UGC 4305, where the star formation in the outskirts might have been triggered by external influence, doesn't show much gradient.
Although most dwarf spirals and irregulars show inside-out evolution, we find stronger evolution in galaxies where ex-situ star formation is observed. The gradient in some galaxies might also depend on how strong the interaction or accretion is. Dwarf irregulars in the sample generally show a moderate gradient or no gradient. Other than accretion and interaction, the local star formation in the outer disk of dwarfs can also be supported by dark matter halo \citep{Das_2023}.

Unlike the strong gradient observed in SFCs supporting inside-out evolution in Figure \ref{fig11} (like in NGC 2541, NGC 4618, and NGC 6822), the radial variation of sSFR over annuli shows a radially decreasing or no specific trend as seen in Figure \ref{alpha}. One of the reasons is that some galaxies are highly extended, and their tidal tails are spread in a very irregular fashion (NGC 6822). These regions in the outer radius are challenging to consider in elliptical annuli, which results in an underestimation of sSFR. The other reason is that the SFCs depend on the local stellar mass. This does not clearly come out on a radial average.

\subsection{The nature of star formation in SFDGs compared to massive galaxies} \label{Nature of star formations in star-forming dwarfs compared to massive galaxies}
One of the main differences between massive star forming galaxies and SFDGs is their gravitational potential and metallicity. These are necessary for retaining the cold gas and supporting star formation. Dwarf galaxies have shallower potentials and are metal-poor compared to massive galaxies. As a result, many of them lack prominent spiral arms and bars, which are the main global disk instabilities. These instabilities can make dense molecular clouds settle in either the spiral arms or bar ends. The lack of such instabilities makes star formation spread all over the galaxies \citep{1996ApJ...467..579E}. Hence, we expect different star forming properties in low-mass systems than in massive galaxies. The only dwarfs that have prominent spiral arms are the dwarf spirals, but they are at most Sd and Sm types. They lack grand spirals and instead have flocculent spiral arms.

In this subsection, We will compare our properties with previous studies of the massive galaxies studied using UVIT so that there should not be any instrumental bias. We are comparing with galaxies studied by \cite{Yadav_2021}, where massive XUV galaxies are analysed with a UVIT telescope. The SFCs are extracted to compare their properties in the inner and outer regions. However, to compare the sizes of SFCs in the massive and dwarf galaxies, the galaxies should be at comparable distances. Hence, we compared the sizes of SFCs of NGC 5474 with NGC 628 and NGC 5457.

 While comparing with NGC 628 and NGC 5457, we find that the smallest SFC detected in NGC 5474 is of the order of \(10^{-2.5}\) \(kpc^2\). And most of the SFCs are between \(10^{-2.5}\) \(kpc^2\) to  \(10^{-1}\) \(kpc^2\). The larger SFCs that we find in NGC 5474 are of size \(10^{-0.75}\) \(kpc^2\), whereas in massive galaxies, there are SFCs with sizes larger than \(10^{0}\) \(kpc^2\). Hence, massive galaxies have SFCs with a larger range than dwarf galaxies such as NGC 5474. They are generally larger in the centre, and they become smaller as we go radially outwards. Some of our samples are farther than these massive galaxies; still, we see that they have SFCs with sizes less than 1 \(kpc^2\). This result is consistent with previous studies. \cite{1996ApJ...467..579E} that had shown that large galaxies form large star-forming complexes that form OB associations for a longer time, whereas dwarf galaxies form small complexes which, in turn, form dense, bright associations for short periods. The reason for the small SFC sizes in dwarfs could be low stellar density and shallower potential, which does not support the formation of larger complexes, although the shear is more for massive galaxies than dwarfs. The smallest complex detected in our sample is from the nearest galaxy, NGC 6822, with an area of around \(10^{-4.5}\) \(kpc^2\).

We can also compare found \(\Sigma(SFR)\) in our galaxy with massive spiral galaxies. We see that dwarf spirals have a higher value of \(\Sigma(SFR)\) than massive spirals from \cite{Yadav_2021}. Hence, the dwarf spirals have much more compact SFCs. Another reason is that the SFR is not corrected for host galaxy extinction in massive galaxies. Hence, the range of \(\Sigma(SFR)\) is similar to dwarf irregulars where the environment for star formation is completely different.

There are some studies such as \cite{1999AJ....117..764E, 2014ApJ...787L..15E}, which discusses the hierarchical structure of these complexes. According to them, large galaxies and LSB dwarfs have a tendency to form uniformly spread-out complexes. The hierarchical structure is seen within these complexes, whereas bright starburst dwarfs have very few large complexes with well-resolved hierarchical star formation inside.

 \section{Conclusions} \label{sec:Conclusions}
In this project, we used UVIT to study 16 star-forming dwarf galaxies and compare the star formation properties of the three SFDG types. This is done by extracting UV-bright star-forming complexes and by analyzing their properties. Our main results are summarised below:\\
1. Most of the sample dwarf spirals have flocculent spiral arms and extended UV disks. Dwarf irregulars have clumpy star formation and some ill-defined arms, whereas BCDs have very compact star forming inner disks and outer LSB disks. This result is consistent with the previous studies.\\
2. For most of the SFDGs in our sample, we have detected more FUV SFCs than NUV SFCs. This could be due to the recent star formation in these galaxies. Some Extended UV galaxies (XUV) like NGC 5474 and NGC 2541 show high star formation in sub-threshold environments, and they have more outer disk complexes (beyond optical radius) than inner disks where the stellar surface density is comparatively high.

3. Dwarf spirals have complexes with a higher area, \(\Sigma(SFR)\), and mass in the inner arms, rings, and at the ends of the bars compared to other SFCs. In some galaxies, it is higher even along the bar. The radial trends of these properties show that the SFR decreases as we go radially outwards. Irregulars and BCDs have no regular pattern in their radial trends. However, in general, outer complexes have smaller sizes, lower \(\Sigma(SFR)\), and mass. 

4. We find that interacting galaxies such as NGC 4618, NGC 5474, and NGC 6822 show an overall high \(\Sigma(SFR)\); they also show star formation in their outer regions that are triggered by interaction. These galaxies show radial trends different from non-interacting galaxies.

5. We see that the lower limit for \(\Sigma(SFR)\) for SFCs of each type is different. Dwarf spirals have a higher threshold or limit for \(\Sigma(SFR)\) ( 0.0078 \(M_{\odot}yr^{-1}\)), and dwarf irregulars have the least (0.0008 \(M_{\odot}yr^{-1}\)). But the mode of log(\(\Sigma(SFR)\)) for SFCs is higher for BCD with the value -1.50 \(M_{\odot}yr^{-1}\). The mode of log(\(\Sigma(SFR)\)) for SFCs of dwarf irregulars is lower than dwarf spirals, which are -2.09 \(M_{\odot}yr^{-1}\), and -1.84 \(M_{\odot}yr^{-1}\) respectively.

6.SFDGs follow the SFMS with the slope and intercept as \(\alpha =0.899 \pm 0.087 \) and \(\beta= -8.59 \pm 0.75\). This fit is shallower than LSB dwarfs and steeper than massive galaxies, indicating recent high star formation in them. The age of stellar disk formation is found to be around 2-4 Gyr.

7. We find that the SFCs of SFDGs also follow the SFMS, but the slope $\alpha$ is different for the different types. The mean slope of the SFMS for dwarf spirals is 0.74. In comparison, dwarf irregulars and BCD have slopes of 0.87 and 0.8, respectively. However, the age of SFCs approximately corresponds to 1 Gyr.

8. In the SFMS plots of mainly dwarf spirals, we find that the outer complexes have higher specific star formation rates compared to the inner ones. This is prominent in the interacting irregular galaxy NGC 6822. This implies that the star formation mechanisms in the inner and outer regions of galaxies are different, which supports the inside-out evolution of galaxies. It is also different in different types of SFDGs.

9. Dwarf spirals have compact SFCs compared to massive galaxies. However, the SFCs of massive galaxies have a larger range of sizes and \(\Sigma(SFR)\).

\section*{Acknowledgements}
  This publication uses data from UVIT, which is part of the AstroSat mission of the Indian Space Research Organisation (ISRO) and is archived at the Indian Space Science Data Centre (ISSDC). We gratefully thank all the members of various teams for supporting the project from the early stages of design to launch and observations in orbit. This research has used Spitzer 3.6 microns and Spitzer MIPS images. This research has made use of the NASA/IPAC Extragalactic Database (NED), which is funded by the National Aeronautics and Space Administration and operated by the California Institute of Technology. MD acknowledges the support of the Science and Engineering Research Board (SERB) Core Research Grant CRG/2022/004531 for this research. Finally, we acknowledge the referee for the valuable suggestions.

\section*{Data Availability}
All the UVIT data used in this paper are publicly available at \url{https://astrobrowse.issdc.gov.in/astro_archive/archive/Home.jsp.}

Spitzer 3.6 and 24 \(\mu \)m data used in this paper are publicly available at
\url{https://irsa.ipac.caltech.edu/applications/Spitzer/SHA/}

Some of the data are also taken from NED images at
\url{ https://ned.ipac.caltech.edu/}

\bibliographystyle{mnras}
\bibliography{Manuscript} 



\appendix

\section{Some extra material}\label{sec:appendix}
Here, we present the sample catalogue of properties of FUV SFCs in NGC 4136. The entire table for all SFCs of SFDGs is available in machine-readable form. We also present images of our sample dwarf galaxies in FUV, NUV, NIR, and Optical B/g bands. 

\begin{landscape}

\begin{table}

    \centering
            \caption{Catalogue of properties of FUV SFCs in NGC 4136}
                \label{table A1}
                \begin{tabular}{cccccccccccccc}
            \hline
            
    Galaxy & R.A.& Decl.& a & b & P.A. & Area  & $\frac{\Delta Area}{Area} $ & \(Flux_{FUV}\) &\(\Delta Flux_{FUV} \) & \(\Sigma (SFR_{FUV)}\) & \(\Sigma (SFR_{FUV+24)}\) & \(\Delta \Sigma (SFR_{FUV+24})\) & \(log(M_{*})\) \\ 
    &(J2000)&(J2000)&&&&&&&&&&&\\
    (name) & (deg) & (deg) & (") & (") & (deg) & (\(arcsec^{2}\)) && (\(\mu Jy\))&(\(\mu Jy\))& (\(M_\odot yr^{-1}kpc^{-2}\)) & (\(M_\odot yr^{-1}kpc^{-2}\)) & (\(M_\odot yr^{-1}kpc^{-2}\)) & (\(M_\odot \))\\
    \hline
    
NGC 4136 & 182.3092913 & 29.95130702  & 0.7  & 0.6 & -39.397  & 1.308   & 0.268                       & 1.67      & 0.02              & 0.004                & 0.042                   & 0.0111                         & 4.453        \\
NGC 4136 & 182.3186966 & 29.95074998  & 2.3  & 1.4 & -71.698  & 10.424  & 0.099                       & 11.65     & 0.11              & 0.004                & 0.039                   & 0.0039                         & 5.502        \\
NGC 4136 & 182.3177649 & 29.94971279  & 2    & 1.8 & -50.788  & 11.545  & 0.118                       & 11.09     & 0.1               & 0.003                & 0.035                   & 0.0041                         & 5.597        \\
NGC 4136 & 182.3306337 & 29.94756966  & 1    & 0.9 & -26.472  & 2.858   & 0.169                       & 3.62      & 0.03              & 0.004                & 0.046                   & 0.0079                         & 5.117        \\
NGC 4136 & 182.3126736 & 29.94640579  & 1.4  & 0.7 & -9.517   & 3.202   & 0.171                       & 3.94      & 0.04              & 0.004                & 0.045                   & 0.0078                         & 5.143        \\
NGC 4136 & 182.3253775 & 29.9471035   & 1.8  & 1.5 & -80.612  & 8.834   & 0.093                       & 11.51     & 0.11              & 0.004                & 0.045                   & 0.0042                         & 5.679        \\
NGC 4136 & 182.3242126 & 29.94569841  & 1.3  & 1   & 4.339    & 4.362   & 0.159                       & 4.76      & 0.05              & 0.004                & 0.035                   & 0.0056                         & 5.425        \\
NGC 4136 & 182.3306297 & 29.94458672  & 1.1  & 0.8 & -34.216  & 2.853   & 0.273                       & 2.3       & 0.02              & 0.003                & 0.026                   & 0.0071                         & 5.243        \\
NGC 4136 & 182.317924  & 29.94340004  & 2.4  & 0.8 & 84.887   & 5.903   & 0.166                       & 5.44      & 0.05              & 0.003                & 0.028                   & 0.0047                         & 5.585        \\
NGC 4136 & 182.3128679 & 29.94262623  & 1.4  & 0.9 & 30.189   & 3.874   & 0.15                        & 4.64      & 0.04              & 0.004                & 0.038                   & 0.0058                         & 5.379        \\
NGC 4136 & 182.3194354 & 29.94513634  & 2.7  & 2   & -26.569  & 17.047  & 0.034                       & 53.68     & 0.5               & 0.011                & 0.101                   & 0.0035                         & 6.02         \\
NGC 4136 & 182.3201493 & 29.94383131  & 3.8  & 3   & -88.112  & 35.84   & 0.022                       & 121.72    & 1.14              & 0.012                & 0.104                   & 0.0025                         & 6.375        \\
NGC 4136 & 182.3219271 & 29.94624008  & 3.1  & 2   & -22.249  & 19.241  & 0.06                        & 27.59     & 0.26              & 0.005                & 0.047                   & 0.0028                         & 6.058        \\
NGC 4136 & 182.3226768 & 29.94741117  & 2.9  & 2.3 & 50.334   & 21.31   & 0.064                       & 25.74     & 0.24              & 0.004                & 0.042                   & 0.0027                         & 6.046        \\
NGC 4136 & 182.3300471 & 29.94308902  & 2.5  & 1.9 & -31.203  & 14.543  & 0.064                       & 22.01     & 0.21              & 0.005                & 0.046                   & 0.0029                         & 5.998        \\
NGC 4136 & 182.3236461 & 29.94185716  & 2.3  & 1.8 & -43.176  & 12.787  & 0.082                       & 15.92     & 0.15              & 0.004                & 0.03                    & 0.0025                         & 6.108        \\
NGC 4136 & 182.2918385 & 29.93957073  & 1    & 0.6 & -86.711  & 2.045   & 0.25                        & 2.2       & 0.02              & 0.004                & 0.035                   & 0.0088                         & 4.647        \\
NGC 4136 & 182.313994  & 29.94079259  & 1.6  & 0.5 & 47.31    & 2.74    & 0.226                       & 2.76      & 0.03              & 0.003                & 0.028                   & 0.0064                         & 5.329        \\
NGC 4136 & 182.3265439 & 29.94152887  & 1.9  & 1.3 & -5.855   & 7.439   & 0.118                       & 8.14      & 0.08              & 0.004                & 0.027                   & 0.0031                         & 5.887        \\
NGC 4136 & 182.2951342 & 29.94018928  & 2.8  & 1.6 & -70.683  & 14.357  & 0.058                       & 23.61     & 0.22              & 0.006                & 0.054                   & 0.0031 & 5.49\\
NGC 4136 & 182.2945662 & 29.93933359  & 2.1  & 1.2 & 62.842   & 7.786   & 0.086                       & 11.71     & 0.11              & 0.005                & 0.049                   & 0.0042                         & 5.228        \\
NGC 4136 & 182.318033  & 29.93979962  & 1.6  & 0.6 & -43.976  & 2.916   & 0.218                       & 2.9       & 0.03              & 0.003                & 0.025                   & 0.0054                         & 5.56         \\
NGC 4136 & 182.339218  & 29.94119165  & 1.9  & 1.5 & -25.62   & 9.099   & 0.084                       & 12.94     & 0.12              & 0.005                & 0.048                   & 0.004                          & 5.719        \\
NGC 4136 & 182.3376932 & 29.9395083   & 1.7  & 1   & 17.996   & 5.151   & 0.141                       & 5.63      & 0.05              & 0.004                & 0.034                   & 0.0048                         & 5.517        \\
NGC 4136 & 182.3269627 & 29.9387716   & 1.9  & 1   & -30.943  & 5.96    & 0.143                       & 6.29      & 0.06              & 0.004                & 0.028                   & 0.004                          & 6            \\
NGC 4136 & 182.3293742 & 29.93842987  & 1.4  & 1.2 & -12.864  & 5.425   & 0.145                       & 5.59      & 0.05              & 0.004                & 0.027                   & 0.0038                         & 5.928        \\
NGC 4136 & 182.3213755 & 29.93731001  & 1.8  & 1.3 & 13.622   & 7.041   & 0.14                        & 6.88      & 0.06              & 0.003                & 0.03                    & 0.0041                         & 6.192        \\
NGC 4136 & 182.2925984 & 29.93477951  & 1.6  & 0.9 & 80.923   & 4.688   & 0.145                       & 5.41      & 0.05              & 0.004                & 0.038                   & 0.0055                         & 5.008        \\
NGC 4136 & 182.3295021 & 29.93671868  & 1.3  & 1.1 & -41.069  & 4.477   & 0.141                       & 5.52      & 0.05              & 0.004                & 0.036                   & 0.0051                         & 5.968        \\
NGC 4136 & 182.3435808 & 29.93697108  & 2.1  & 0.8 & -69.21   & 5.064   & 0.162                       & 4.94      & 0.05              & 0.003                & 0.033                   & 0.0054                         & 5.442        \\
NGC 4136 & 182.341325  & 29.93669755  & 1.7  & 1.2 & -23.192  & 6.295   & 0.135                       & 6.55      & 0.06              & 0.004                & 0.033                   & 0.0045                         & 5.592        \\
NGC 4136 & 182.3060872 & 29.93384785  & 1.5  & 0.9 & -78.655  & 4.384   & 0.146                       & 5.26      & 0.05              & 0.004                & 0.037                   & 0.0054                         & 5.456        \\
NGC 4136 & 182.3181183 & 29.93429779  & 1.8  & 0.7 & -10.293  & 4.086   & 0.197                       & 3.98      & 0.04              & 0.003                & 0.033                   & 0.0065                         & 6.053        \\
NGC 4136 & 182.3125184 & 29.93779459  & 3.2  & 2.6 & 10.288   & 26.708  & 0.022                       & 108.21    & 1.02              & 0.014                & 0.098                   & 0.0023                         & 6.438        \\

            \hline
            \end{tabular}
            
            \begin{minipage}{220mm}
             \textit{Note}: Here, R.A.(J2000) and Decl. (J2000) are the right ascension and declination of the SFCs. 'a' and 'b' are the semi-major and semi-minor axis of an ellipse fitted to SFCs in arcsec. P.A. is the position angle of the SFC. Area is the area of the SFCs found using equation \ref{eq1}. $\frac{\Delta Area}{Area} $ is the relative error associated with the estimated area of SFCs.  \(Flux_{FUV}\) and \(\Delta Flux_{FUV} \) are the FUV flux and error associated with the FUV flux, respectively, in \(10^{-6} Jy\).  \(\Sigma (SFR_{FUV})\) and \(\Sigma (SFR_{FUV+24})\) are the FUV and total star formation density of SFCs respectively in \(M_\odot yr^{-1}kpc^{-2}\). \(\Delta \Sigma (SFR_{FUV+24})\) is error associated with total star formation density in \(M_\odot yr^{-1}kpc^{-2}\). \(log(M_{*})\) is log of stellar mass of SFCs in \(M_\odot \).  (The entire table for all SFCs of SFDGs is available in machine-readable form.)
              \end{minipage}
            
\end{table}
\end{landscape}

\begin{table}
    \centering
                    \caption{Calculated parameters for extinction-free SFR}
                        \label{table A2}
                        \begin{tabular}{cccccccccc}
                    \hline

                    Galaxy & \(F_{FUV}\) & \(F_{24}\)  & \(\alpha\) \\ 
                     &\(mJy\) & \(mJy\)  &  \\
                    \hline
                    NGC 4136 &10.76\(\pm\)0.10&215.88\(\pm\)2.44 &20.06\(\pm\)0.29\\
                    NGC 4618 &24.75\(\pm\)0.23&384.47\(\pm\)1.74 &15.53\(\pm\)0.16\\
                    NGC 4625 &6.81\(\pm\)0.06 &146.46\(\pm\)1.08 &21.50\(\pm\)0.26\\
                    NGC 5474 &27.55\(\pm\)0.31&166.14\(\pm\)1.15&6.03\(\pm\)0.08\\
                    NGC 5832 &6.49\(\pm\)0.06 &49.16\(\pm\)0.62&7.57\(\pm\)0.12\\
                    NGC 4395 &73.96\(\pm\)0.69&486.02\(\pm\) 1.9&6.57\(\pm\)0.07 \\
                    NGC 2541 &14.36\(\pm\)0.16 &0.1419\(\pm\) &21.87\(\pm\)0.29\\
                    \hline
                    UGC 7608 &4.87\(\pm\)0.04 &27.88\(\pm\)0.47 &5.72\(\pm\)0.11\\
                    WLM  &36.15\(\pm\)0.34 &112.59\(\pm\)1.12 &3.11\(\pm\)0.04\\
                    UGC 4305 &40.18\(\pm\)0.38 &198.18\(\pm\) 1.25&4.93\(\pm\)0.06\\
                    NGC 6822 &201.41\(\pm\)13.48&2223.49\(\pm\) 4.2&11.04\(\pm\)0.74\\ 
                    IC 2574 &53.3\(\pm\)0.5 &343.89\(\pm\)1.65 &6.45\(\pm\)0.08\\
                    \hline
                    NGC 3738 &11.31\(\pm\)0.13 &96.7\(\pm\)0.88 &8.55\(\pm\)0.12\\
                    VIIZw403 &3.16\(\pm\)0.04 &24.43\(\pm\)0.44&7.73\(\pm\)0.17\\ 
                    Haro 36 &2.52\(\pm\)0.00 &7.7\(\pm\)0.24 &3.05\(\pm\)0.05\\ 
                    Mrk 5&1.12\(\pm\)0.01 &4.592\(\pm\)0.5 &4.1\(\pm\)0.07\\ 
                    \hline
                        \end{tabular}
                        \begin{minipage}{80mm}
                   \textit{Note}: \(F_{FUV}\) and \(F_{24}\) are the FUV and 24 microns integrated flux in mJy estimated within the outer ellipse of the elliptical fitting (see Figure \ref{fig5}) to the SFDGs. \(\alpha\) is ratio of \(F_{FUV}\) and \(F_{24}\).
                    \end{minipage}

\end{table}

\begin{figure*}

\fbox{\includegraphics[scale=0.3]{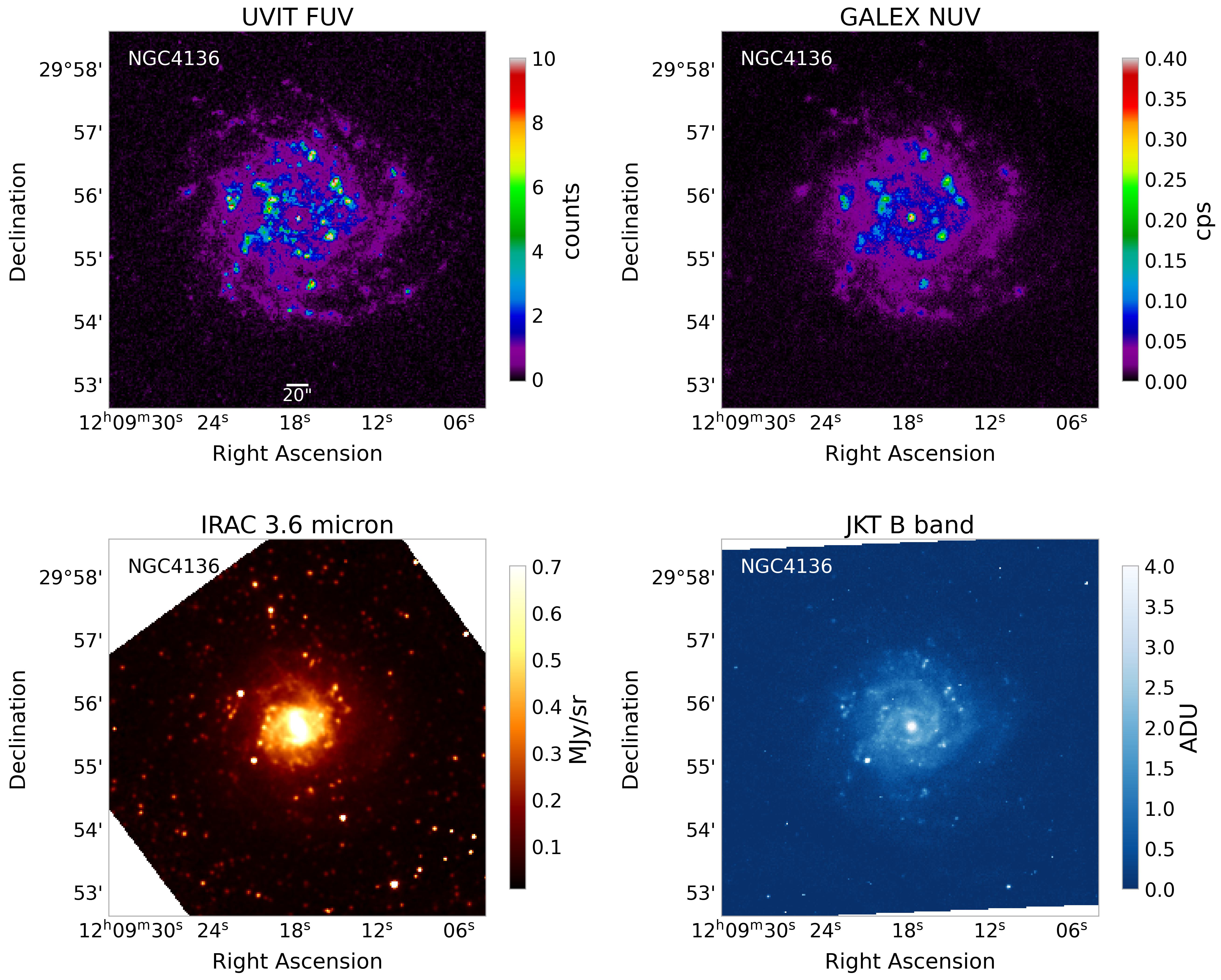}}
\end{figure*}

\vspace{0.2cm}

\begin{figure*}

\fbox{\includegraphics[scale=0.3]{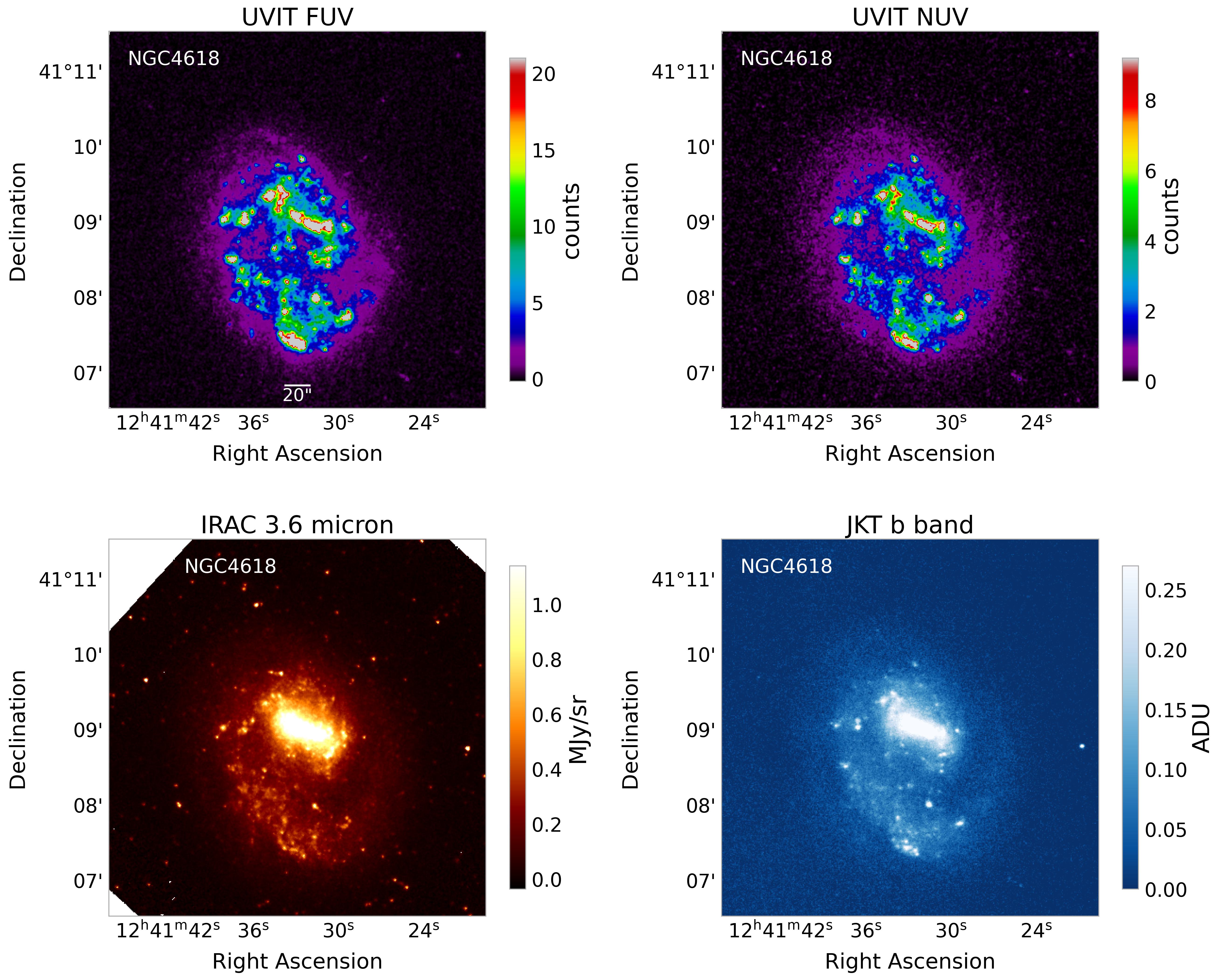}}
\end{figure*}

\begin{figure*}
\fbox{\includegraphics[scale=0.3]{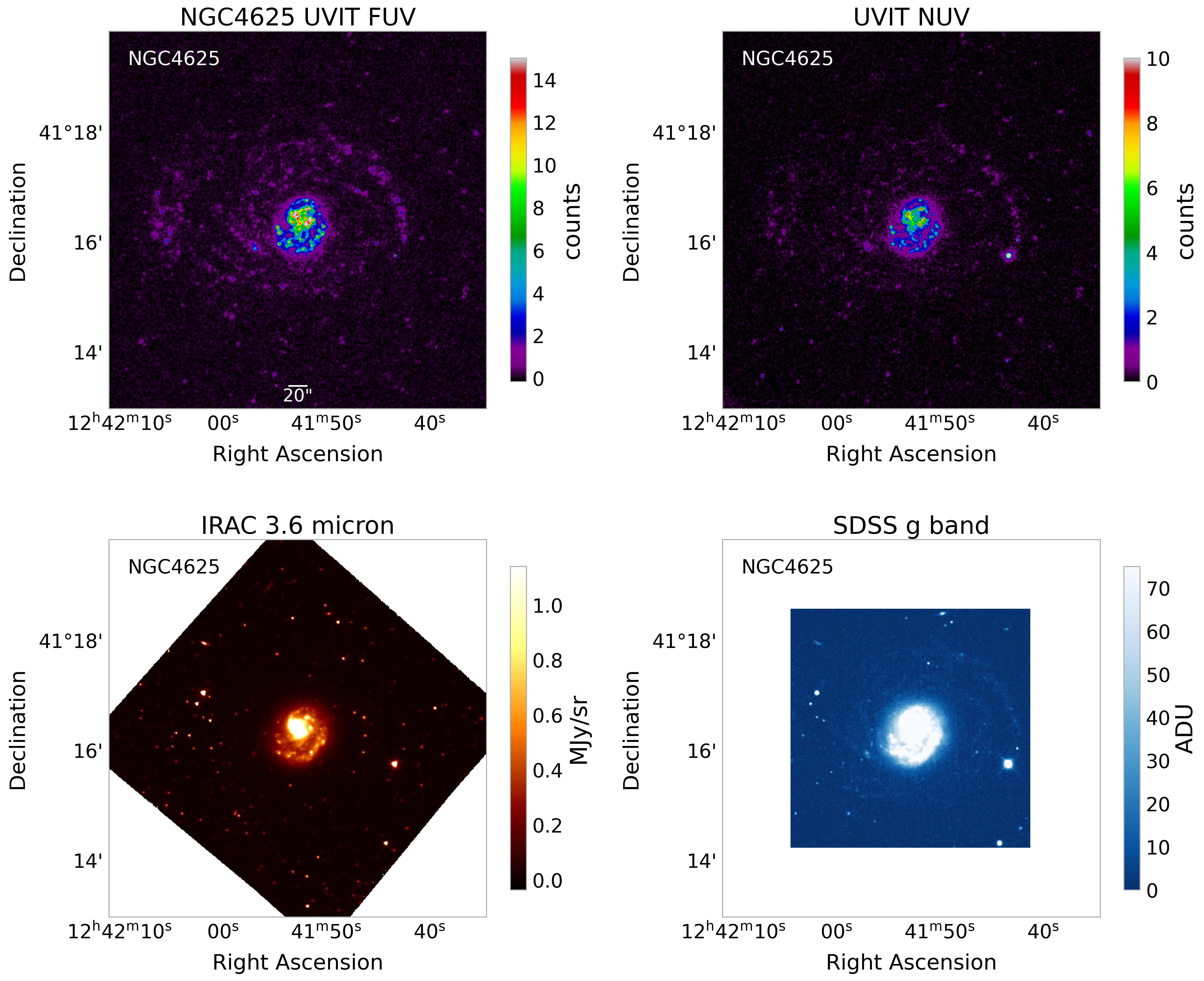}}
\end{figure*}

\vspace{0.2cm}

\begin{figure*}

\fbox{\includegraphics[scale=0.3]{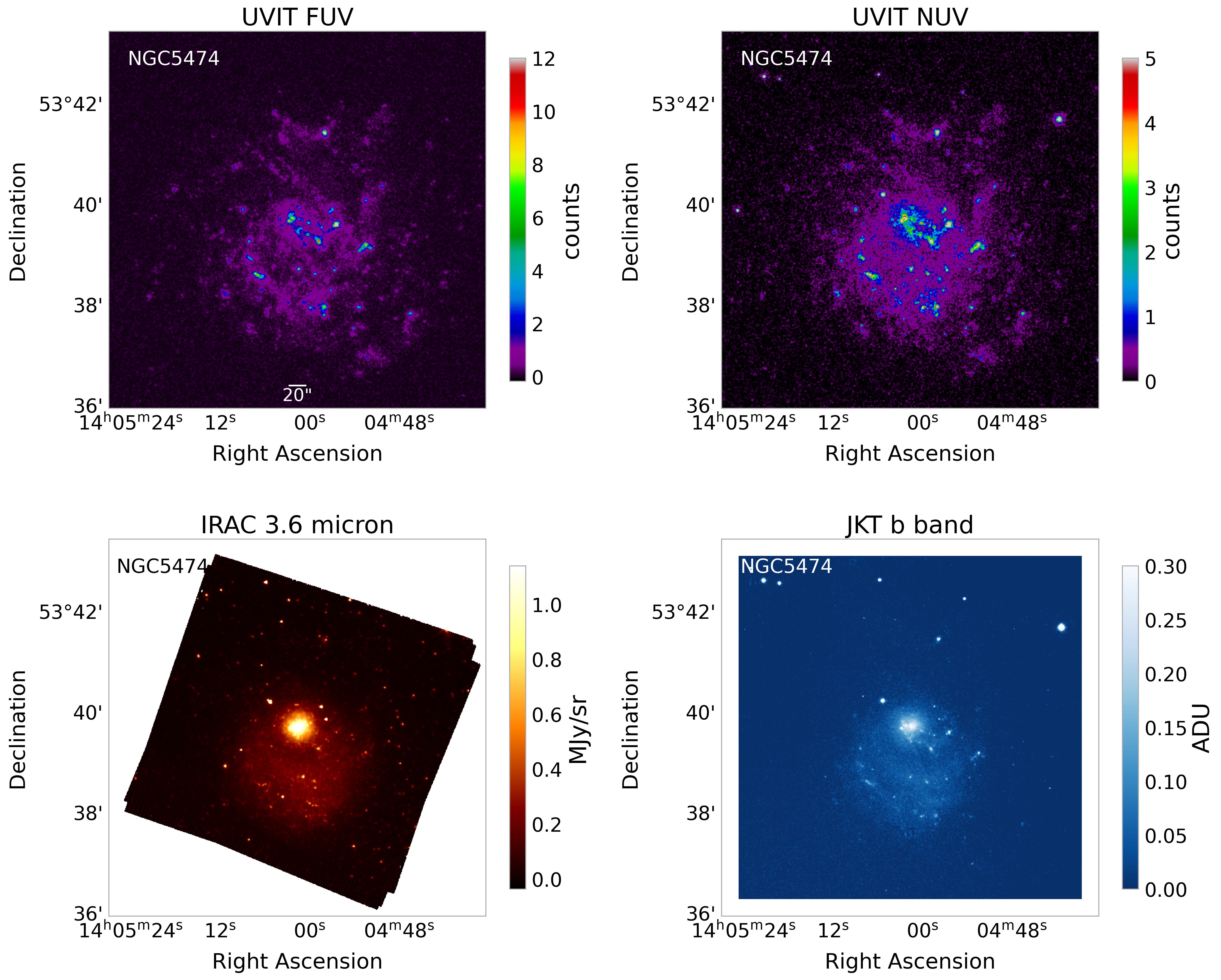}}
\end{figure*}

\begin{figure*}
\fbox{\includegraphics[scale=0.3]{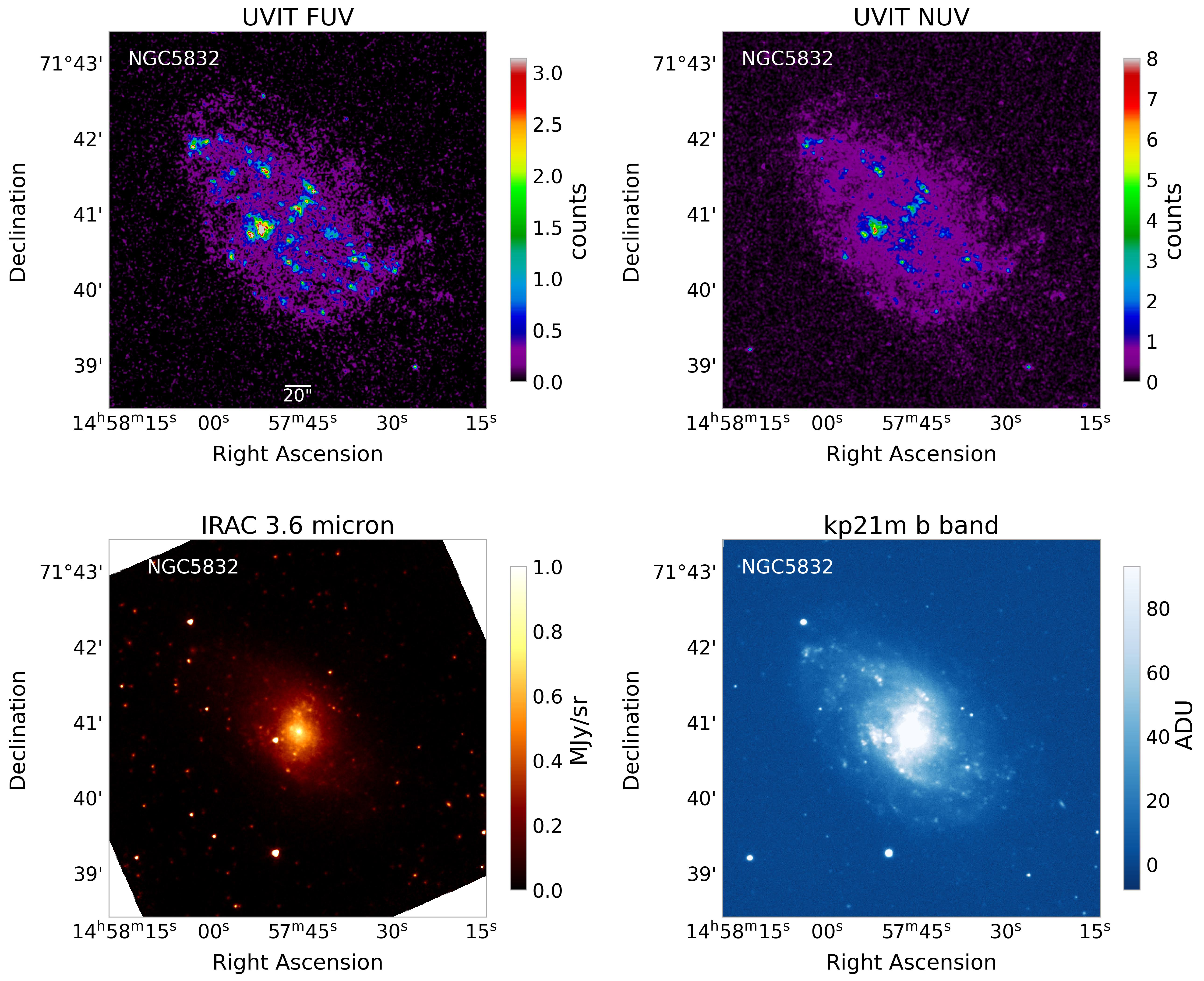}}
\end{figure*}

\vspace{0.2cm}

\begin{figure*}

\fbox{\includegraphics[scale=0.3]{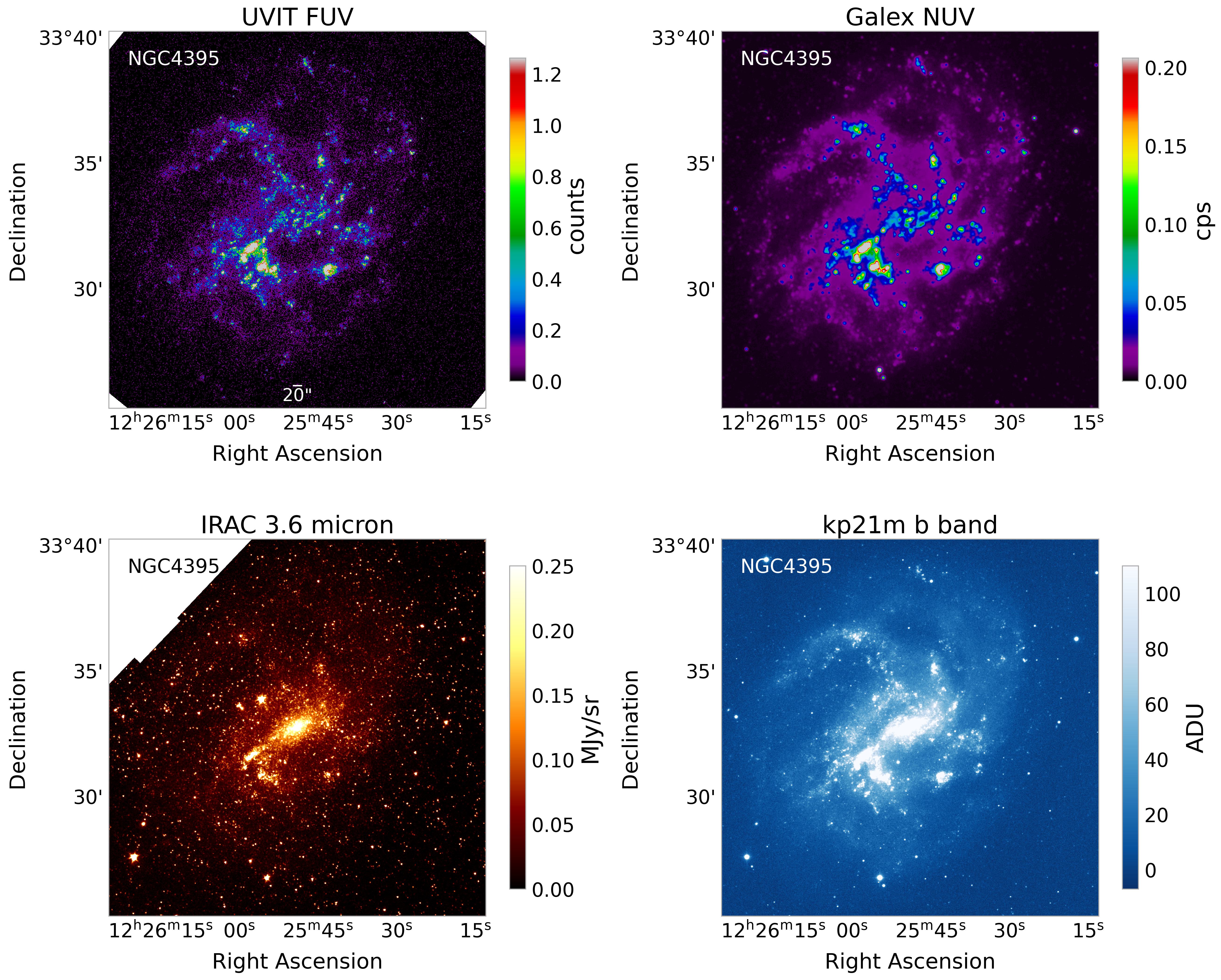}}
\end{figure*}

\vspace{1cm}

\begin{figure*}
 
\fbox{\includegraphics[scale=0.3]{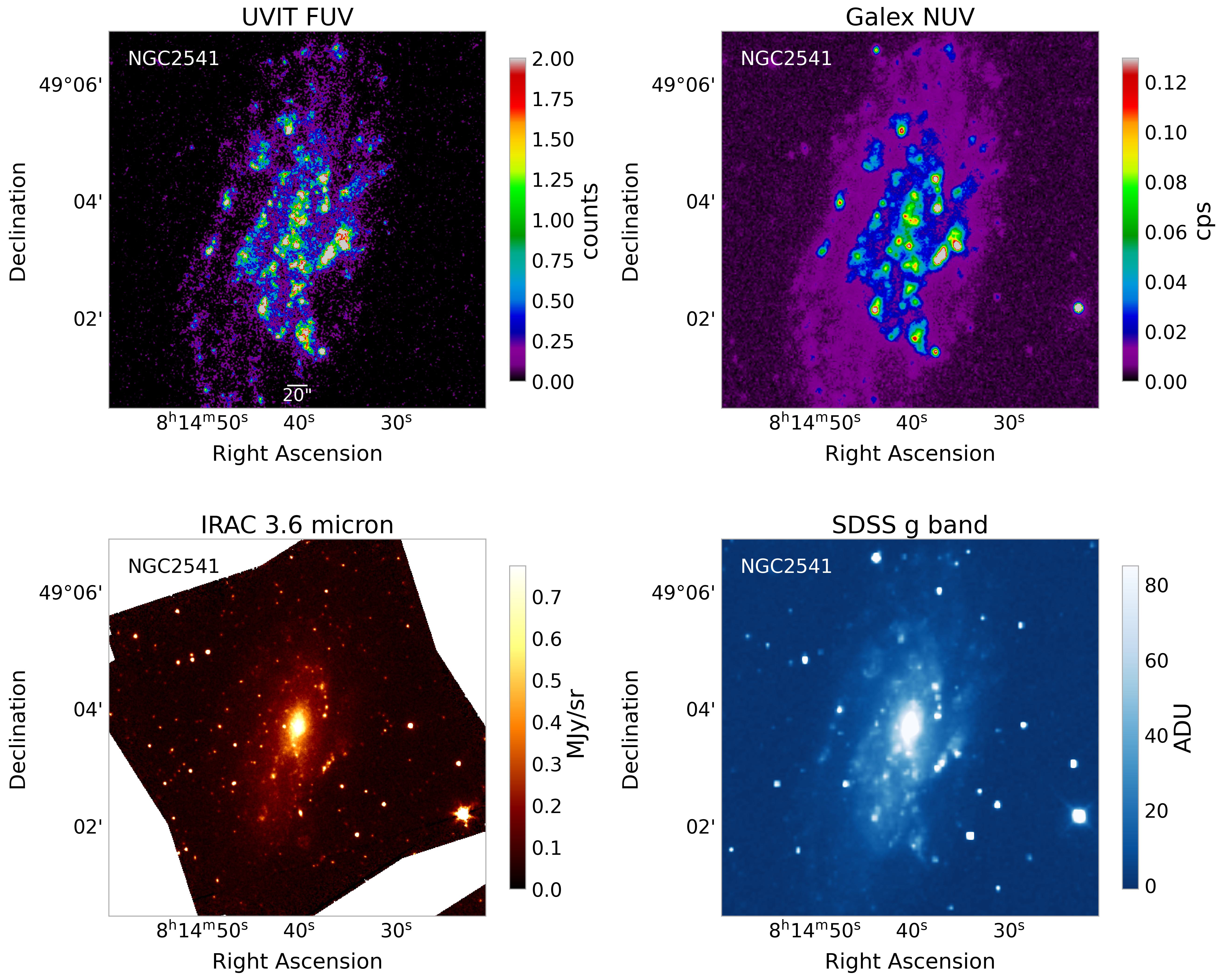}}
\end{figure*}

\vspace{0.2cm}

\begin{figure*}

\fbox{\includegraphics[scale=0.3]{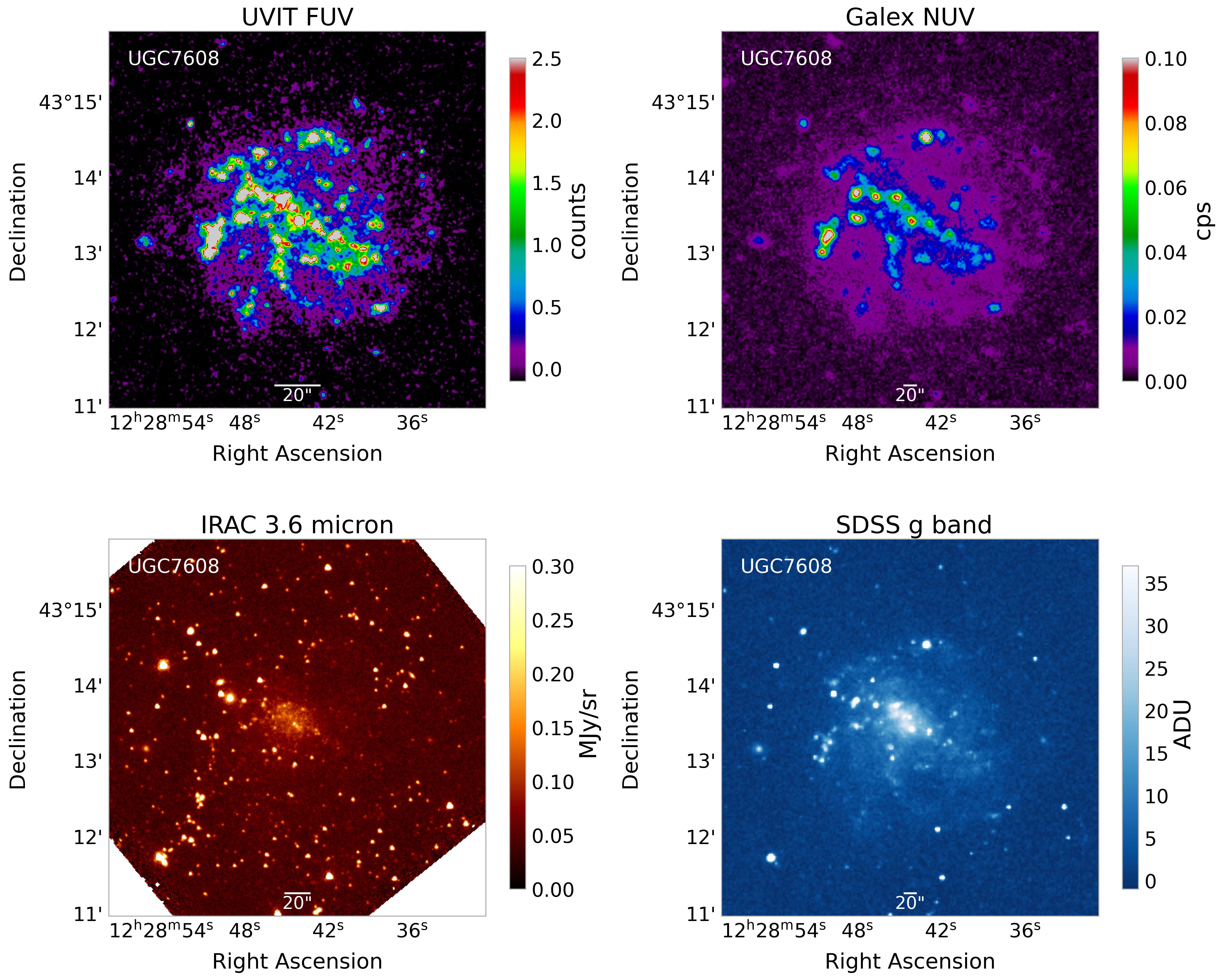}}
\end{figure*}

\vspace{0.2cm}

\begin{figure*}
 
\fbox{\includegraphics[scale=0.3]{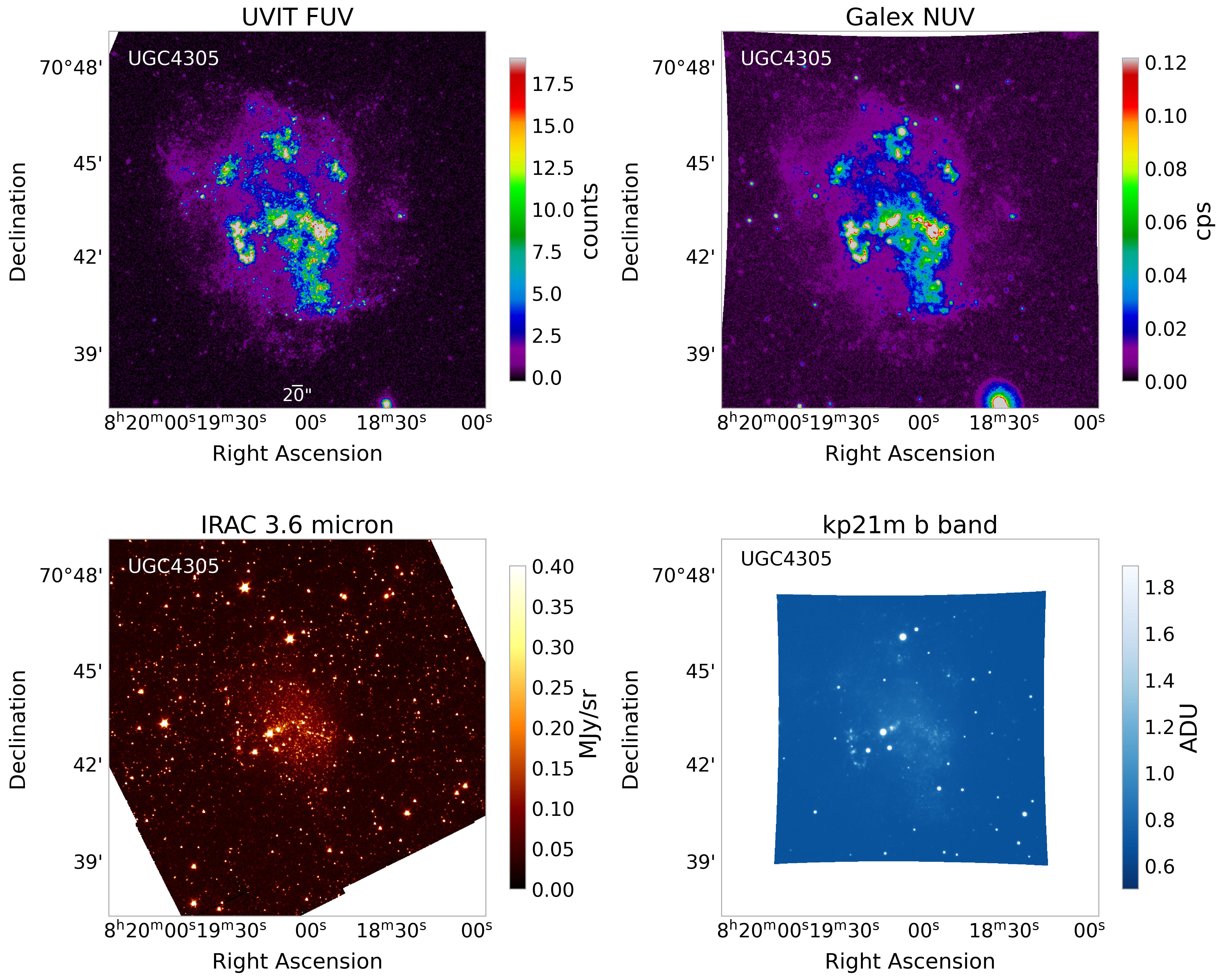}}
\end{figure*}

\vspace{0.2cm}

\begin{figure*}

\fbox{\includegraphics[scale=0.3]{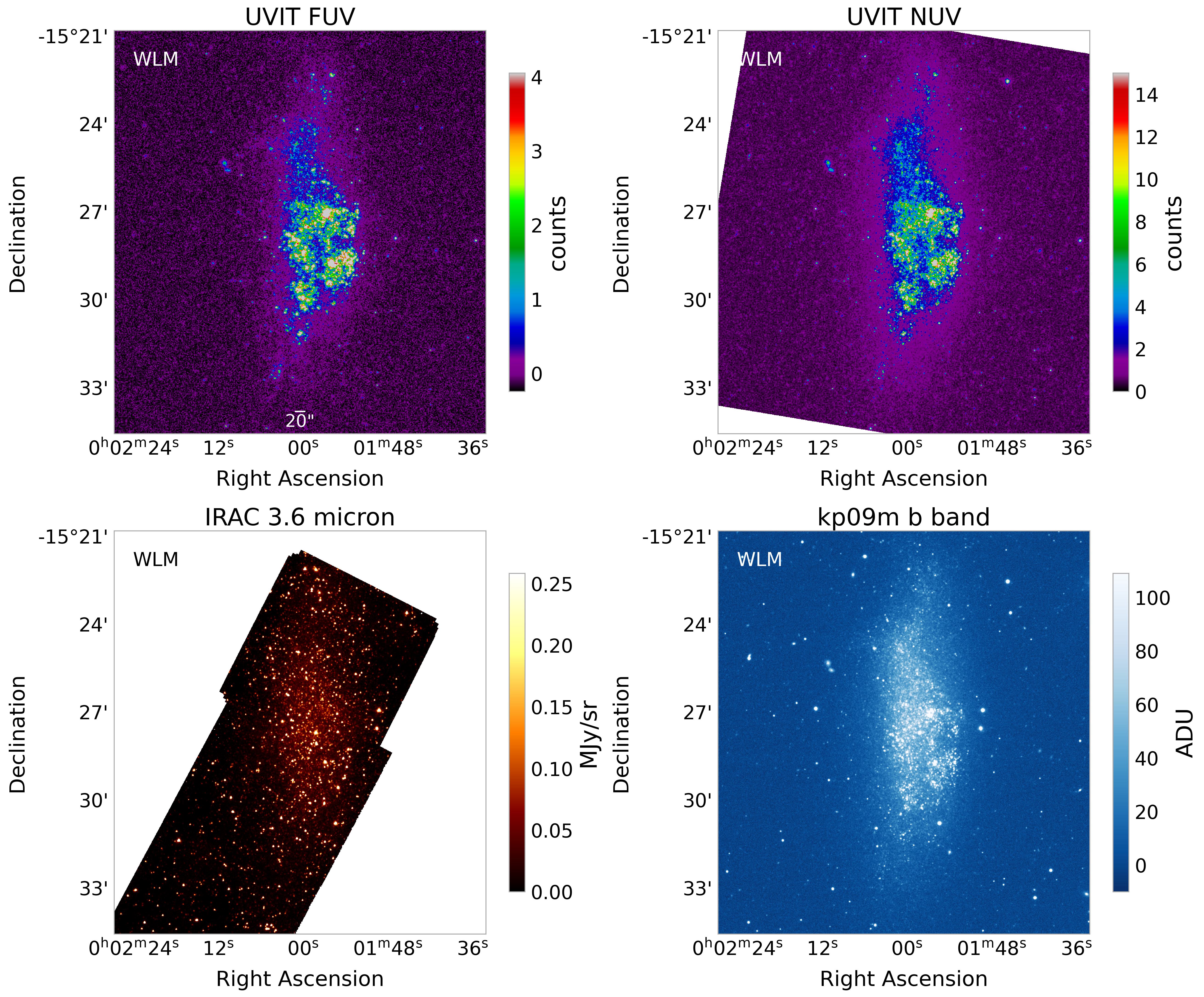}}
\end{figure*}

\vspace{0.2cm}

\begin{figure*}

\fbox{\includegraphics[scale=0.3]{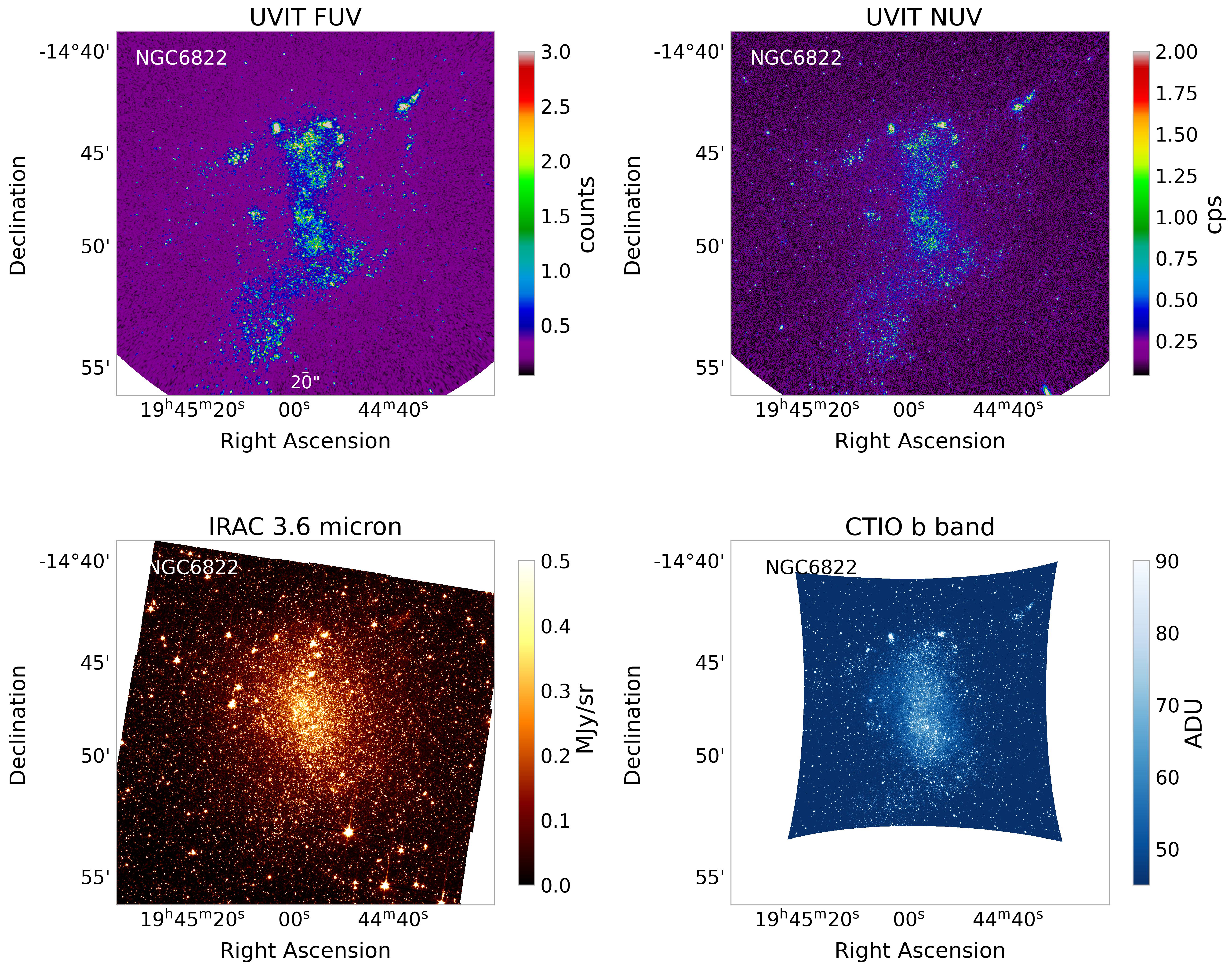}}
\end{figure*}

\vspace{0.2cm}

\begin{figure*}

\fbox{\includegraphics[scale=0.3]{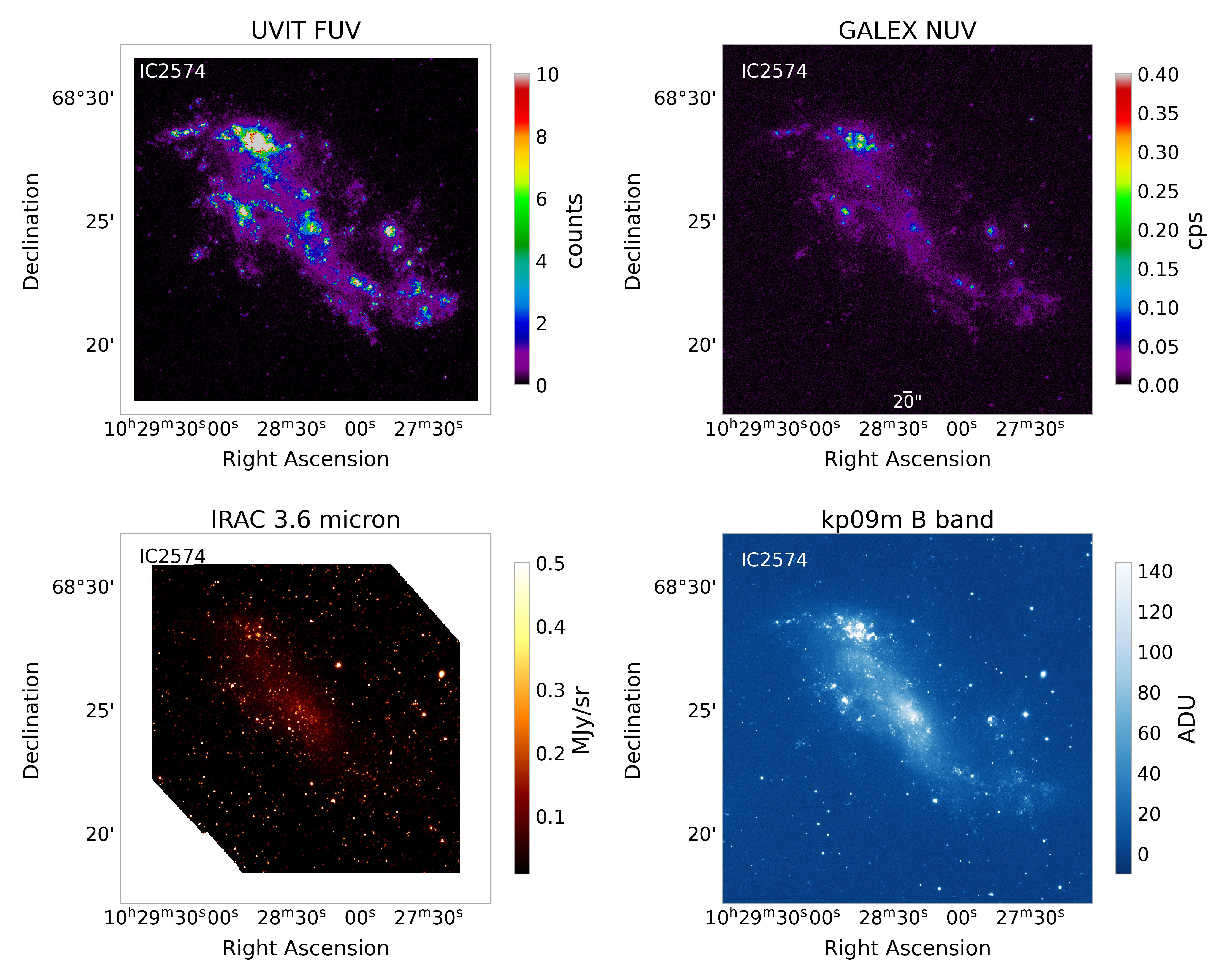}}
\end{figure*}

\vspace{0.2cm}

\begin{figure*}

\fbox{\includegraphics[scale=0.3]{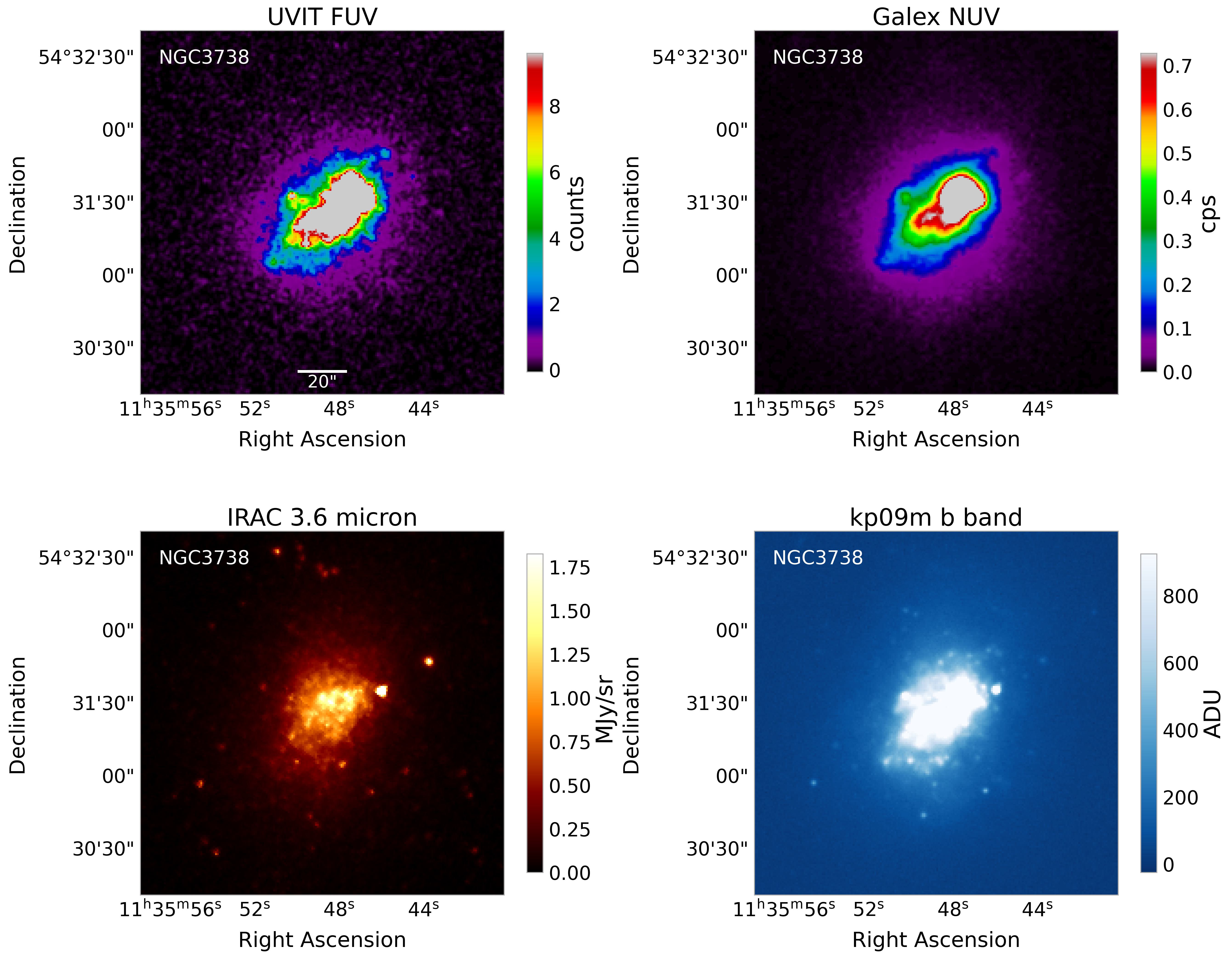}}
\end{figure*}

\vspace{0.2cm}

\begin{figure*}

\fbox{\includegraphics[scale=0.3]{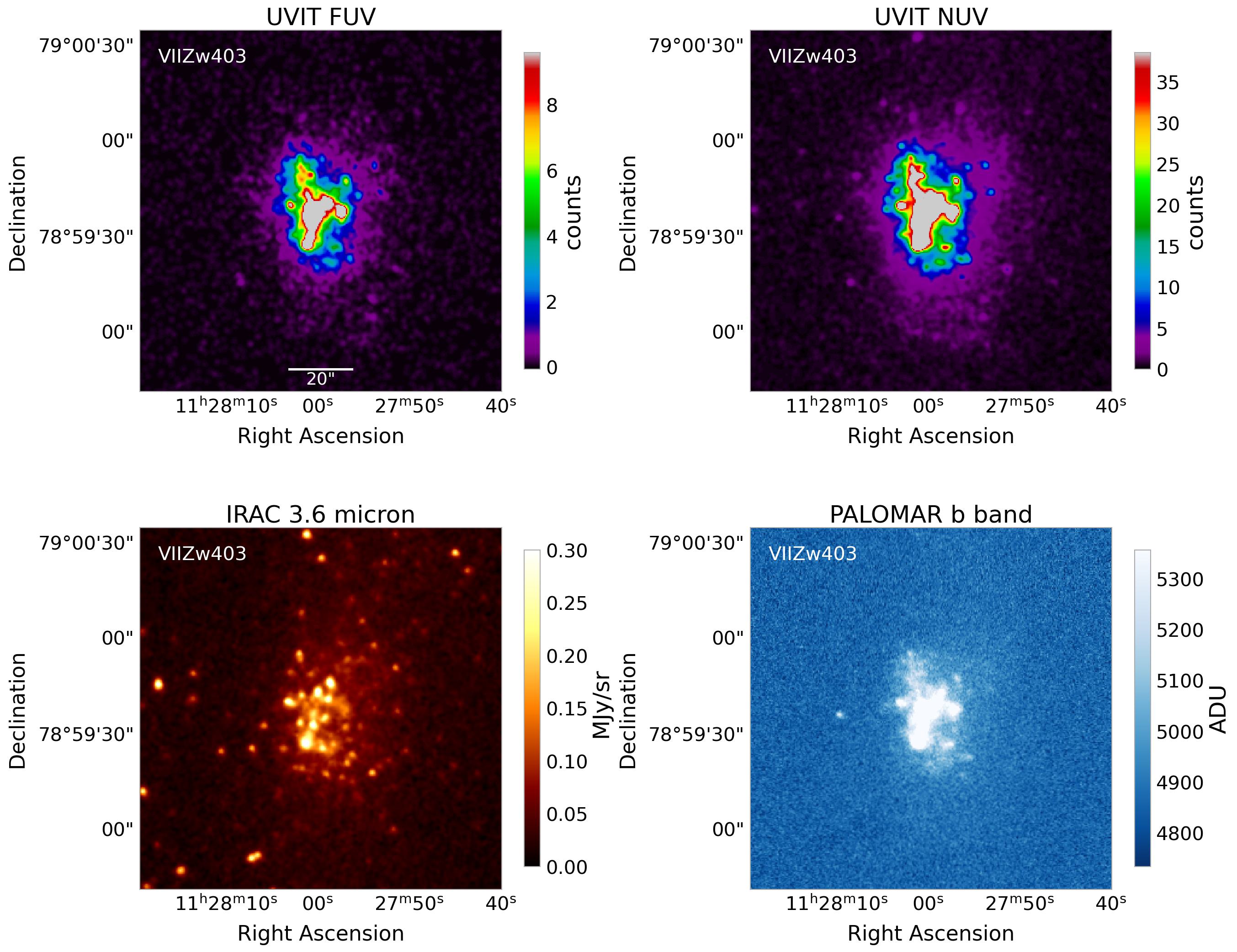}}
\end{figure*}

\vspace{0.2cm}

\begin{figure*}

\fbox{\includegraphics[scale=0.3]{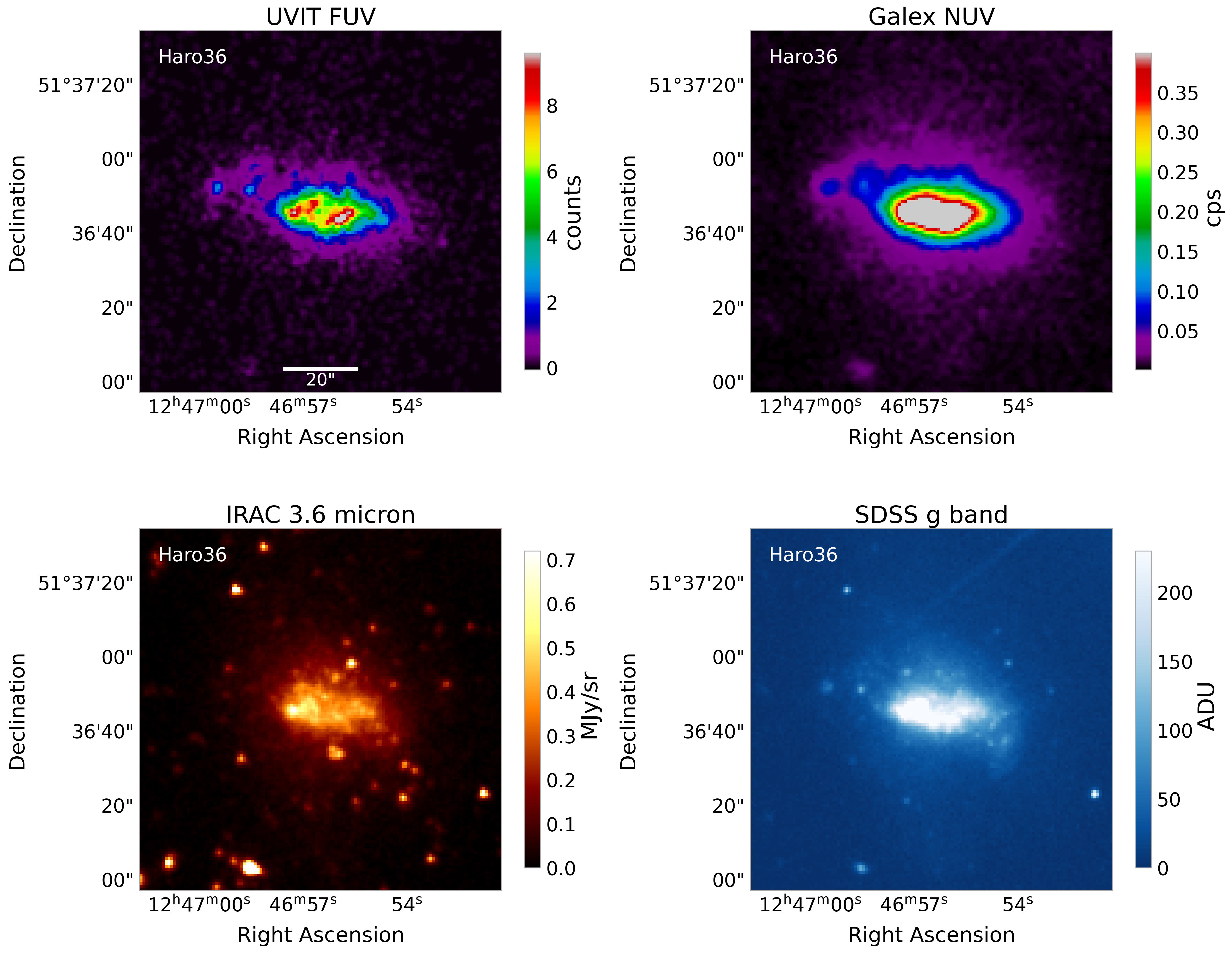}}
\end{figure*}

\vspace{0.2cm}

\begin{figure*}

    \fbox{\includegraphics[scale=0.3]{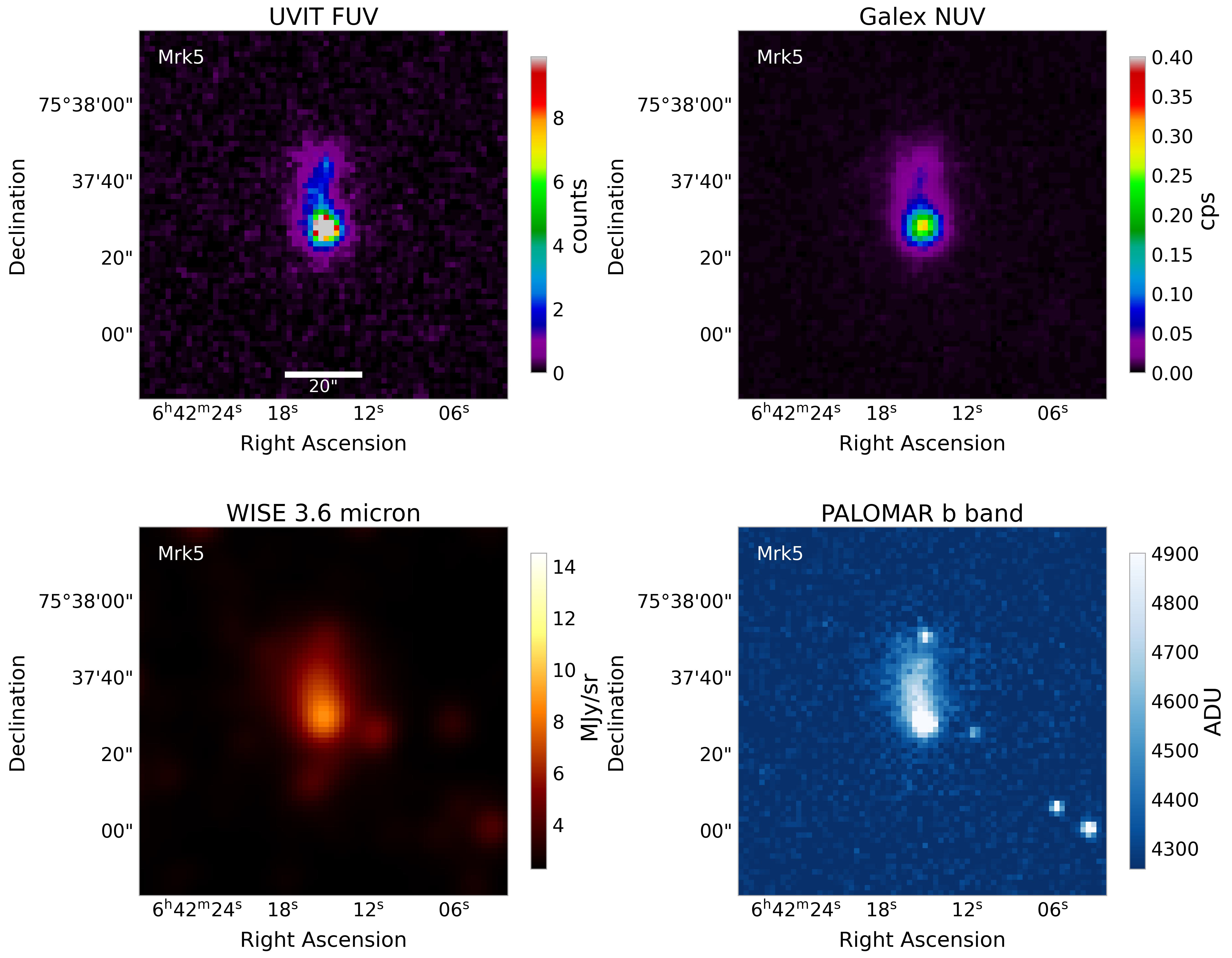}}
\end{figure*}



\bsp	
\label{lastpage}
\end{document}